\documentclass[a4paper,12pt,twoside]{article}

\usepackage[utf8]{inputenc}
\usepackage[greek,english]{babel}
\usepackage{graphicx}
\usepackage{lmodern}
\usepackage{mathptmx}
\usepackage{color}
\usepackage[linkbordercolor={0 0.8 0.8}]{hyperref}
\usepackage{listings}
\usepackage{xcolor}
\usepackage{subfigure}
\usepackage{lscape}
\usepackage{amsfonts}
\usepackage[explicit]{titlesec}
\usepackage{amsmath}
\usepackage{setspace}
\usepackage[subfigure]{tocloft}
\usepackage{fancyhdr} 
\usepackage{parskip}
\usepackage{titlesec}
\usepackage{makecell}
\usepackage{booktabs}
\usepackage{siunitx}
\usepackage{blindtext}
\usepackage{cite}
\usepackage[font=footnotesize,labelfont=bf,skip=12pt]{caption}

\hypersetup{
  citebordercolor=cyan,
  filebordercolor=cyan,
  linkbordercolor=cyan
}

\usepackage{geometry}
\numberwithin{equation}{section}

\colorlet{ctcolorchapterline}{cyan}
\colorlet{ctcolorchapternum}{cyan}

\titleformat{\section}[display]%
  {\normalfont\Large\bfseries}
  {\vspace{-8em}\raggedleft{%
    {\color{ctcolorchapterline}%
        \rule[-5pt]{2pt}{5cm}}\quad%
    {\color{ctcolorchapternum}
        \fontsize{60}{60}\selectfont\thesection}%
    }%
  }%
  {-2.1em}%
  {\parbox[b]{\dimexpr\textwidth-3em\relax}{\raggedright#1}}%
  [\phantomsection]

\begin{document}

\begin{titlepage}
\begin{center}
\Large\textbf{Synthesis of a Semiconducting 2D Material for Novel Strong PUFs} \\
\vspace*{18mm}
\includegraphics[width=6cm, height=6cm]{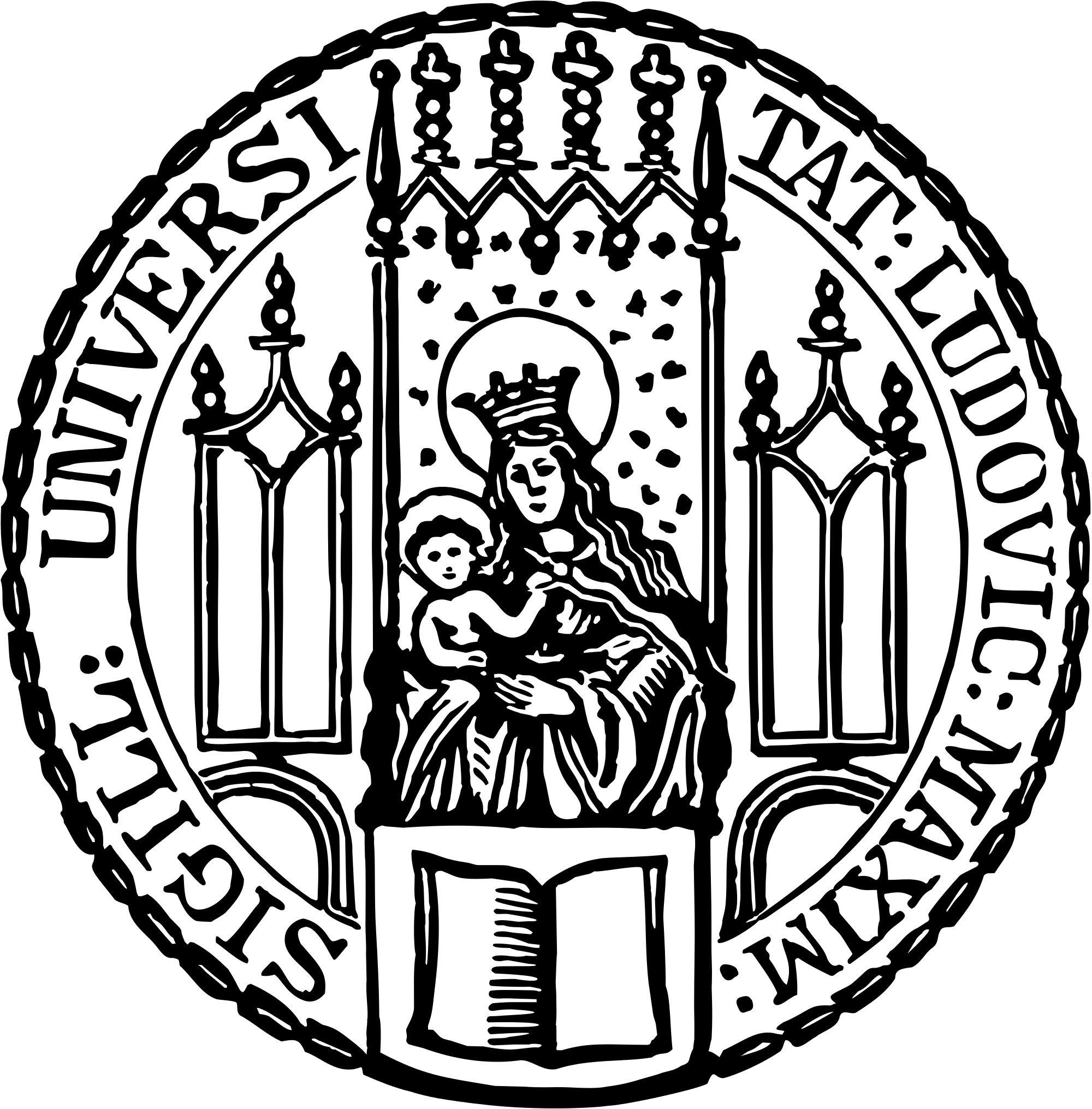} \\
\vspace*{18mm}
\includegraphics[width=10cm]{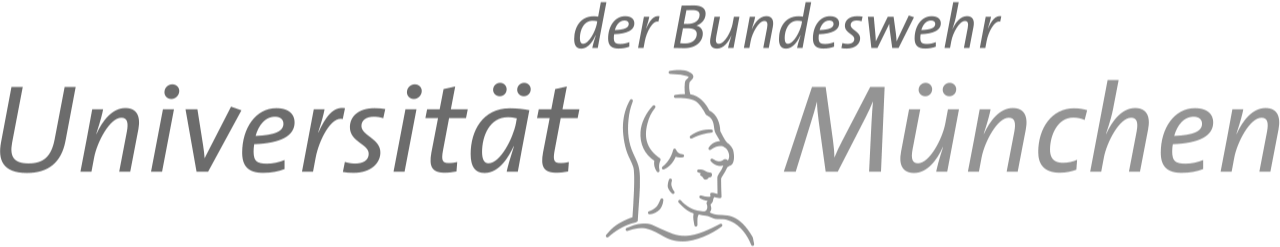}
\vspace*{18mm} \\
\noindent Master Thesis \\
\noindent Faculty of Physics
\vspace*{18mm} \\
\noindent Submitted by \\
\noindent \textbf{Peter Eder} \\
\noindent Munich, 04\textsuperscript{th} October 2022 \\
\vspace*{12mm}
\small{With minor additions and modifications in December 2022 before publication.}
\end{center}
\end{titlepage}

\clearpage
\newpage
\null
\thispagestyle{empty}
\clearpage
\newpage

\begin{titlepage}
\begin{center}
\Large\textbf{Synthese eines halbleitenden 2D-Materials für neuartige Strong PUFs} \\
\vspace*{18mm}
\includegraphics[width=6cm, height=6cm]{Images/LMU_Siegel.png} \\
\vspace*{18mm}
\includegraphics[width=10cm]{Images/UniBW_Logo-modified.png}
\vspace*{18mm} \\
\noindent Masterarbeit \\
\noindent Fakultät für Physik
\vspace*{18mm} \\
\noindent Vorgelegt von \\
\noindent \textbf{Peter Eder} \\
\noindent München, den 04. Oktober 2022 \\
\vspace*{12mm}
\small{Mit kleineren Ergänzungen und Modifikationen im Dezember 2022 vor Veröffentlichung.}
\end{center}
\end{titlepage}

\clearpage
\newpage

\pagenumbering{gobble}

\noindent Supervisor: Prof. Dr. Harald Weinfurter \\
\noindent Advisors: Prof. Dr. rer. nat. Georg Düsberg, M. Sc. Stefan Heiserer, Prof. Dr. Dr. Ulrich Rührmair \\

\clearpage
\newpage

\pagenumbering{Roman}

\section*{Abstract}
\addcontentsline{toc}{section}{Abstract}

The exponentially growing number of interconnected devices in the Internet of Things (IoT) poses an increasing amount of challenges to the field of cyber security and encryption. For authenticated use and communication, each device must securely store and apply secret keys without giving a potential attacker the possibility to access them neither by software nor hardware attacks. In 2002, the concept of Physical Unclonable Function (PUF) emerged\textsuperscript{\cite{Pappu.2002}} to counteract the inherent threats, associated with this trend. Physical disordered systems are excited by external stimuli and can thereby be uniquely identified. Since then, various PUF implementations have evolved, based on optical scattering\textsuperscript{\cite{Pappu.2002}}, electrical circuits\textsuperscript{\cite{Herder.2014}}, Carbon-nanotubes (CNT)\textsuperscript{\cite{Burzuri.2019,Moon.2019,Hu.2016,JonasSchroder.2022}}, or, for instance, two-dimensional (2D) materials.\textsuperscript{\cite{Im.2021,Cao.2017,Dodda.2021,Dimitrakopoulosetal.2014,Shao.2021}}

2D materials have profoundly different characteristics when isolated from their bulk form to few- or monolayers. Mechanical properties, such as strength, flexibility, and thermal conductivity, as well as electrical properties, such as a tunable bandgap, high mobility, or strong light-matter interaction, can be exploited. Transition metal dichalcogenides (TMDs) represent a potential PUF candidate since, unlike graphene, they feature a bandgap and are thus semiconducting. Furthermore, a random percolation network can be formed from them, which is expected to possess an electrical response that is uniquely identifiable yet unpredictable.

This work has built the foundation for implementing a PUF consisting of a TMD, namely tungsten disulfide (WS\textsubscript{2}). The material has been synthesized using chemical vapor deposition (CVD) in the close proximity approach. In order to fabricate a large-scale interconnected surface consisting of WS\textsubscript{2}, the growth parameters are selected and enclosed by a preliminary trial and error optimization and ultimately optimized by Design of Experiments (DoE). This yields a potentially available PUF area of several millimeters, which was then patterned and contacted with initial trials.

\clearpage
\newpage
\null
\thispagestyle{empty}

\clearpage
\newpage

\section*{Zusammenfassung}
\addcontentsline{toc}{section}{Zusammenfassung}

Die exponentiell wachsende Anzahl an miteinander verbundener Geräte im Internet of Things (IoT) stellt den Sektor der Cybersecurity vor immer größer werdende Herausforderungen. Für eine authentifizierte Nutzung und Kommunikation muss jedes Gerät geheime Schlüssel sicher speichern und verarbeiten, ohne dass ein potenzieller Angreifer die Möglichkeit hat, auf diese zuzugreifen, weder durch Software- noch durch Hardware-Angriffe. Im Jahr 2002 entstand das Konzept der Physical Unclonable Function (PUF)\textsuperscript{\cite{Pappu.2002}}, um den mit diesem Trend verbundenen Bedrohungen entgegenzuwirken. Physikalisch ungeordnete Systeme werden durch äußere Reize angeregt und können dadurch eindeutig identifiziert werden. Seitdem wurden verschiedene PUF-Implementierungen entwickelt, die auf optischer Streuung\textsuperscript{\cite{Pappu.2002}}, elektrischen Schaltkreisen\textsuperscript{\cite{Herder.2014}}, Carbon-Nanotubes\textsuperscript{\cite{Burzuri.2019,Moon.2019,Hu.2016,JonasSchroder.2022}} (CNT) oder beispielsweise zwei-dimensionalen (2D) Materialien\textsuperscript{\cite{Im.2021,Cao.2017,Dodda.2021,Dimitrakopoulosetal.2014,Shao.2021}} basieren.

2D-Materialien haben völlig unterschiedliche Eigenschaften, wenn sie von ihrer Bulkform isoliert werden und in wenigen Schichten oder als Monolagen vorliegen. Mechanische Eigenschaften wie Festigkeit, Flexibilität und Wärmeleitfähigkeit sowie elektrische Eigenschaften wie eine veränderbare Bandlücke, hohe Mobilität oder eine starke Licht-Materie-Wechselwirkung können ausgenutzt werden. Übergangsmetalldichalcogenide sind ein potenzieller PUF-Kandidat, da sie im Gegensatz zu Graphen eine Bandlücke aufweisen und somit halbleitende Eigenschaften vorweisen. Außerdem kann aus ihnen ein zufälliges Perkolationsnetzwerk gebildet werden, dessen elektrisches Signal eindeutig identifizierbar, aber nicht vorhersagbar ist.

In dieser Arbeit wurde die Grundlage für die Implementierung eines PUF geschaffen, der aus einem TMD, nämlich Wolframdisulfid (WS\textsubscript{2}), besteht. Das Material wurde durch eine spezielle Art der Gasphasenabscheidung, die sogenannte Close Proximity Methode, synthetisiert. Um eine großflächig vernetzte Oberfläche aus WS\textsubscript{2} herzustellen, werden die Wachstumsparameter durch eine vorläufige Trial-and-Error-Optimierung eingegrenzt und schließlich durch ein Design of Experiments (DoE) optimiert. Daraus ergibt sich eine potenziell verfügbare PUF-Fläche von mehreren Millimetern, die dann mit ersten Versuchen strukturiert und kontaktiert wurde.

\clearpage
\newpage
\null
\thispagestyle{empty}
\clearpage
\newpage
\null

\section*{Acknowledgements}
\addcontentsline{toc}{section}{Acknowledgements}

First of all, I would like to thank my supervisor Stefan Heiserer for his constant support, passion, encouragement, brilliant suggestions, and the work that we have done together every day. Also thanks for all the useful things that you have taught me. Without you, the results of this work would not have nearly been as good as they are now and my future plans wouldn't be as clear.

Thank you Georg Düsberg for the opportunity to join your group and work on this interesting topic. Thank you for all the valuable advice, constant encouragement, and discussions throughout my time in the group. I really like how down-to-earth you are about science, which I think is the best approach to making a difference on the one hand and treating people well on the other.

I would also like to thank Prof. Ulrich Rührmair for all the discussions about PUFs, and the constant support that you gave me on my thesis. In addition, you brought me to this interesting topic that combines nanophysics and cryptography, both of which are ideal topics for me. I think you are one of the most pleasant professors for a student because the way you explain things is extremely intuitive and understandable, and because you always respond quickly to emails and other things and take your time even though you are very occupied.

Also thanks to my friends, the whole Düsberg group, for your helpfulness, pieces of advice, and sympathy. Even outside of working hours it was nice to do sports or otherwise spend time with you. Please all stay as you are.

Last but not least, thank you to my family and to my girlfriend Sofia Eyzaguirre, who has accompanied me through all this time, always motivating and cheering me up. Thank you for keeping my back free and for your support.

\clearpage
\newpage
\null
\thispagestyle{empty}
\clearpage
\newpage
\null
\thispagestyle{empty}

\tableofcontents

\clearpage
\newpage

\listoffigures

\clearpage
\newpage

\listoftables

\clearpage
\newpage
\null
\thispagestyle{empty}
\clearpage
\newpage

\section*{Abbreviations}
\addcontentsline{toc}{section}{Abbreviations}

\begin{tabular}{ll}
\textbf{IoT} & Internet of Things \\
\textbf{PUFs} & Physical Unclonable Functions \\
\textbf{CNT} & Carbon Nanotube \\
\textbf{2D} & Two-Dimensional \\
\textbf{TMD} & Transition Metal Dichalcogenide \\
\textbf{WS\textsubscript{2}} & Tungsten disulfide \\
\textbf{CVD} & Chemical Vapor Deposition \\
\textbf{NVM} & Non-Volatile Memory \\
\textbf{AES} & Advanced Encryption Standard \\
\textbf{CMOS} & Complementary Metal-Oxide Semiconductor \\
\textbf{DoE} & Design of Experiments \\
\textbf{PL} & Photoluminescence \\
\textbf{RGO} & Reduced Graphene Oxide \\
\textbf{CRPs} & Challenge-Response Pairs \\
\textbf{Si} & Silicon \\
\textbf{MUX} & Multiplexer \\
\textbf{FET} & Field-Effect Transistor \\
\textbf{VdW} & Van-der-Waals \\
\textbf{DFT} & Density Functional Theory \\
\textbf{ME} & Mechanical Exfoliation \\
\textbf{LPE} & Liquid Phase Exfoliation \\
\textbf{ALD} & Atomic Layer Deposition \\
\textbf{TAC} & Thermally Assisted Conversion \\
\textbf{PVD} & Physical Vapor Deposition \\
\textbf{VW} & Volmer-Weber \\
\textbf{SK} & Stranski-Krastanov \\
\textbf{FvdM} & Frank-van-der-Merwe \\
\textbf{HPLC} & High-Performance Liquid Chromatography \\
\textbf{IPA} & Isopropanol \\
\textbf{DI} & Distilled \\
\textbf{N\textsubscript{2}} & Nitrogen gas \\
\textbf{PLT} & Photolithography \\
\textbf{EBL} & Electron Beam Lithography
\end{tabular}

\begin{tabular}{ll}
\textbf{SEM} & Scanning Electron Microscopy \\
\textbf{FWHM} & Full Width at Half Maximum \\
\textbf{AFM} & Atomic Force Microscopy \\
\textbf{IDE} & Interdigitated-electrode \\
\textbf{Ni} & Nickel \\
\textbf{Au} & Gold \\
\textbf{LBM} & Layer Breathing Mode \\
\textbf{SM} & Shearing Mode \\
\textbf{PL} & Photoluminescence
\end{tabular}

\clearpage
\newpage

\clearpage
\newpage
\pagenumbering{arabic}

\pagestyle{fancy} 
\fancyhf{} 
\fancyhead[LO,RE]{\leftmark}
\thispagestyle{plain} 
\fancyfoot[C]{\thepage} 

\section{Introduction \& Overview}

\subsection{Internet of Things and Hardware Security}

Our modern societies are increasingly striving for digital connectivity, giving rise to megatrends, such as the IoT. From a cryptographic perspective, the ever-growing number of devices to deal with is becoming more and more challenging. It is estimated that the number of connected devices will increase to 31 billion by 2025.\textsuperscript{\cite{SatyajitSinha.2021}} In this era of swift digitization, cryptographic encryption techniques and authentication schemes are more important than ever.

Security nowadays is mainly based on a secret key - a random digital bitstring - stored in non-volatile memory (NVM). In theory, secret keys should not be accessible to any adversary. The widely used Advanced Encryption Standard (AES) for instance, processes its input and private key by byte substitutions, permutations, and linear transformations to encrypt data.\textsuperscript{\cite{Nechvatal.2001}} Despite the fact that this algorithm is not information-theoretically secure, no known attack is practically feasible regarding today's computing power.\textsuperscript{\cite{Bogdanov.2011}} While several protocols are currently considered secure, it must be assumed that the devices can permanently store and employ a digital key that can not be retrieved by a potential attacker - neither through software nor hardware attacks.

However, reality shows that this assumption is not always fulfilled. Physical attacks such as invasive, semi-invasive, or side-channel attacks can lead to data exposure with intellectual or economic value or result in counterfeits.\textsuperscript{\cite{Anderson1997LowCA,260552,Skorobogatov2005SemiinvasiveAA}} For a high level of security, the hardware needs tamper-sensing circuitry with continuous power supply. This is expensive and difficult to implement, due to the spatial scarcity in mobile and resource-limited devices such as RFIDs.\textsuperscript{\cite{4261134}}

Around two decades ago, the idea of Physical Unclonable Functions emerged to solve the aforementioned challenges. This new hardware-based security primitive does not rely on known computationally difficult problems but physics. The main ingredient is the random physical disorder that results from uncontrollable manufacturing variations on microscopic length scales. The goal is to fabricate unique systems that are unclonable - even for the original manufacturer. PUFs provide low-cost authentication regarding economy, power consumption, and space. Furthermore, they are able to generate cryptographic keys, which are only present when needed, avoiding the necessity of NVM. An attacker is additionally challenged by the high tamper sensitivity, where an invasive attack will lead to a variation of the resulting responses.\textsuperscript{\cite{Burzuri.2019,Moon.2019,BautistaAdames.2016,4261134,Herder.2014,6800561,Kumar.2019,Hu.2016,Pappu.2002,Scholz.2020,Ruhrmair.2011,jaeger2010random,Ruhrmair.2010,Chen.2011}}

\subsection{Goals and Contributions of this Thesis}

The goal of this thesis is to work towards an implementation of a novel semiconductor-based Strong PUF. It is based on a 2D TMD, namely WS\textsubscript{2}. A random percolation network from mono- to few-layer WS\textsubscript{2} is created and the first electrical measurements are conducted. This thesis focuses on the design and manufacturing of the PUF, while \cite{PatrickSchuster.2022} starts to measure and evaluate it. The main contributions of this thesis are listed in the following:

\begin{itemize}
 \item Benchmarking of state-of-the-art 2D PUFs in chapter \ref{sec:StateOfTheArt}.
 \item Introduction of a novel WS\textsubscript{2} CVD growth strategy by patterning the sputtered metal-precursor and investigating the influences of the pattern, which can be found in figure \ref{fig:4.2_MasksResults}.
 \item Investigation of the inter-sample distance behavior of WS\textsubscript{2} growth in close proximity CVD approach in section \ref{sec:DistanceBehaviour}.
 \item Successful optimization of WS\textsubscript{2} growth on the order of milimeters by performing a Design of Experiments. The results can be found in chapter \ref{sec:5.3_Results}.
 \item Setting all the fundamental building blocks for the implementation of the first-ever WS\textsubscript{2} PUF prototype. A sketch of the process steps is provided in \ref{sec:PUFFabricationSteps}.
\end{itemize}

This thesis is divided into four main sections. The two following chapters \ref{sec:CryptographicBackground} and \ref{sec:FabricationalBackground} deal with the cryptographic background and the theory of the PUF design. It is followed by a part about the experimental methods in \ref{sec:ExperimentalAnalyticalMethods}, used to achieve the results finally presented in chapter \ref{sec:ResultsDiscussion}.

\clearpage
\newpage
\thispagestyle{plain}

\section{Cryptographic Background}
\label{sec:CryptographicBackground}

The design of a 2D material requires two basic building blocks. First, that of cryptography for the basic principles, and second, solid state physics for fabrication. This chapter deals with the theory on cryptography.

\subsection{Physical Unclonable Functions}

In 2002, Pappu et al.\textsuperscript{\cite{Pappu.2002}} published the design of the first cryptographic device that exploits physical disorder. The apparatus consisted of random optical speckle patterns, and they titled it, due to its non-invertible input-output behavior, "physical one-way function". In the meantime, the term Physical Unclonable Function has gained acceptance in literature, and numerous ideas and realizations have emerged.

\begin{figure}[b!]
\centering
\includegraphics[width=13cm]{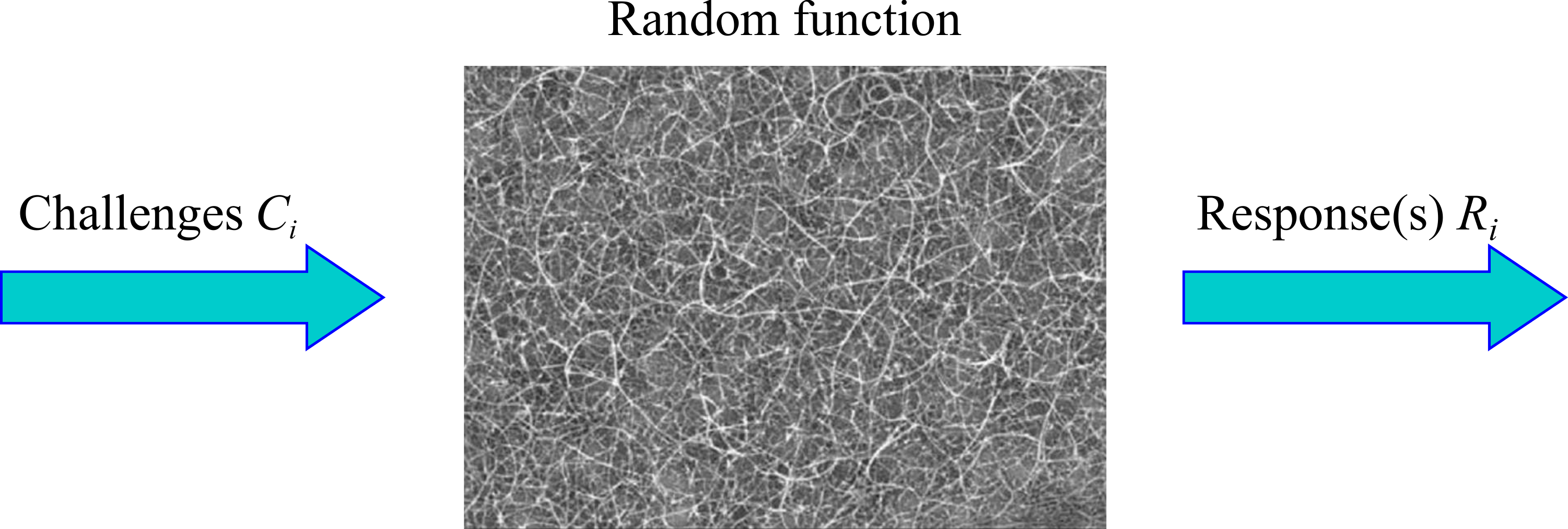}
\caption[The working principle of a PUF]{The working principle of a PUF.\textsuperscript{\cite{Cheng.2011}} External challenges $C_i$, in this case, electrical impulses, stimulate a random CNT percolation network - $f_{PUF}(C_i)$ - causing unpredictable responses $R_i$.}
\label{fig:2.1_PUF_Priciple}
\end{figure}

As illustrated in figure \ref{fig:2.1_PUF_Priciple}, a PUF consists of a disordered physical system, whose inputs are external stimuli or challenges $C\textsubscript{i}$, leading to one or multiple corresponding response(s) $R\textsubscript{i}$, where $i$ denotes the index for the challenge and the corresponding response. The responses can be, for example, continuous or currents or optical interference patterns, which require conversion into a bitstring by a selected algorithm to generate a secret key.

\newpage

The tuples $(C\textsubscript{i}, R\textsubscript{i})$ are called challenge-response pairs (CRPs) and can be mathematically linked via
\begin{align}
 R_i = f_{PUF}(C_i)
\end{align}
with some PUF function $f_{PUF}$, which is given by the physical disorder. On the one hand, this system must be irreproducible, even for the original manufacturer. On the other hand, the output needs to be stable under various environmental conditions, such as different temperatures, humidities, or simply after multiple measurements of the same CRP, which often requires error correction or soft-decision coding.\textsuperscript{\cite{6800561}}

As more and more PUF constructions have evolved over the years, the question arises of how to compare them to each other. A convenient variable to classify PUFs is the number of CRPs, which led to the categorization into Weak and Strong PUFs. Their distinctive properties result in different pros and cons in their applications.\textsuperscript{\cite{LukasZimmermann.06.11.2020}}

\textit{Weak} PUFs hold very few or, in the extreme case, precisely one challenge. This leads to the requirement that adversaries do not have access to the CRPs, even if they physically hold the PUF. The primary use of Weak PUFs is generating secret keys, which requires perfect error correction of possible noise in the output.\textsuperscript{\cite{6800561}} One of the most famous implementations is the SRAM PUF. It consists of two cross-coupled inverters, which are brought into an unstable state during power-up, dropping into a stable, preferred state when released.\textsuperscript{\cite{BautistaAdames.2016}}

\textit{Strong} PUFs, on the other hand, possess an ample, ideally exponentially-sized CRP space. This prevents a potential attacker from reading out the entire device if he gains physical access. There is no more necessity to assume an access-restricted response mechanism. The adversary can now collect vast amounts of CRPs, but will only reach a negligible percentage of the total space. However, this construction leads to the need for unpredictability, which means that if the attacker has read out a large sub-space, he should not be able to extrapolate all CRPs.\textsuperscript{\cite{6800561}}

A Strong PUF must meet several criteria, most of which counteract each other. On the one hand, it must be unclonable, unpredictable, and unique. On the other hand, it should also be stable, easy to evaluate and fabricate. On top of that, it needs to be lightweight, space and power-saving, and cheap. The art in the design of a PUF is to keep these properties in balance, thus producing an innovative cryptographic security primitive. The applications and implementations of Strong PUFs are discussed in the following two sections.

\newpage

\subsection{PUF-Based Protocols}
\label{sec:PUFBasedProtocols}
\subsubsection{Authentication and Identification}

\begin{figure}[b!]
\centering
\includegraphics[width=13cm]{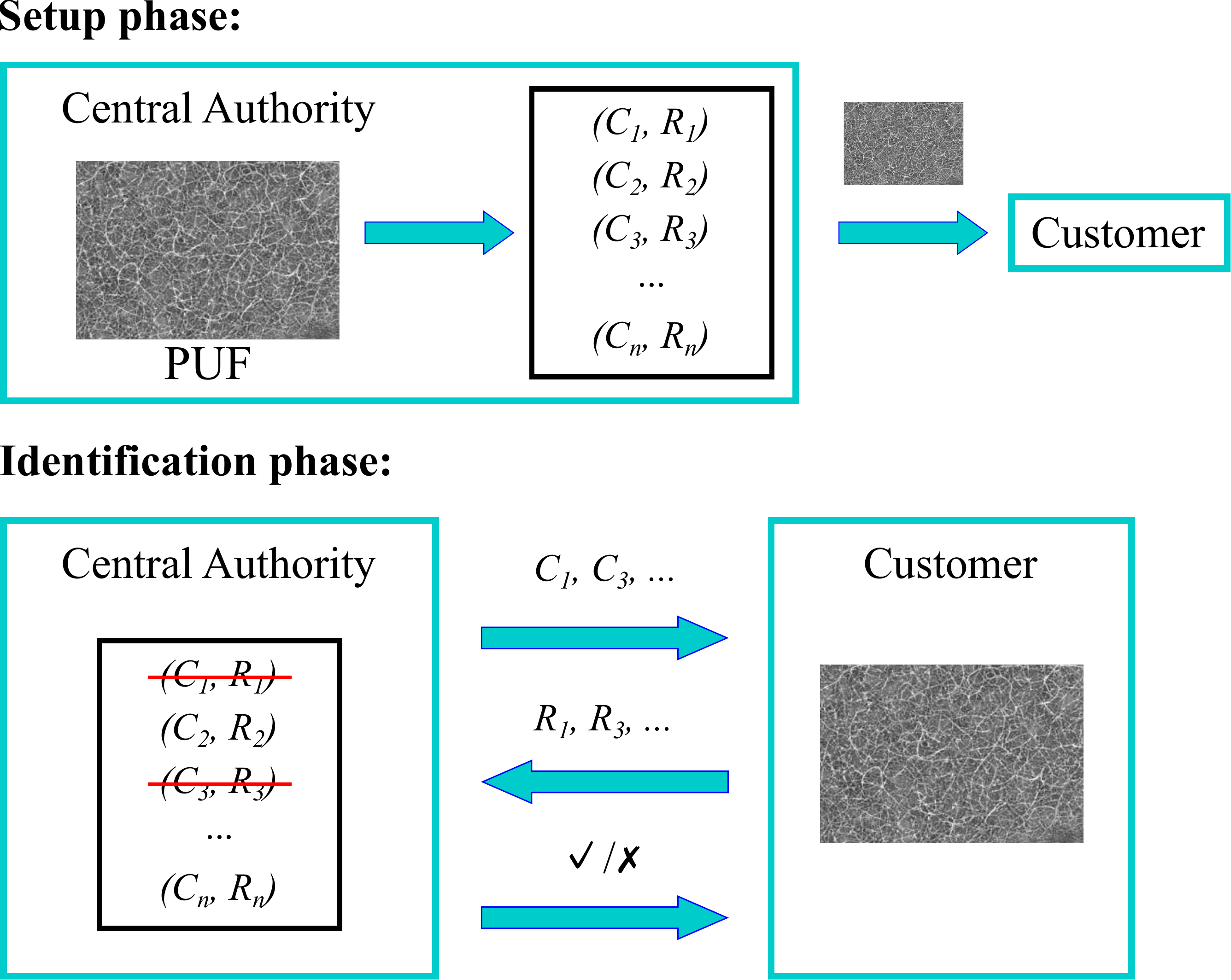}
\caption[Strong PUF identification protocol]{Strong PUF identification protocol, adapted from \cite{6800561}. The protocol is divided into a secure setup phase and an identification phase. In the former, a central authority stores a large number of CRPs in a list and subsequently sends the PUF to a customer. When the client utilizes the PUF, he or she can be uniquely authenticated by the protocol of the identification phase.}
\label{fig:2.2_PUF_Authentication}
\end{figure}

The main use cases of Strong PUFs are authentication and identification. To accomplish this, one can construct a protocol that is divided into a setup- and an identification phase, as described in \cite{Pappu.2002}. The individual steps are outlined in figure \ref{fig:2.2_PUF_Authentication}.

The scheme can best be illustrated using a smart card or credit card withdrawal as an example.\textsuperscript{\cite{6800561,Gassend.2002}} During a secure setup phase, a central authority, in this case, the bank, reads out a large number of CRPs from the PUF and stores them in a secret list. This PUF is then implemented into the card sent to the customer. A potential attacker can only intervene in the process flow once the credit card has been dispatched.

When the customer wants to withdraw money, the identification phase begins. He has to insert his card into the terminal, sending a request to the central authority over an authenticated channel. The bank responds with a sufficient number of challenges randomly selected from the list. Note that the necessary amount of challenges depends on the particular PUF implementation. If the PUF's response consists of only one bit, an adversary has a $50\,\%$ chance to guess the response correctly when only a single challenge is given. Thus, the response should comprise a long enough bit string. In the case of an optical PUF, a single challenge may be sufficient.

In the next step, the terminal applies the given challenges and returns the responses to the bank. If the received responses match those from the stored list, the customer is granted to charge his card. However, the CRPs used must be erased from the list since they are now potentially known to an adversary. Note that due to the demanded properties, an attacker cannot clone the PUF, read out all CRPs, or extrapolate and recreate them using machine learning algorithms.

This protocol provides an authentication mechanism without making unproven computational assumptions, such as the discrete logarithm problem\textsuperscript{\cite{WHITFIELDDIFFIEANDMARTINE.HELLMAN.}}, which is employed in public key exchange and assumed to possess no efficient solution, except for special cases. Furthermore, there is no need for a secret digital key stored permanently in the hardware, and there are no resources except the lightweight PUF on the card.

\clearpage
\newpage

\subsubsection{Key Exchange}

\begin{figure}[b!]
\centering
\includegraphics[width=15cm]{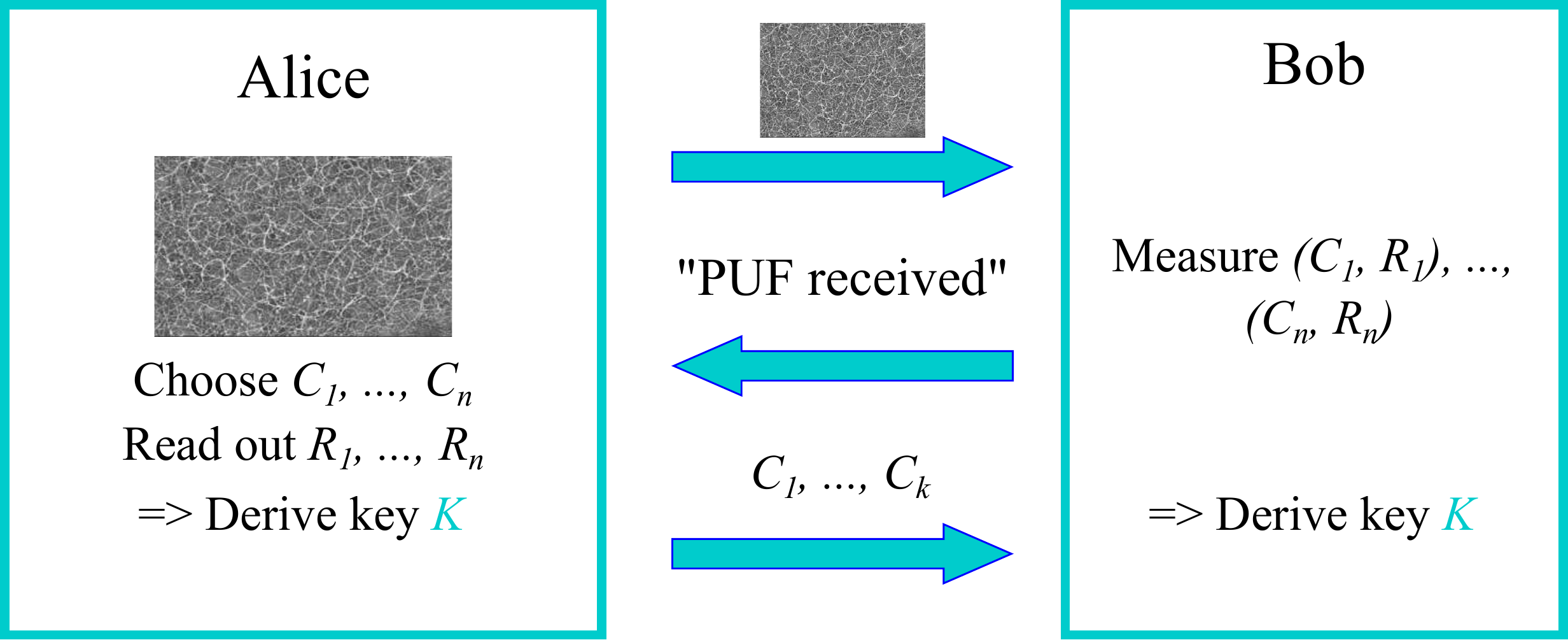}
\caption[Strong PUF key exchange protocol]{Strong PUF key exchange protocol, adapted from \cite{UlrichRuhrmair.18.01.2021}. Alice first derives a key $K$ based on arbitrarily chosen challenges $C_1$, ..., $C_n$. Next, she sends the PUF to Bob, who signals to her that he has received the PUF. Finally, Alice transmits the selected challenges and Bob is able to deduce the same key $K$.}
\label{fig:2.2_PUF_Key_Exchange}
\end{figure}

Besides authentication in the absence of a secret key, it is even possible to exchange a key with the help of Strong PUFs.\textsuperscript{\cite{UlrichRuhrmair.18.01.2021,ruhrmair2011physical}} The protocol is outlined in figure \ref{fig:2.2_PUF_Key_Exchange}. One party A, called Alice, holds the PUF at the beginning of the protocol. She chooses a large enough set of random challenges and measures their responses. Alice can then derive a key $K$ using any publicly known algorithm.

Subsequently, the PUF is handed over to party B, named Bob. At this stage, a potential attacker can theoretically gain access. Once Bob has received the PUF, he sends a signal to Alice over an authenticated channel. She then shares the selected challenges with him. Notice that an adversary now can also see the challenges.

Finally, Bob measures the necessary CRPs and can derive the same key $K$ from the responses and the publicly known algorithm. An attacker has no chance to generate the same key. While he possessed the PUF, he did not know the selected challenges. Furthermore, due to the properties of a Strong PUF, no cloning or complete read-out is possible.

In 2004, a key exchange protocol based on Strong PUFs was patented.\textsuperscript{\cite{M.vanDijk.}} It has further been shown in \cite{ruhrmair2010oblivious} that Strong PUFs can be used to implement an Oblivious Transfer, which makes it possible to securely evaluate any function computable in polynomial time without additional primitives. Formal security proofs and further protocol analyses have been provided in \cite{ruhrmair2013pufs,ruhrmair2010strong,ruhrmair2013practical,ruhrmair2012practical,van2014protocol,ruhrmair2016security,dachman2014feasibility,ostrovsky2013universally}.  Other PUF protocols and applications, including so-called Virtual Proofs of Reality and Sensor PUFs, have been given in \cite{ruhrmair2015virtual,ruhrmair2012simpl,ruhrmair2009simpl,ruhrmair2011simpl,ruhrmair2010towards,rosenfeld2010sensor}. 

\clearpage
\newpage

\subsection{State-of-the-Art on PUF implementation}

After illustrating the use cases of PUFs, the following section summarizes the most relevant PUF implementations.

\subsubsection{Optical PUF}

The first optical Strong PUF was implemented by Pappu et al.\textsuperscript{\cite{Pappu.2002}} in 2002 and set the foundation for PUF research. It uses imperfect speckle patterns as the source of entropy.

\begin{figure}[h!]
\centering
\includegraphics[width=15cm]{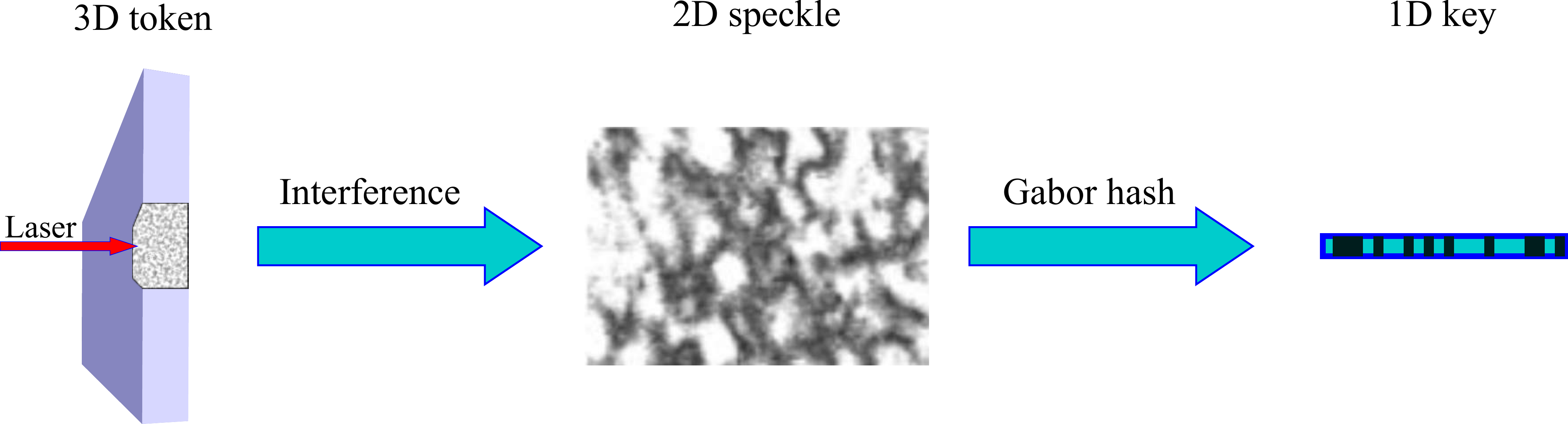}
\caption[Optical PUF working principle]{Optical PUF working principle, adapted from \cite{Pappu.2002}. The 2D interference pattern of a laser, scattered at a 3D token can be transformed into a 1D key by some hash algorithm.}
\label{fig:2.3_Optical_PUF}
\end{figure}

The mode of operation is shown in detail in figure \ref{fig:2.3_Optical_PUF}. A 3D token consisting of uncontrollably arranged glass spheres in epoxy is irradiated with a laser. This creates a 2D speckle pattern that can be transformed into a 1D key using, for example, a Gabor hash. The input challenge is a laser XY location and polarization, and the response is the associated speckle pattern. This pattern is strongly dependent on the input location/polarization since multiple scattering events occur inside the scattering medium. This principle can be used to implement an authentication mechanism as described in section \ref{sec:PUFBasedProtocols}.

The main advantage of the above optical PUF construction is that it has neither been cloned nor modeled\textsuperscript{\cite{ruhrmair2022secret,ruhrmair2020sok,ruhrmair2019towards}} and it is tough to deduce the speckle from the output. The 3D token only possesses a value of around $0.01\,\$$ and further exhibits a high tamper resistivity. Pappu et al. drilled $1\,mm$ deep into the epoxy token with a 533\textit{\,µm}-diameter drill and observed a strong alteration in the output behavior. Unfortunately, this also makes the device susceptible to irreversible accidental damage.

The optical PUF above has several drawbacks: It is hard to measure and calibrate. Besides, it is hardly chip-integratable in contrast to the Arbiter PUF in section \ref{sec:SiliconPUF}. Moreover, it possesses relatively few CRPs for a Strong PUF, making it potentially more vulnerable to full read-out attack.\textsuperscript{\cite{UlrichRuhrmair.18.01.2021}} As countermeasure, this quantity can be further scaled by varying the phase, wavelength, and amplitude. Furthermore, one could use a non-linear scattering medium and/or create an exponential CRP space.\textsuperscript{\cite{ruhrmair2013optical,ruehrmair2012method,horstmeyer2015physically,pavanello2021recent,ruehrmair2012method,eliezer2022exploiting,langhuth2011strong}}.   

\subsubsection{Silicon PUF}
\label{sec:SiliconPUF}

A more practical approach is to design a PUF that can be integrated directly on-chip with a conventional complementary metal-oxide-semiconductor (CMOS) process. It would be created in a way that makes it impossible to separate the PUF from the integrated circuit. An example of a silicon (Si) PUF is the Arbiter PUF \textsuperscript{\cite{suh2007physical}}.

\begin{figure}[h!]
\centering
\includegraphics[width=15cm]{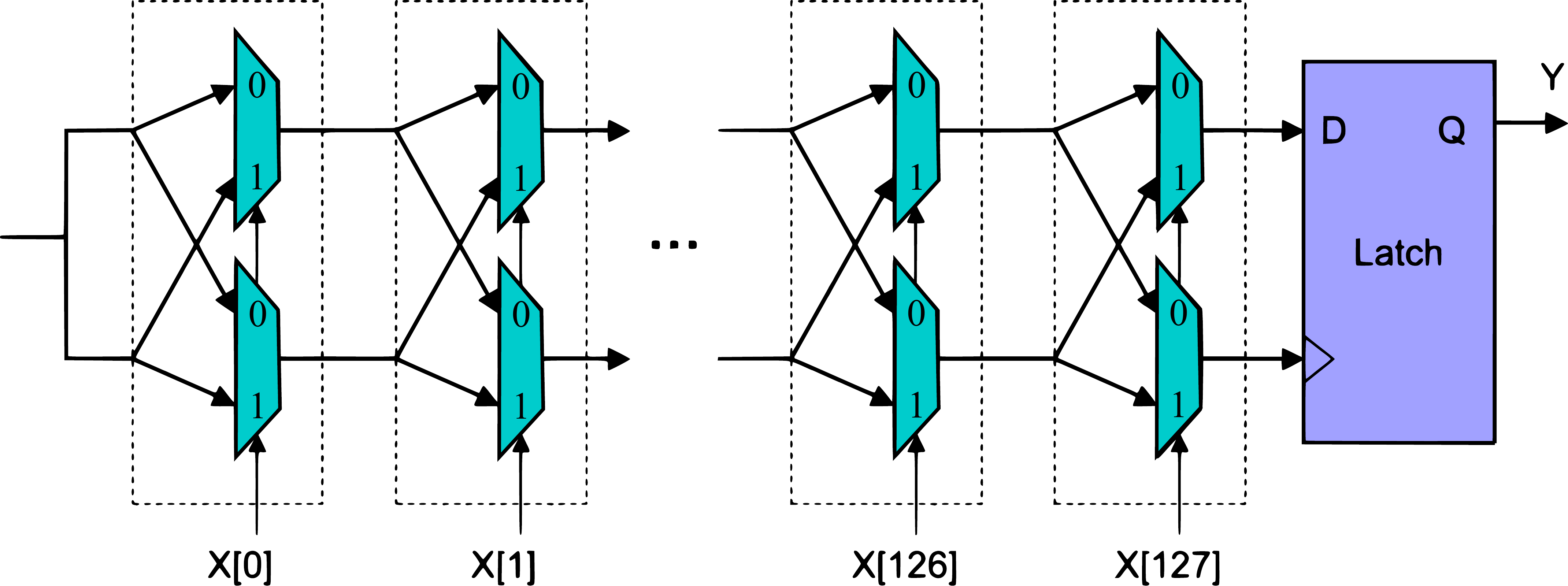}
\caption[Arbiter PUF working principle]{Arbiter PUF working principle, adapted from \cite{Herder.2014}. A signal is divided and inserted into a pair of MUXes, controlled by a challenge X[0]. The challenge decides whether the signals swap paths in the next step or not. The process is repeated until the signals reach the latch, which acts as an arbiter.}
\label{fig:2.3_Arbiter_PUF}
\end{figure}

It exploits manufacturing variations in the gate delay of multiplexers (MUXes) as the source of unclonable randomness. The setup is shown in figure \ref{fig:2.3_Arbiter_PUF}. Initially, a signal is divided and inserted into a pair of MUXes, controlled by a challenge $X[0]$. The challenge decides whether the signals swap paths in the next step or not. The process is repeated until the signals reach the latch, which acts as an arbiter. Thus, the output Y depends on which signal wins the race.

The security assumptions that must be met here are first that a manufacturer will not be able to control the gate delays and second, that an invasive attacker will vary the MUXes' properties when trying to measure them. Since the Arbiter PUF can relatively easily be broken by simple linear analysis\textsuperscript{\cite{LimDaihyun.2004,ruhrmair2009foundations,ruhrmair2010modeling,ruhrmair2013puf}}, several more complicated silicon Strong PUF architectures have been proposed \textsuperscript{\cite{chen2011bistable,jaeger2010random,lugli2013physical,csaba2009chip,csaba2010application,chen2011circuit,chen2009analog,sauer2017sensitized,gao2018efficient,jin2015playpuf,jin2017fpga,jin2020erasable,jin2022programmable,ruhrmair2010applications,majzoobi2010fpga,majzoobi2009techniques,majzoobi2008lightweight,xi2022provably,xi2017strong,vijayakumar2016machine,vijayakumar2015novel}}. For instance, the response of an arbiter PUF can be used as the input for a later MUX of another arbiter chain, or the outputs of several arbiter PUFs can be XOR'ed to complicate machine learning attacks.\textsuperscript{\cite{Herder.2014,ruhrmair2012security,ruhrmair2014pufs,suh2007physical}}  For completeness, it should be mentioned that also silicon Weak PUFs have been investigated intensively \textsuperscript{\cite{guajardo2007fpga,kumar2008butterfly,holcomb2008power,vskoric2005robust,tuyls2006read,kim2018dram,tehranipoor2016dram,maes2012pufky,rahman2014aro,xiao2014bit,cao2015low,lofstrom2000ic}}.  

\subsubsection{Carbon Nanotube PUF}

A more sophisticated PUF approach, which appears most similar to that of 2D materials, is by exploiting the randomness of CNT networks. Several devices have already been implemented and analyzed in the past.\textsuperscript{\cite{Burzuri.2019,Moon.2019,Hu.2016,JonasSchroder.2022}} The challenge is an electrical signal propagating through the network and generating an output current that requires conversion to a key.

\begin{figure}[h!]
\centering
\includegraphics[width=15cm]{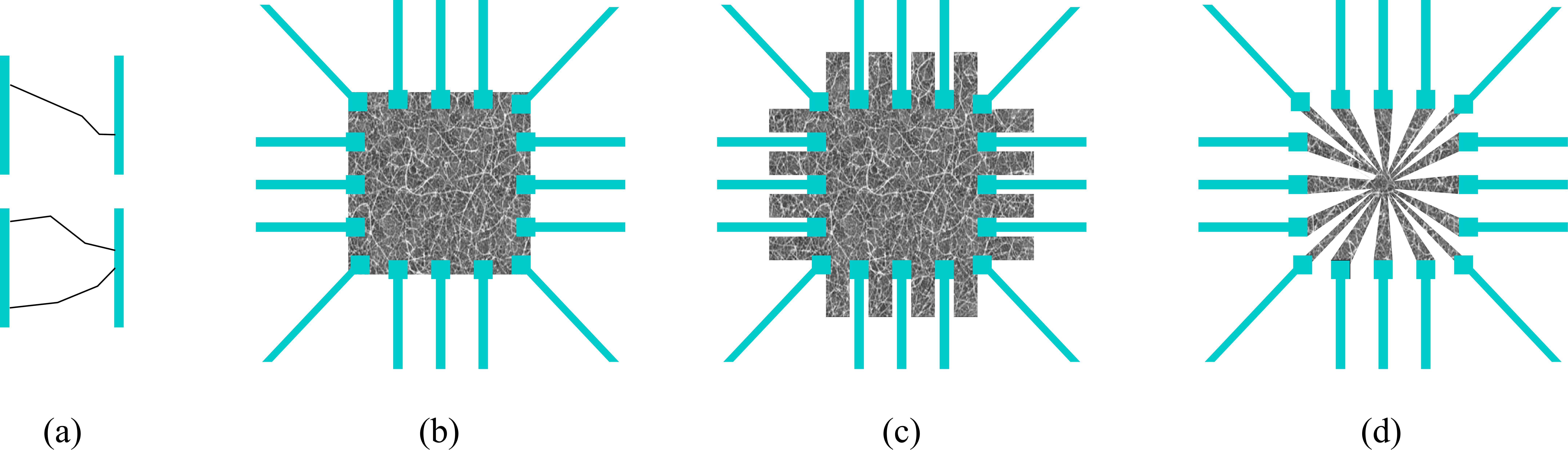}
\caption[CNT PUF structures]{CNT PUF structures. Simple contact pairs, connected by a few CNTs in (a).\textsuperscript{\cite{Burzuri.2019}} Square structure to enlarge the output space in (b).\textsuperscript{\cite{Moon.2019}} Optimized square structure to ensure that the adjacent contacts do not influence each other significantly in (c), adapted from \cite{JonasSchroder.2022}. A square structure with nearly equidistant contacts in (d).}
\label{fig:2.3_CNT_PUF}
\end{figure}

In \cite{Burzuri.2019}, CNTs were assembled by dielectrophoresis, which is an unpredictable liquid-phase-based fabrication method. The PUFs simply consist of multiple contact pairs, connected by a few on one to two CNTs, which can be seen in figure \ref{fig:2.3_CNT_PUF} (a).

Hu et al.\textsuperscript{\cite{Hu.2016}}, on the other hand, generated a random double-binary-key bit map. An array of self-assembled CNT networks with optimized contact distance is processed and evaluated according to the following threshold rule: no current between the two contacts: (00), semiconducting response: (01), metallic response: (11). By the division into three sections, they provide an elegant way to scale up the CRP space and establish a Strong PUF.

Yet another CNT-random-percolation network implementation can be found in \cite{Moon.2019}. All-printed CNT networks are placed on a flexible substrate. This network was tested for randomness and stability. The results reveal that the PUF is satisfyingly stable after $10\,k$ electrical measurements, different temperatures, under radiation exposure, and different light conditions. Furthermore, the PUF consisted of a square structure with several contacts around the outside, all connected to the CNT network in the center, as illustrated in figure \ref{fig:2.3_CNT_PUF} (b).

This approach allowed the CRP space to be scaled up further but may also result in linear behavior in the individual contacts. If the addition of the responses of two triggered contacts yields the same as the simultaneous driving of the contacts, then the PUF shows a linear behavior and is susceptible to machine learning attacks.\textsuperscript{\cite{ruhrmair2010modeling}} The more two resistors $R_1$ and $R_2$ follow the following formula for the addition of two parallel resistors $R_{||}$, the less favorable it is.

\begin{align}
  R_{||} = \frac{R_1 \cdot R_2}{R_1+R_2}
\end{align}

This problem was further investigated in \cite{JonasSchroder.2022} and different device structures were analyzed. The PUF also consisted of multiple contacts, all arranged around the center. This time though, the percolation network was patterned such that the adjacent contacts do not influence each other significantly. The optimized shape is depicted in figure \ref{fig:2.3_CNT_PUF} (c).

First, a median-based threshold voltage is set, and the responses, which in this case are resistances normalized to the distance between the contacts, are thus evaluated. Eventually, a secret key is extracted. In addition, a double-gated CNT-PUF was fabricated to scale the CRP space further. It possesses a conventional gate and two layers of CNTs on top, one of which serves as an additive gate. However, this PUF suffered from electrical and mechanical instability.

In a final step, the device structure could be modified so that the network is star-shaped around the center and the contacts all have approximately the same distance to each other, as shown in \ref{fig:2.3_CNT_PUF} (d). This can easily be extended to 2D materials that also form a random percolation network.

\subsubsection{2D-Material-Based PUFs}
\label{sec:StateOfTheArt}

An intent to further increase the randomness in comparison to the CNT PUF is to substitute the CNTs with 2D materials. As of today, there are only a few published implementation examples of PUFs based on 2D materials. Two instances of optical 2D PUFs can be found in the literature. In \cite{Cao.2017}, the nanometer defects in WS\textsubscript{2}, introduced during growth, are captured by PL analysis and exploited in terms of randomness. In contrast, Im et al.\textsuperscript{\cite{Im.2021}} grow the TMD MoS\textsubscript{2}, and drop cast phosphorescent organic crystals on top. The 2D material acts as an atomic seed to introduce disorder in the crystals, which are subsequently analyzed by microscopic-scale image analysis.

In another publication\textsuperscript{\cite{Dodda.2021}}, the electrical properties of graphene are exploited. A reconfigurable graphene-PUF is implemented, which reads out the disorder in charge carrier transport of graphene field-effect transistors (FETs). Machine learning analysis shows that these are resilient to simple neural network attacks. Furthermore, a patent has been filed that utilizes the properties of graphene as the source of entropy to establish a PUF.\textsuperscript{\cite{Dimitrakopoulosetal.2014}}

The implementation that comes closest to the one presented in this thesis is the MoS\textsubscript{2}-based 2D-PUF in \cite{Shao.2021}. 448 FETs with channels composed of MOCVD-grown MoS\textsubscript{2} in bilayer form were fabricated. Subsequently, secret keys were generated by splitting drain currents at specific levels to evaluate the response performance.

In \cite{Shao.2021}, MoS\textsubscript{2} PUFs were compared to a CMOS implementation and a CNT-PUF in terms of uniformity, uniqueness, and autocorrelation. The results are shown in Table \ref{tab:CNT_vs_2D}.

\begin{table}[t!]
\begin{center}
\begin{tabular}{cccc} \toprule
 & Uniformity & Uniqueness & Autocorrelation \\ \midrule
CMOS PUF\textsuperscript{\cite{Yang201783A5}} & N/A & $0.4989$ & Low \\
CNT PUF\textsuperscript{\cite{Hu.2016}} & $0.5047$ & $0.5000$ & $0.9572$ \\
MoS\textsubscript{2} PUF \textsuperscript{\cite{Shao.2021}} & $0.5023$ & $0.5091$ & $1.03058$ \\ \bottomrule
\end{tabular}
\caption[Performance comparison of different PUF implementations]{Performance comparison of different PUF implementations, adapted from \cite{Shao.2021}.}
\label{tab:CNT_vs_2D}
\end{center}
\end{table}

The uniformity describes the number of bit-substitutions necessary to change a given key to an all-zero one. The uniqueness is given by the number of bits that vary in the output between different PUF instances, and an autocorrelation of 1 indicates a device with true randomness. The performance of these three PUF instances seems comparable concerning these parameters.

\subsection{Attacks on PUFs}

All featured PUFs have weaknesses that can potentially be exploited by adversaries. The attacks on PUFs can generally be divided into two groups \footnote{Please note that we are only considering attacks on PUFs in this section, not attacks that exploit PUF-like structures for new security attack vectors, such as \cite{orosa2022spyhammer}.}: Physical attacks or digital attacks.\textsuperscript{\cite{ruhrmair2022secret,LukasZimmermann.06.11.2020}} In the case of physical attacks, an adversary tries to invade the PUF physically in one way or the other.  The class of physical attacks can be further sectioned into non-invasive, semi-invasive, and invasive physical attacks.

Non-invasive physical attacks leave no evidence, as no direct access to the internal components is necessary. They include, for instance, side-channel analysis, fault attacks, and power analysis. Invasive physical attacks, on the other hand, require intervention in the device, modifying the internal structure.\textsuperscript{\cite{LukasZimmermann.06.11.2020}} Helfmeier et al.\textsuperscript{\cite{Helfmeier.2013}} have successfully cloned an SRAM PUF by invasive decapsulation coupled with micro probing. However, this form of attack is the most costly form of eavesdropping. The category that stands between these two groups is semi-invasive. The typical approaches here are electromagnetic or optical analyses.

Nevertheless, the currently most simple, effective, and thus dangerous type of attack on Strong PUFs are digital attacks, such as machine learning-based PUF-modeling.\textsuperscript{\cite{ruhrmair2010modeling,ruhrmair2014efficient,ruhrmair2013power,ruhrmair2014special,sehnke2010policy}} Here, an adversary reads out a large CRP subspace in a reasonable amount of time from the PUF, and/or collects many CRPs by eavesdropping on PUF protocols. Then he designs a parametric model, mostly for machine learning, to extrapolate the overall behavior of the PUF. This often causes PUFs to be broken and thus the design to be strengthened.

The most intensive back and forth between codemakers and codebreakers has appeared on the Arbiter PUF family. The linear behavior of the delays resulted in many successful attempts to model the PUF in the past.\textsuperscript{\cite{Herder.2014}} Rührmair et al.\textsuperscript{\cite{ruhrmair2010modeling}} managed to use linear regression to break an Arbiter PUF with 128 stages in $0.06\,s$ through $1350$ measured CRPs with a prediction range of $95\,\%$. Nevertheless, for the same approach on an XOR Arbiter PUF, exponentially more resources and time were needed for extrapolation. As mentioned above, the original optical PUF construction by Pappu et al.\ \textsuperscript{\cite{pappu2002physical}} has been excluded from all model attacks so far, even though other, integrated and linear versions of optical Strong PUFs have been attacked successfully \textsuperscript{\cite{ruhrmair2013optical}}. 

Excessive machine learning analysis is irreplaceable. Nevertheless, by simply measuring the response sensitivity to small challenge perturbations, as described in \cite{kappelhoff2022strong}, one can obtain helpful information about whether the PUF can be broken by machine learning easily. One or more bits are flipped simultaneously, and the reaction of the response is observed and evaluated. Using this simple form of analysis, one can deduce that the Arbiter PUF can be easily broken without designing complex machine learning algorithms.

All in all, it is crucial to ensure high entropy, i.e., statistical independence of the CRPs in the PUF design, which is an indicator of high resilience against modeling attacks.\textsuperscript{\cite{LukasZimmermann.06.11.2020}} The goal of the present thesis is to achieve this by using random percolation networks consisting of WS\textsubscript{2}, a 2D material, which is introduced in the next section.

\clearpage
\newpage
\thispagestyle{plain}

\section{Fabrication Background}
\label{sec:FabricationalBackground}

Having covered the cryptographic background, the theory of fabrication is now discussed. The next chapter starts with a brief history of 2D materials.

\subsection{2D Materials}

In 1965, Intel co-founder Gordon Moore postulated a law, stating that the number of circuit components on a computer chip doubles at regular intervals, varying from 12 to 24 months.\textsuperscript{\cite{Moore.2006}} As of today, this has proven to be true. Nevertheless, it is estimated that the current yet optimized silicon technology and thus the transistor downscaling are reaching their physical limits. When the devices approach atomic dimensions, unwanted reactions, such as short channel effects or heat dissipation, occur increasingly.\textsuperscript{\cite{Ovchinnikov.2014,L.M.J.J.Peters.June2021,BerndKling.2019}}

Up to now, Moore's law could still be maintained due to newer, emerging devices, such as tri-gate or ultra-thin-body transistors.\textsuperscript{\cite{Fiori.2014}} When an ultra-thin-body transistor is pushed to its limit, the channel consists of only one atomic layer, i.e., a 2D material. While 2D materials have been known for centuries\textsuperscript{\cite{Liu.2015}}, it was assumed that they are thermally unstable and could not exist under ambient conditions. This was the case until the beginning of the millennium, when A. Geim and K. Novoselov managed to isolate few-layer graphite in 2004, and narrowed it down, only one year later, to monolayer graphene.\textsuperscript{\cite{Novoselov.2004,Novoselov.2005}}

2D materials have profoundly different characteristics when isolated from their bulk form to few- or monolayers. Mechanical properties, such as strength, flexibility, and thermal conductivity, as well as electrical properties, such as a tunable bandgap, high mobility, or strong light-matter interaction, can be exploited.\textsuperscript{\cite{L.M.J.J.Peters.June2021,Akinwande.2014,Liu.2015}} In addition, unlike conventional silicon, they possess a low density of open atomic bonds, known as dangling bonds, which act as traps for electrons and thus reduce the performance of transistors. Lastly, due to the minimized spacial expansion, the transistor can be much better controlled by gate electrostatics.\textsuperscript{\cite{ManishChhowallaDebdeepJenaandHuaZhang.2016}}

Akinwande et al.\textsuperscript{\cite{Akinwande.2019}} provided an estimate in a publication from 2019. They guessed that 2D materials are currently in the mature \& benchmarking phase and will gradually enter production, starting around 2025. Cryptographic measures based on 2D materials cannot be missing for this technology to be fully exploited.

\begin{figure}
\centering
\includegraphics[width=13cm]{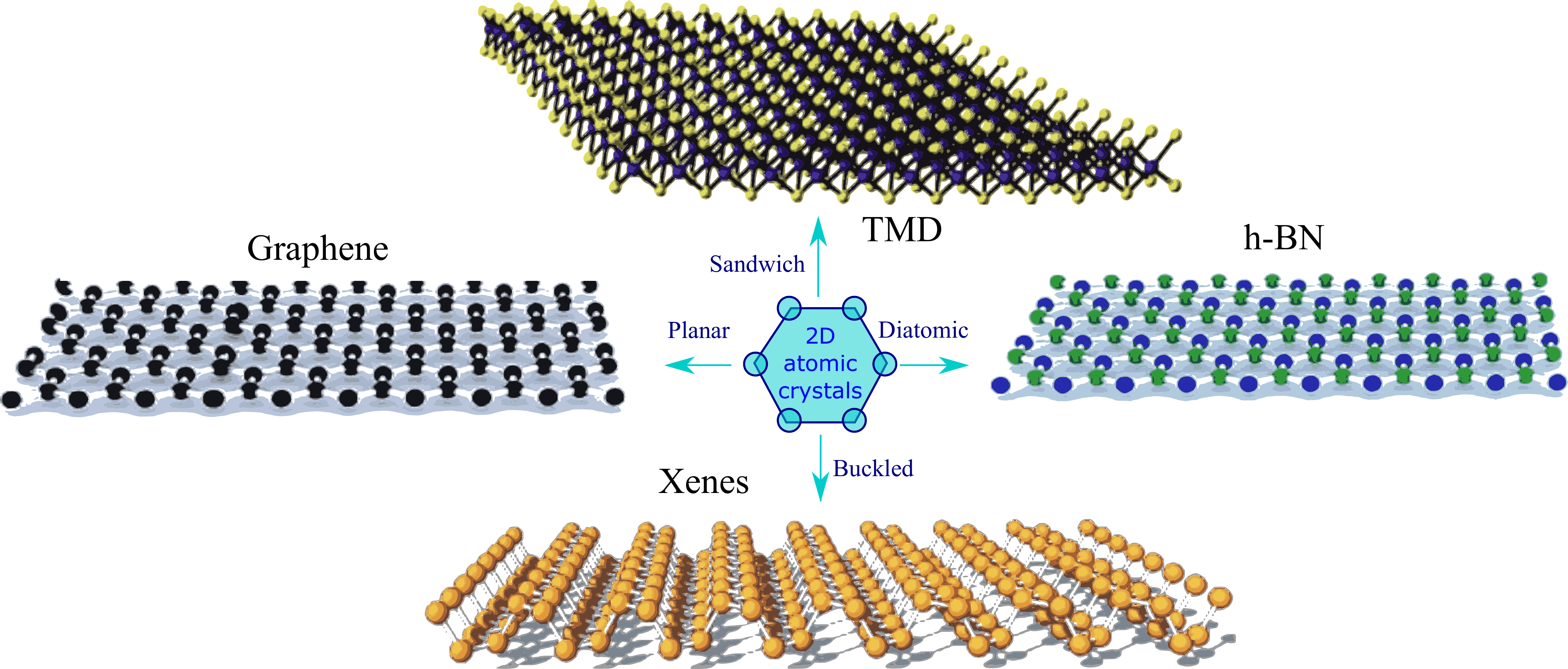}
\caption[2D materials overview]{An overview on 2D materials, adapted from \cite{Akinwande.2014}. The 2D material WS\textsubscript{2} examined in this work belongs to the group of TMDs.}
\label{fig:3.1_2D_Materials}
\end{figure}

As illustrated in figure \ref{fig:3.1_2D_Materials}, various 2D materials have been discovered, each exhibiting different electrical properties. Diatomic boron nitride is an insulator and can be used e.g. as a gate dielectric, whereas graphene is highly conductive. In addition, there are emerging monoatomic buckled crystals, including silicene, germanene, and phosphorene, collectively called Xenes.

Transistors usually comprise two contacts, source and drain, which are connected to each other by a channel material. This channel material is generally a semiconductor, since it allows the current to be brought below a threshold value by a third contact, the gate, and the transistor can thus be switched off. Although graphene has been studied the most and exhibits outstanding properties, it is unsuitable as a channel material due to its lack of a bandgap. At this point, the semiconducting TMDs, utilized in this thesis come into play.

\subsubsection{Transition Metal Dichalcogenides}

\begin{figure}[t!]
 \centering
 \subfigure[]{\includegraphics[width=0.55\textwidth]{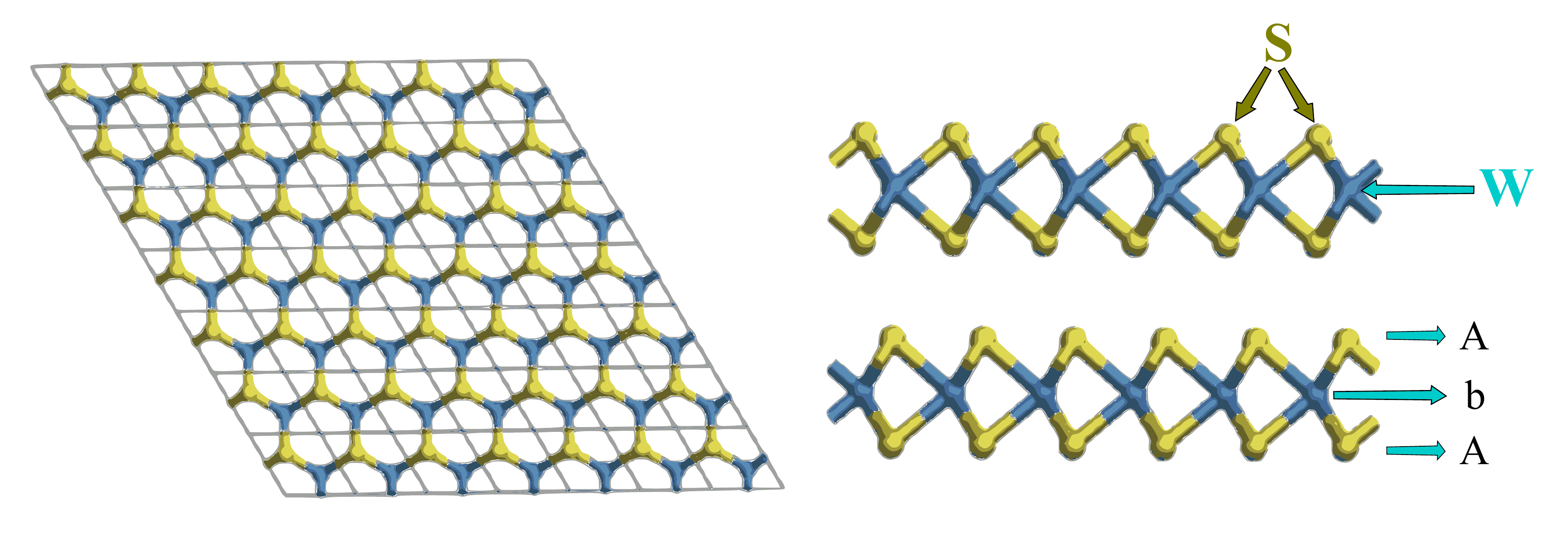}}\quad
 \subfigure[]{\includegraphics[width=0.22\textwidth]{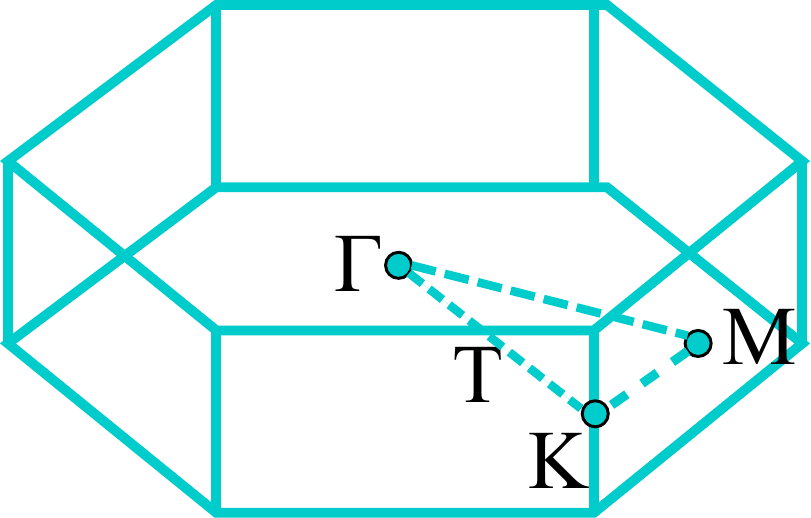}}\qquad
 \subfigure[]{\includegraphics[width=0.4\textwidth]{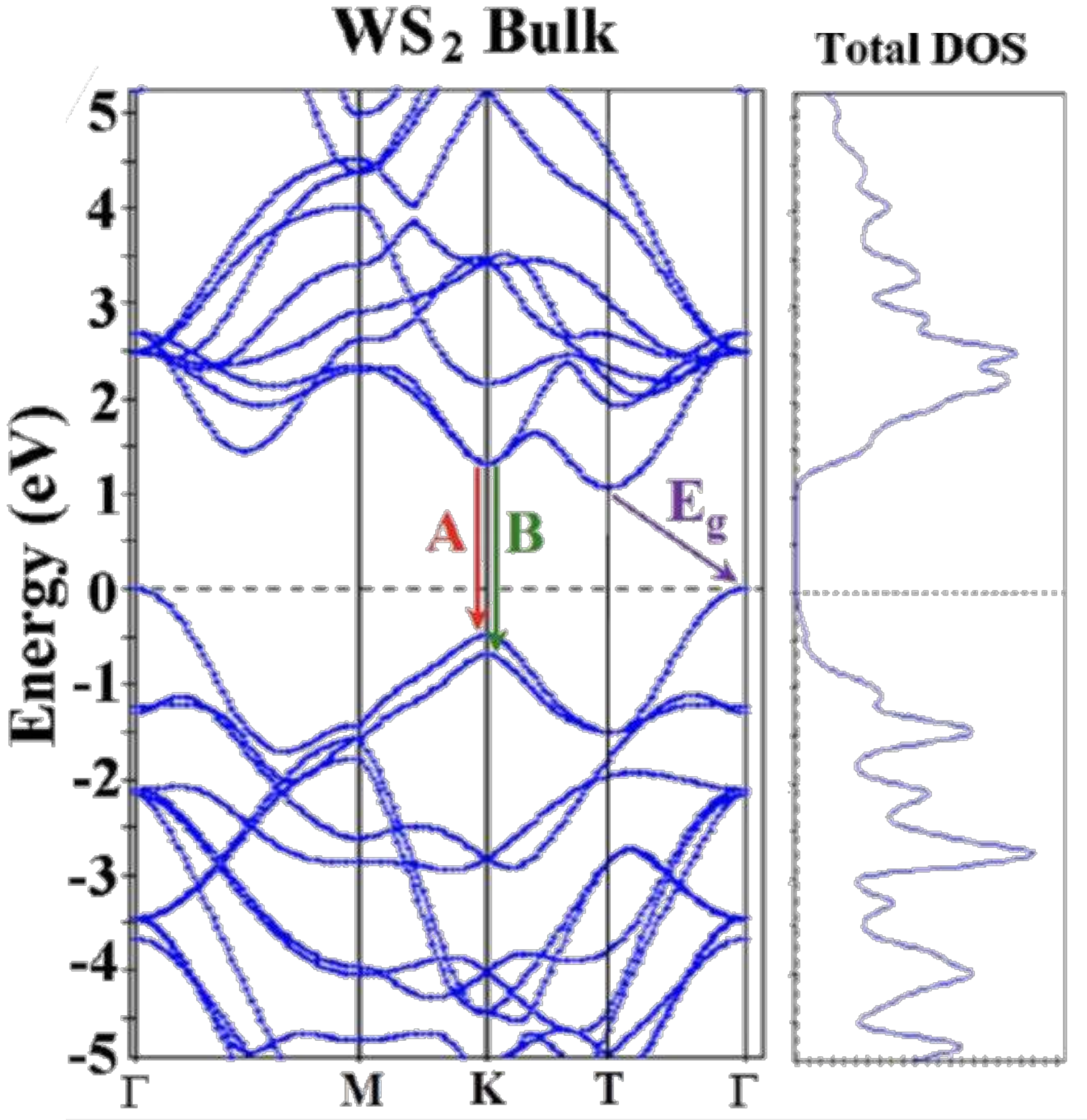}}\quad
 \subfigure[]{\includegraphics[width=0.4\textwidth]{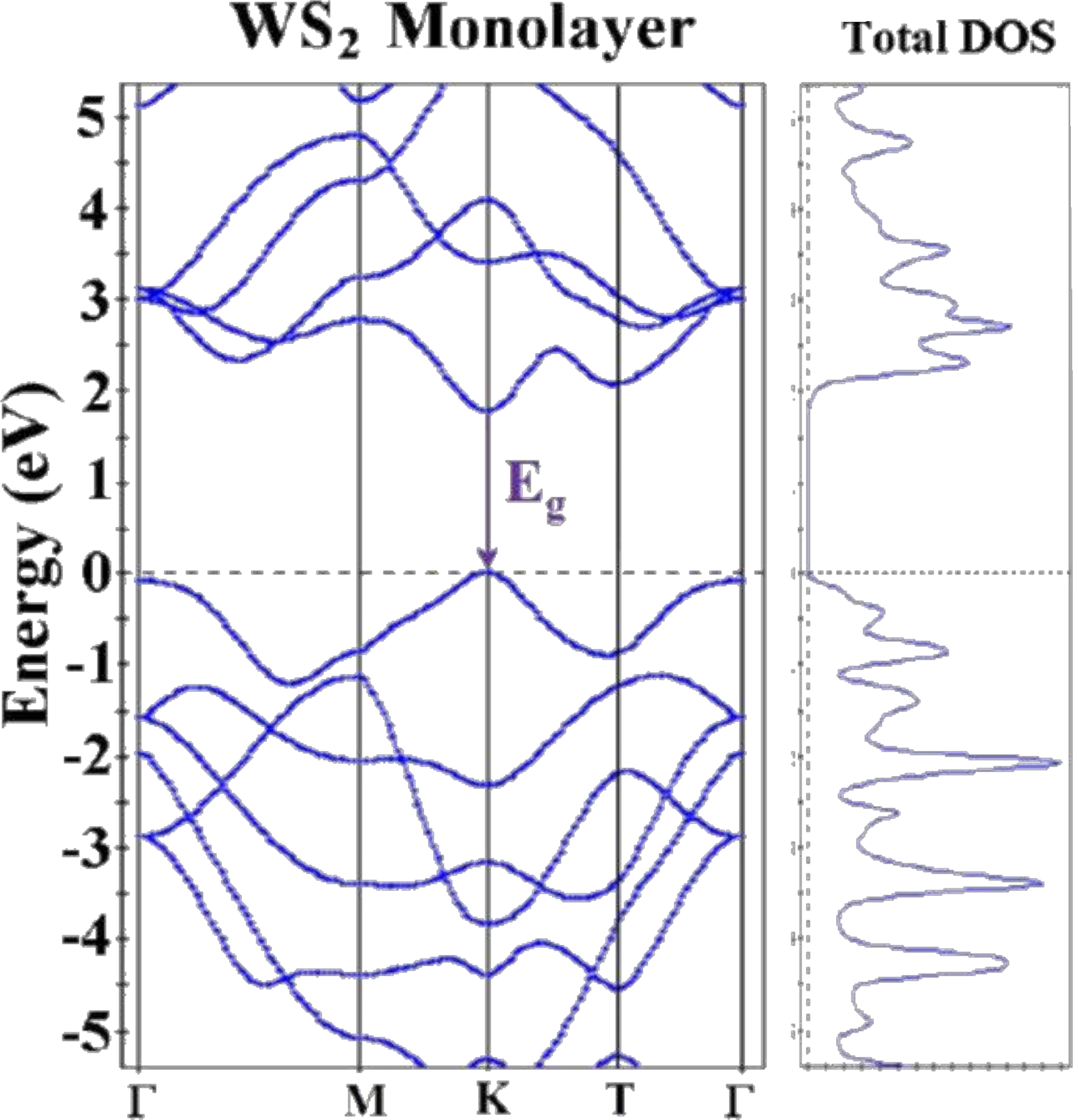}}
 \caption[TMD crystal structures]{TMD crystal structures in (a), the reciprocal space with lattice points in (b), and WS\textsubscript{2} band diagrams, simulated with density functional theory (DFT) in (c) \& (d), adapted from \cite{Gutierrez.2013}.}
 \label{fig:3.1_WS2CrystalStructureAndBandDiagram}
\end{figure}

TMDs can be represented by the formula MX\textsubscript{2}, where M is a transition metal from the 4-7 or 9-10 group and X a chalcogen, i.e., either sulfur, selenium, or tellurium. There are currently more than 40 different known TMD types.\textsuperscript{\cite{Chhowalla.2013}} As shown in figure \ref{fig:3.1_WS2CrystalStructureAndBandDiagram} a), TMDs feature a six-fold coordination symmetry, hexagonally packed between two trigonal atomic layers, in which, when viewed from a cross-sectional perspective, a metal atom is sandwiched between two chalcogenide atoms.\textsuperscript{\cite{Gutierrez.2013}} The metal atoms are bound to the chalcogens by strong-covalent bonds, while the individual layers are being kept apart by weak Van-der-Waals (VdW) forces.\textsuperscript{\cite{Wang.2012,Radisavljevic.2011}}

\clearpage
\newpage

The semiconducting 2D materials can crystallize into three different packaging structures: 1T, which features an octahedral symmetry, 2H with an AbA trigonal prismatic structure as indicated in figure \ref{fig:3.1_WS2CrystalStructureAndBandDiagram} (a), and the 3R phase, showing a rhombohedral symmetry.\textsuperscript{\cite{Song.2015}} In the case of WS\textsubscript{2}, the 2H phase, which is the most abundant one, possesses the hexagonal space group $P6_3/mmc$ along with the lattice parameters $a = \SI{3.1532}{\angstrom}$ and $c=\SI{12.323}{\angstrom}$.\textsuperscript{\cite{Lauritsen.2007}}

From an electronic perspective, TMDs exhibit either semimetallic or semiconducting behavior.\textsuperscript{\cite{L.M.J.J.Peters.June2021}} WS\textsubscript{2} is found to be an n-type semiconductor, and the theoretical charge carrier mobility at room temperature in the monolayer form amounts to $1103\,\frac{cm^2}{Vs}$, which is the highest among all TMDs. This elevated value stems from the relatively low electron and hole effective mass, reading $0,34\,m_0$ and $0,45\,m_0$ respectively, where $m_0$ represents the free-electron mass.\textsuperscript{\cite{Liu.2011}} However, only up to about $50\,\frac{cm^2}{Vs}$ could be measured experimentally so far\textsuperscript{\cite{Sebastian.2021,Zhang.2013,Lee.2013,Jo.2014}}, with one exception. Ovchinnikov et al.\textsuperscript{\cite{Ovchinnikov.2014}} achieved increased mobility of $140\,\frac{cm^2}{Vs}$ by post-annealing of up to $145\,h$, which is still a tiny fraction of the theoretically possible value. In comparison, the realized charge carrier mobility of MoS\textsubscript{2} reaches approximately $60\,\frac{cm^2}{Vs}$.\textsuperscript{\cite{L.M.J.J.Peters.June2021}}

WS\textsubscript{2} is still quite far from its theoretically possible characteristic values in terms of performance, as it has not been studied as excessively in recent years as, for instance, MoS\textsubscript{2}. Hence, there is still a lack of required knowledge, which sometimes makes it necessary to extrapolate the results from MoS\textsubscript{2} to WS\textsubscript{2}. The on-off-current ratios of WS\textsubscript{2} transistors range from $10^5$\textsuperscript{\cite{Tang.2019}} to $10^6$\textsuperscript{\cite{Ovchinnikov.2014}}, whereas MoS\textsubscript{2} transistors already achieve around $10^8$, including high switching speeds.\textsuperscript{\cite{Sebastian.2021}}

On the optical side, the most important feature of TMDs is their tunable bandgap, which depends strongly on the number of layers. As one thins down WS\textsubscript{2}, from bulk to monolayer, the bandgap increases, and indirect transitions become direct instead.\textsuperscript{\cite{L.M.J.J.Peters.June2021}} This shift comes from quantum confinement effects in combination with spatial confinement of electron-hole pairs and the loss of crystal symmetry.\textsuperscript{\cite{Splendiani.2010,Mak.2010}}

In particular, in the case of WS\textsubscript{2}, the band gap has a magnitude of only $E_G=1.3\,eV$ and features an indirect transition in the bulk state. Nevertheless, two additional direct transitions occur at $E_A=1,95\,eV$ and $E_B=2,36\,eV$, as it is plotted in figure \ref{fig:3.1_WS2CrystalStructureAndBandDiagram} b)-d). These results were calculated by DFT, but will be proven by measurements in section \ref{sec:GrowthCharacterization}. \textsuperscript{\cite{Frey.1998,Ballif.1996}} The valence band splitting at the K-point stems from strong spin-orbit coupling in WS\textsubscript{2}. In the case of monolayer material, a much higher photoluminescence intensity can be measured, as the lowest transition becomes direct, with a bandgap value of $E_G=2,1\,eV$.\textsuperscript{\cite{Gutierrez.2013}}

From all these unique properties, various application cases emerge. These include, from an electronic perspective, the use in transistors, where conventional silicon technology could be replaced or complimented by TMDs - even on flexible substrates.\textsuperscript{\cite{Akinwande.2014}} Furthermore, the semiconducting 2D materials can be utilized in optoelectronics due to their tunable bandgap, strong light-matter interaction and viability for use in heterostructures.\textsuperscript{\cite{L.M.J.J.Peters.June2021}} There are existing implementations of photovoltaics\textsuperscript{\cite{Britnell.2013,Wi.2014}}, photodetectors\textsuperscript{\cite{Tahersima.2015,PereaLopez.2013,Zhang.2013b}} and LEDs \textsuperscript{\cite{Baugher.2014}}, containing TMDs. In addition, their properties have been exploited to facilitate resistive switching\textsuperscript{\cite{Ge.2018}}, energy storage, sensing\textsuperscript{\cite{Lee.2013b,Loan.2014,Bollella.2017}} and catalysis.\textsuperscript{\cite{Lukowski.2013}}

There have been many different approaches to growing TMDs on a large scale in the past. However, it can indeed be said that fabrication still faces various problems. Following the introduction of 2D materials, the next section elaborates on producing TMDs.

\clearpage
\newpage

\subsection{Synthesis of 2D Materials}

A wide variety of methods for the synthesis of 2D materials are known, suitable for plenty of applications, depending on the resulting quality and film size. One can generally divide these into bottom-up and top-down methods. The former starts with the nucleation of atoms and then stacks up layer by layer, whereas in the latter, the few-layer material is isolated from the bulk. 

The original and most intuitive top-down method is mechanical exfoliation (ME). It is also known as the scotch tape method and is the approach reported by Novoselov et al.\textsuperscript{\cite{Novoselov.2005}} to isolate the first graphene monolayer in 2005. This method uses tape to fracture the weak VdW forces between the individual layers.
 Although it seems very simple, it is difficult to control the film size and usually results in small flakes, on the order of $1-50\, µm$.\textsuperscript{\cite{Das.2014}}
 
A more advanced method is liquid phase exfoliation (LPE), in which bulk material is dispersed in a liquid and subsequently thinned by, for instance, sonication\textsuperscript{\cite{CanoMarquez.2009}} or thermal shock detachment\textsuperscript{\cite{Kuo.2013}}. This approach is more scalable but does not yield individual flakes and usually results in a few-layered material.\textsuperscript{\cite{L.M.J.J.Peters.June2021}}

An alternative approach is provided by atomic layer deposition (ALD), in which typically two precursors are being used. These react with the surface in a sequential, self-limiting manner. Finally, a thin film is created by repeated deposition of the precursors.\textsuperscript{\cite{Johnson.2014}} It has been successfully used to synthesize atomically thin MoS\textsubscript{2} and WS\textsubscript{2} at low temperatures, though ALD is proving difficult due to a large number of parameters involved.\textsuperscript{\cite{Delabie.2015,Vos.2019,Mane.2018}}

Lastly, there is thermally assisted conversion (TAC), which is a simplified version of CVD. In TAC, a metal precursor is deposited onto a substrate by physical vapor deposition (PVD), which includes, for instance, sputtering or ALD. Afterward, the metal layer is converted into a TMD by heating in a furnace under chalcogen-vapor flow.\textsuperscript{\cite{McManus.2019}} A significant practical difference between TAC and CVD is that in the latter, the metal precursor is deposited on an additional substrate, allowing the resulting films to be used directly for electronic applications without the need for a transfer.

\subsubsection{Chemical Vapor Deposition in Detail}
\label{sec:ChemicalVapourDepositionInDetail}

Chemical vapor deposition is a widely used process technology. The majority of applications are the production of high-purity bulk materials, the fabrication of composite materials via infiltration techniques, and the deposition of solid thin films.\textsuperscript{\cite{J.R.CreightonandP.Ho.2001}} A disadvantage of CVD is that silicon technologies are only compatible with processes involving temperatures below $450\,$°$C$ and CVD usually takes place at higher temperatures. However, it is used frequently since it yields high throughput and pure material quality.\textsuperscript{\cite{Ansari.2019}}

\begin{figure}[t!]
\centering
\includegraphics[width=15cm]{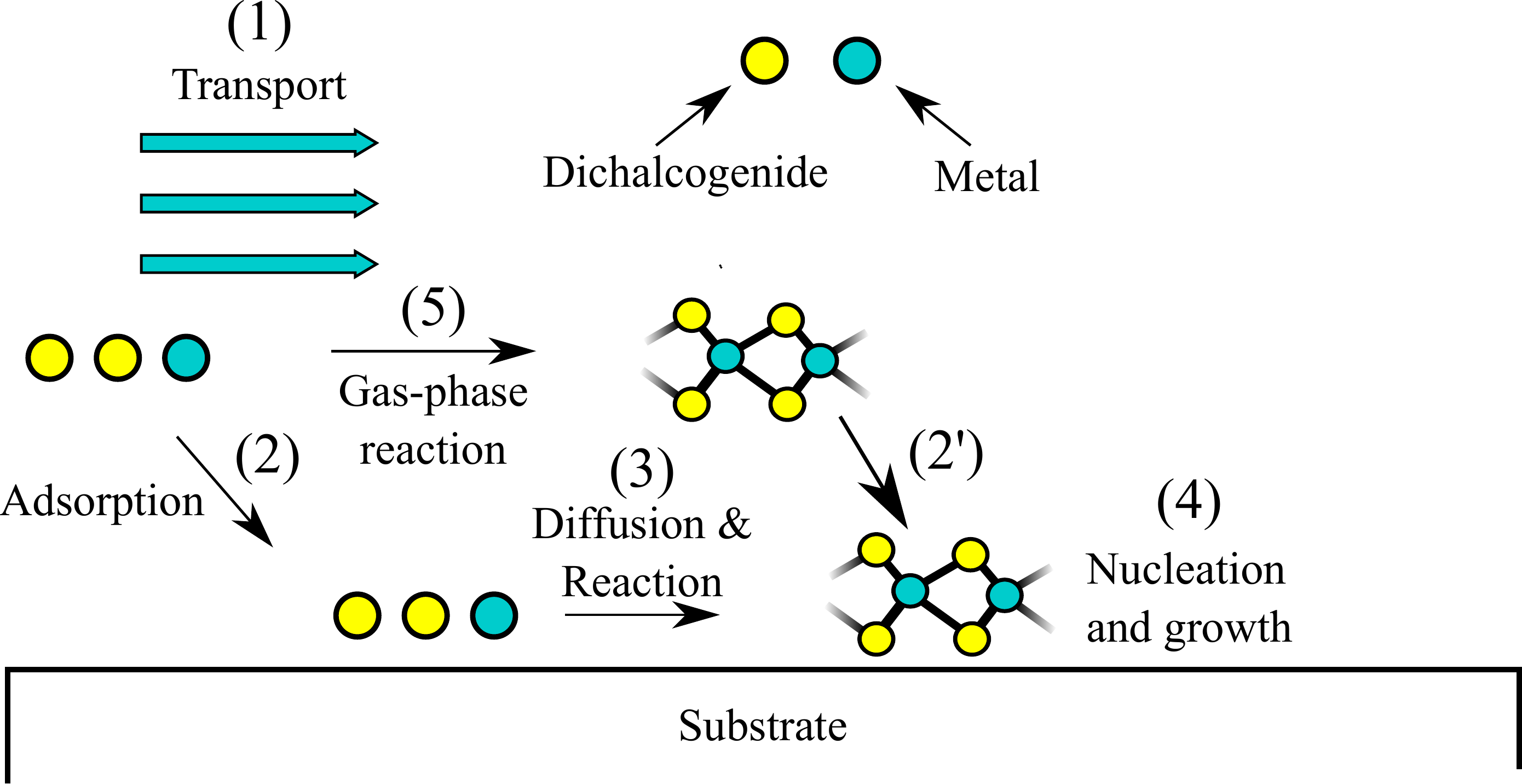}
\caption[Key steps involved in CVD]{Key steps involved in CVD, adapted from \cite{HampdenSmith.1995}. The precursor can be either directly adsorbed (2) and subsequently diffuse and react on the surface (3) or react in the gas phase (5) followed by adsorption (2'). Finally, a TMD crystal can be formed on the surface (4).}
\label{fig:3.2_CVDProcess}
\end{figure}

The individual steps of a CVD process are outlined in figure \ref{fig:3.2_CVDProcess}. The precursors are initially introduced into the reactor (1), where they are then adsorbed (2), diffuse, and react on the surface (3). Eventually, the metal atoms diffuse to form a stable nucleus in the form of a TMD crystal and subsequently grow laterally. In parallel, gas-phase reactions can take place under certain circumstances (5).\textsuperscript{\cite{HampdenSmith.1995}} Note, that the individual processes cannot be influenced directly. Hence, controllable parameters have to be found and optimized to obtain a suitable growth process for the respective application. The chosen parameters for this project are listed in section \ref{sec:CloseProximityCVD}.

\begin{figure}[t!]
\centering
\captionsetup{justification=centering}
\includegraphics[width=15cm]{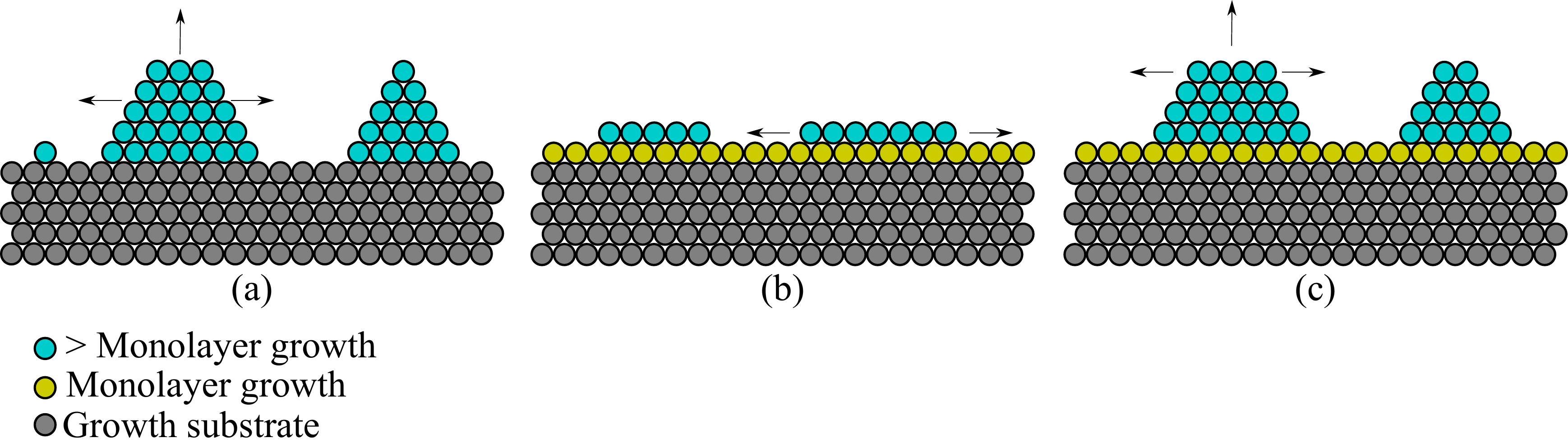}
\caption[The different growth modes in CVD]{Outline of the different growth modes in CVD, adapted from \cite{L.M.J.J.Peters.June2021}. With (a) Volmer-Weber island formation growth, (b) Frank-van-der-Merwe layer-by-layer growth and (c) Stranski-Krastanov monolayer with island growth.}
\label{fig:3.2_GrowthModes}
\end{figure}

In the case of TMDs, one can generally categorize the growth at step (4) into three distinct types, as illustrated in figure \ref{fig:3.2_GrowthModes}. Volmer-Weber\textsuperscript{\cite{Volmer.1926}} (VW) and Stranski-Krastanov\textsuperscript{\cite{I.N.StranskiandL.Krastanov.1938}} (SK) growth are the two more common types, albeit one usually strives for Frank-van-der-Merwe\textsuperscript{\cite{.1949}} (FvdM) growth. Individual islands grow independently within the VW growth, forming separate few-layer flakes. These islands can also coalesce and form a continuous film, which leads to grain boundaries in case of an orientation mismatch.

During FvdM growth, one layer is formed at a time. When the bottom layer has achieved complete surface coverage, the overlying growth begins. The SK growth, on the other hand, is a mixture of both, in which first a monolayer grows as in the FvdM case, and then VW-like islands form on top.

If one inspects the growth in greater detail, it is possible to distinguish the individual flakes in terms of their shape. In theory, the 2D flakes that eventually coalesce into one large layer can take a dodecagonal, a hexagonal, or a triangular shape. This depends on the chemical potential, with the flakes taking on more and more of a triangular shape as the growth becomes increasingly sulfuric.\textsuperscript{\cite{Cao.2015}} This was proven by DFT calculations in \cite{Cao.2015} and additional experiments for MoS\textsubscript{2} in \cite{Wang.2014}, whereby the Mo/S-ratio was varied by changing the spatial position of the silicon substrate in the furnace. A further study on WS\textsubscript{2} has been carried out, in which changes in the growth temperature altered the flake structure. Between $1000$ and $1100\,$°$C$, the individual shapes took a triangular structure, whereas below they trended towards a hexagonal shape.\textsuperscript{\cite{Dong.2019}}

A further important influence, especially for electronic applications of TMDs, carry the defects that can occur inside these individual flakes. The most common defects are sulfur vacancies\textsuperscript{\cite{Addou.2015}} and grain boundaries\textsuperscript{\cite{vanderZande.2013}}. They result in scattering events and charge traps that reduce the carrier mobility. Up to date, it is difficult to produce non-polycrystalline large films with CVD\textsuperscript{\cite{L.M.J.J.Peters.June2021}}, which is appropriate and desired for PUFs. Most growth attempts aim to yield a homogeneous film, whereby reverse engineering is necessary for PUF production to introduce as much randomness as possible.

\clearpage
\newpage

\subsubsection{CVD in the case of WS\textsubscript{2}}

As stated previously, CVD-grown WS\textsubscript{2} is utilized to fabricate the PUF in this work. There exists only few documentation about the growth of WS\textsubscript{2} as opposed to MoS\textsubscript{2}.\textsuperscript{\cite{Pam.2019,Gutierrez.2013,Sebastian.2021,Cui.2015}} Yet, a few tips and tricks are known to yield large-scale WS\textsubscript{2} thin films. Adding a reducing agent such as hydrogen seems to enhance growth. Further, in the case of \cite{Das.2014}, WO\textsubscript{3} is used as a metal precursor since it is easier to vaporize than pure tungsten. The WO\textsubscript{3} reacts with the reducing agent to form substoichiometric WO\textsubscript{3-x}, according to the reducing reaction \ref{equ:CVD1}. It represents an intermediate step in the entire reaction equation, provided in \ref{equ:CVD}, and plays a crucial role in nucleation and crystallization, as it is even more volatile and even easier to vaporize.\textsuperscript{\cite{Das.2014}}

\begin{gather}
  WO_3+xH_2 \rightarrow WO_{3-x}+xH_2O
  \label{equ:CVD1} \\
  WO_3+4S+2H_2 \rightarrow WS_2 + H_2O + SO_2 + H_2S
  \label{equ:CVD}
\end{gather}


In order to obtain reliable growth, the entire synthesis must be systematically optimized. Since there are generally many different input parameters that influence the growth in an unpredictable, non-linear way, a DoE strategy is chosen in this work to find the ideal parameter subspace.

\subsection{Design of Experiments in a Nutshell}

When faced with a product or process that demands optimization, one can use a methodology called Design of Experiments (DoE) to maximize the efficiency of the optimization process. In a DoE, a series of experiments is conducted, in which the input variables of a system are varied, and the effects on response variables are measured. It provides an elegant way to minimize the effort and maximize the amount of information collected. Thereby, the factors are changed simultaneously, and a predictive model is determined.\textsuperscript{\cite{Cox.2000}} DoE was initially employed in an agricultural research center around 1930\textsuperscript{\cite{FisherR.A..1935}} but has been successfully applied in the military and industry since 1940.\textsuperscript{\cite{article}}

\begin{figure}[h!]
\centering
\includegraphics[width=13cm]{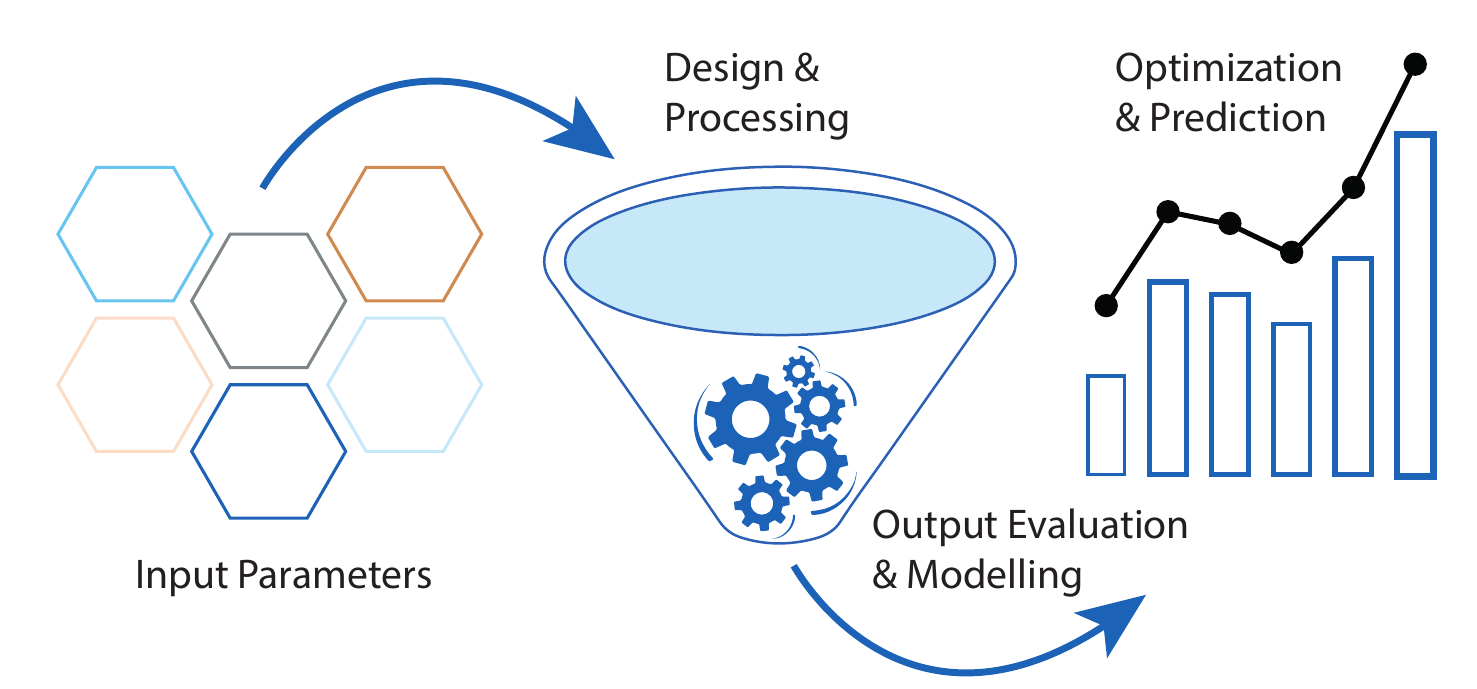}
\caption[A DoE process scheme]{A DoE process scheme. Once the input parameters have been defined, the DoE is designed and executed. Subsequently, the results are evaluated and a potential model is established. Finally, further optimizations can be performed and predictions can be derived.}
\label{fig:3.3_DoEProcedure}
\end{figure}

The structure and the individual steps that must be performed to conduct a DoE are listed in figure \ref{fig:3.3_DoEProcedure}. First, input parameters must be defined and narrowed down. Then the experiment needs to be designed and progressed, which is described in the case of this thesis in section \ref{sec:DesignOfExperiment}. Finally, one has to evaluate, create a model, and optimize the product or process. This is done in section \ref{sec:5DesignOfExperiment}.

\newpage

There are a handful of fundamental principles that must be adhered to in order to achieve optimal results.\textsuperscript{\cite{article}} One is randomization, which is intended to protect against an unknown bias. The bias can be, for instance, an instrument drift, in which the random order of experiments may average out. The degree to which the experimental conditions, such as an unknown bias, change throughout the DoE, can be determined by a consistency experiment, repeated several times in intervals.

Furthermore, replications, i.e., complete repetitions of the same experimental conditions, can prevent nuisance factors. Automatic reproducibility already takes place in a DoE by factors that change the response variable insignificantly. Another confounding variable in an optimization process is batch-to-batch variability, which could distort the results. It is essential that in advance, all samples are prepared in the same process and that they are all of the same type. This requirement is usually referred to as blocking.

To finally derive the predictive model, one must apply a quantitative evaluation method. For this purpose, one generally uses an approximation in the form of a Taylor series in order to estimate the real form of the response variable:

\begin{equation}
  Y=\beta_0 + \sum\limits_{i=1}^{p} \beta_i X_i + \sum\limits_{i=1}^{p} \sum_{\substack{j=1 \\ i\neq j}}^p \beta_{ij} X_i X_j + \sum\limits_{i=1}^{p} \beta_{ii} X_i^2 + ... 
\label{equ:DoE}
\end{equation}

\begin{minipage}[t]{.5\textwidth}
  \begin{itemize}
    \item $Y$: Response variable
    \item $X_i$: Input factor i out of p
    \item $\beta_0$: Overall mean response
  \end{itemize}  
\end{minipage}%
\begin{minipage}[t]{.5\textwidth}
    \begin{itemize}
    \item $\beta_i$: Main effect of factor i
    \item $\beta_{ij}$: Two-way interaction between the ith and jth factor
    \item $\beta_{ii}$: Quadratic effect of factor i
  \end{itemize}
\end{minipage}

Both the influence of the individual parameters and the impact of the parameters on each other are considered. Interactions between two factors are present when the combined effect of these two factors cannot be predicted from the individual ones. Depending on the desired resolution level, any number of terms can be appended to equation \ref{equ:DoE} to consider three-way interactions or other more sophisticated dependencies.

\begin{figure}[t!]
\centering
\includegraphics[width=15cm]{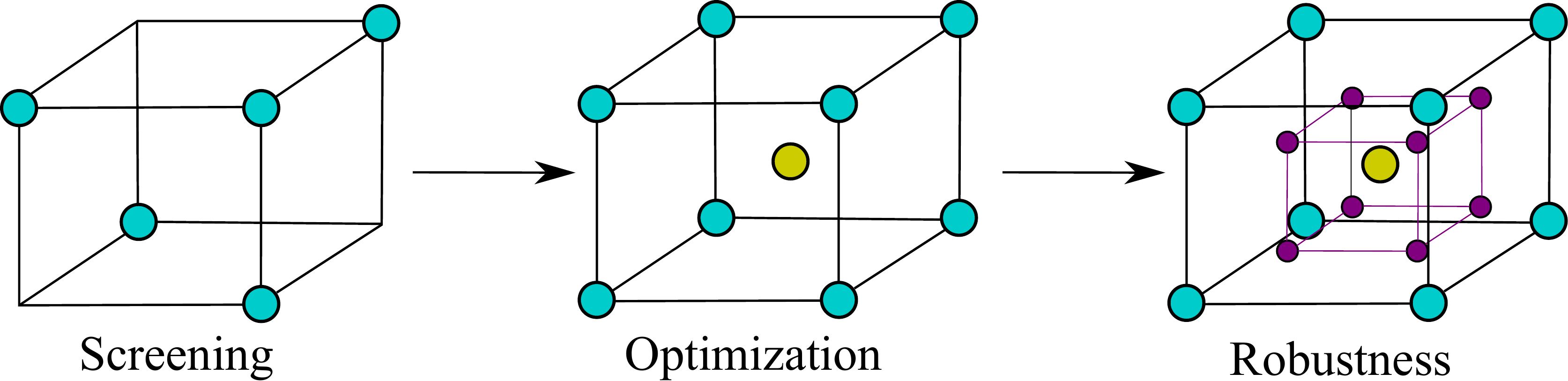}
\caption[The different depths of DoE]{Illustration of the different depths of DoE, adapted from \cite{JeremySpringall.2018}.}
\label{fig:3.3_DoE}
\end{figure}

A DoE can be used for various purposes. It can be employed to determine the factor interactions, screening, establish and maintain quality control, optimize a process, or design robust products. Depending on the application, the experiment becomes more or less complex, as illustrated graphically in figure \ref{fig:3.3_DoE}. In the production of the PUF, an objective is to perform a robust DoE to achieve precise control of the growth. A significant advantage is that even if the optimization process fails, it still provides valuable information about which input variables contribute to the majority of variability.

\clearpage
\newpage
\thispagestyle{plain}

\section{Experimental \& Analytical Methods}
\label{sec:ExperimentalAnalyticalMethods}

In order to fabricate the 2D-material-based PUF in the laboratory, an experimental setup to realize a CVD process is needed. This setup and all the experimental details on the PUF device fabrication are explained in this chapter.

\subsection{Chemical Vapor Deposition}

At first, a unique technique of CVD is discussed, namely, the close proximity approach, in which the growth takes place in a so-called microreactor.

\subsubsection{The Close Proximity Approach}
\label{sec:CloseProximityCVD}

Since 2D films cannot exist free-standing, growing them on a substrate is necessary. To fabricate a PUF, the substrate must be further suitable for electronic applications. Hence, a semiconducting Si substrate with thermally grown SiO\textsubscript{2} on top is adopted in this work. SiO\textsubscript{2} is an amorphous insulator and thus allows electronic use-cases. By using a semiconducting Si substrate, the responses can additionally be influenced by a gate voltage. The substrate dimensions amount $10$x$10\,mm^2$ in width and length, with the Si substrate being 500\textit{\,µm} tall, including $300\,nm$ SiO\textsubscript{2} on top. Prior to each CVD process, the substrate is cleaned in an ultrasonic bath in high-performance liquid chromatography (HPLC) acetone, followed by HPLC isopropanol (IPA), and then rinsed in DI (Distilled) water. The steps are performed for ten minutes each. Finally, the sample is dried with nitrogen gas (N\textsubscript{2}).

A further benefit of a $300\,nm$ thick SiO\textsubscript{2} layer is that the TMD on the surface can be easily identified optically, as the interference color and material opacity of the incident light shift in phase, which even allows concluding the layer number. Due to this phenomenon, it frequently occurs that samples with different heights of the SiO\textsubscript{2} layer show different colors in optical images. This, however, should neither influence the growth nor the behavior of the PUF device remarkably.

In order to carry out a CVD process, two precursors are required - one that contains the tungsten and another one including sulfur. To ensure good control over the amount of metal released during the process, the transition metal precursor was deposited on an additional Si/SiO\textsubscript{2} sample by either sputtering or drop-casting. The specific details are provided in the next section \ref{sec:PrecursorSynthesis}. The chalcogen precursor is precipitated sulfur powder from Alfa Aesar with a purity of $99.5\,\%$.

\begin{figure}[b!]
 \centering
 \subfigure[]{\includegraphics[width=0.7\textwidth]{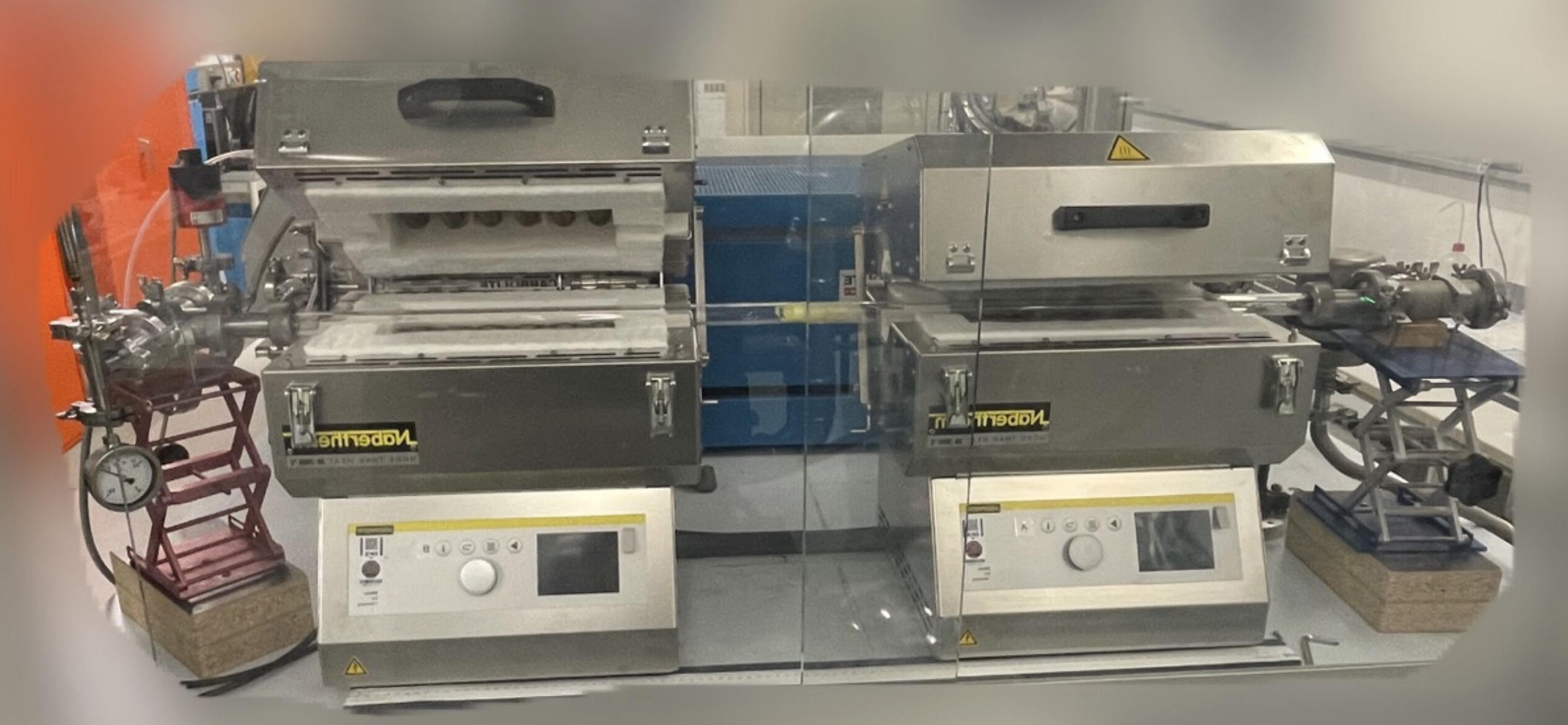}}\qquad
 \subfigure[]{\includegraphics[width=0.8\textwidth]{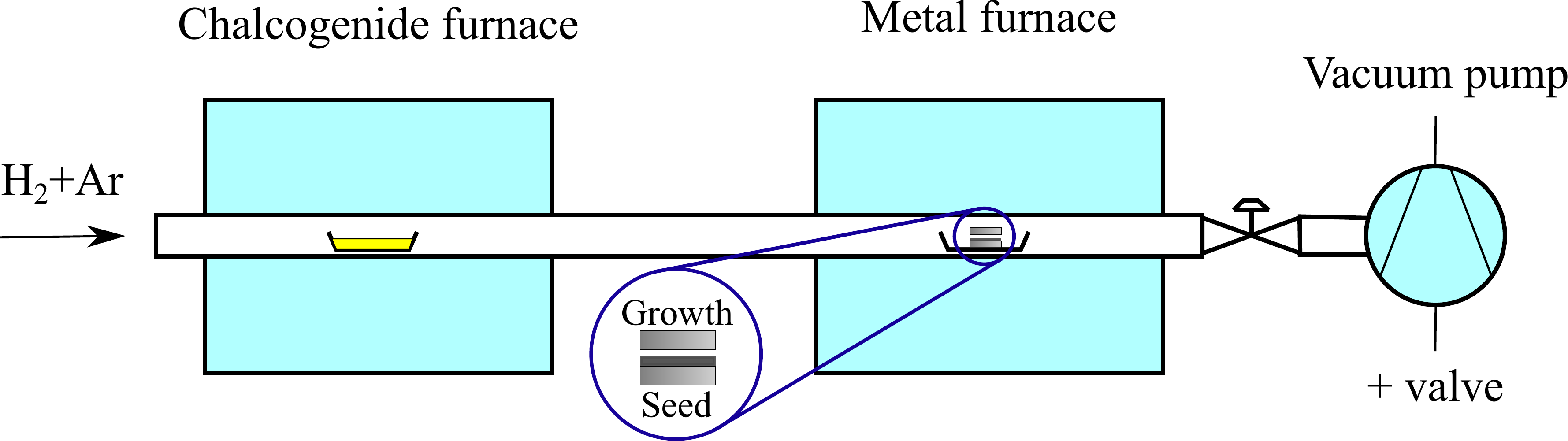}}
 \caption[WS\textsubscript{2} experimental growth setup]{Image of the experimental WS\textsubscript{2} growth setup from the laboratory in (a) with the corresponding labelling in (b). The seed and growth substrate are superimposed in a microreactor style.}
 \label{fig:4.1_WS2ExperimentalSetup}
\end{figure}

A picture of the experimental setup and an illustrative sketch of the same is given in figure \ref{fig:4.1_WS2ExperimentalSetup}. The growth took place in a quartz tube at low pressure, which is surrounded by two furnaces, that are used to heat the sulfur and tungsten precursor separately to control their ratio individually. Prior to each process, the sulfur powder was placed in a ceramic boat and inserted into the low-temperature chalcogenide oven, which was located upstream. Approximately $35-40\,cm$ further downstream, in the high-temperature metal furnace, the growth and seed substrates were located in a microreactor style along the lines of \cite{OBrien.2014}, in another ceramic boat. A microreactor consists of the seed sample, including the deposited metal precursor at the bottom and a growth substrate placed on top.

After a heating-up phase, the vaporized sulfur was carried to the high-temperature zone by a gas flow containing a specific ratio of argon and hydrogen, where the growth occurred according to equation \ref{equ:CVD}. The pressure was controlled by an electric valve connected to a vacuum pump at the end of the tube. The experiments were performed under low-pressure conditions in the millibar range to introduce as few impurities as possible.

This specific experimental setup results in various parameters that can be more or less easily tuned and optimized to influence the individual processes from figure \ref{fig:3.2_CVDProcess} indirectly. These are enumerated in the following:

\begin{minipage}[t]{.33\textwidth}
  \begin{itemize}
    \item Heating-up time
    \item Process time
    \item Cool-down time
    \item Pressure
  \end{itemize}  
\end{minipage}%
\begin{minipage}[t]{.33\textwidth}
    \begin{itemize}
    \item Boat positions
    \item Total flow
    \item Ar/H\textsubscript{2} ratio
    \item Metal precursor type and amount
  \end{itemize}
\end{minipage}
\begin{minipage}[t]{.33\textwidth}
    \begin{itemize}
    \item Chalcogenide temperature
    \item Metal temperature
    \item Microreactor vertical distance
  \end{itemize}
\end{minipage}

\subsubsection{Microreactor Distance Control}
\label{sec:MicroreactorDistanceControl}

\begin{figure}[h!]
\centering
\includegraphics[width=15cm]{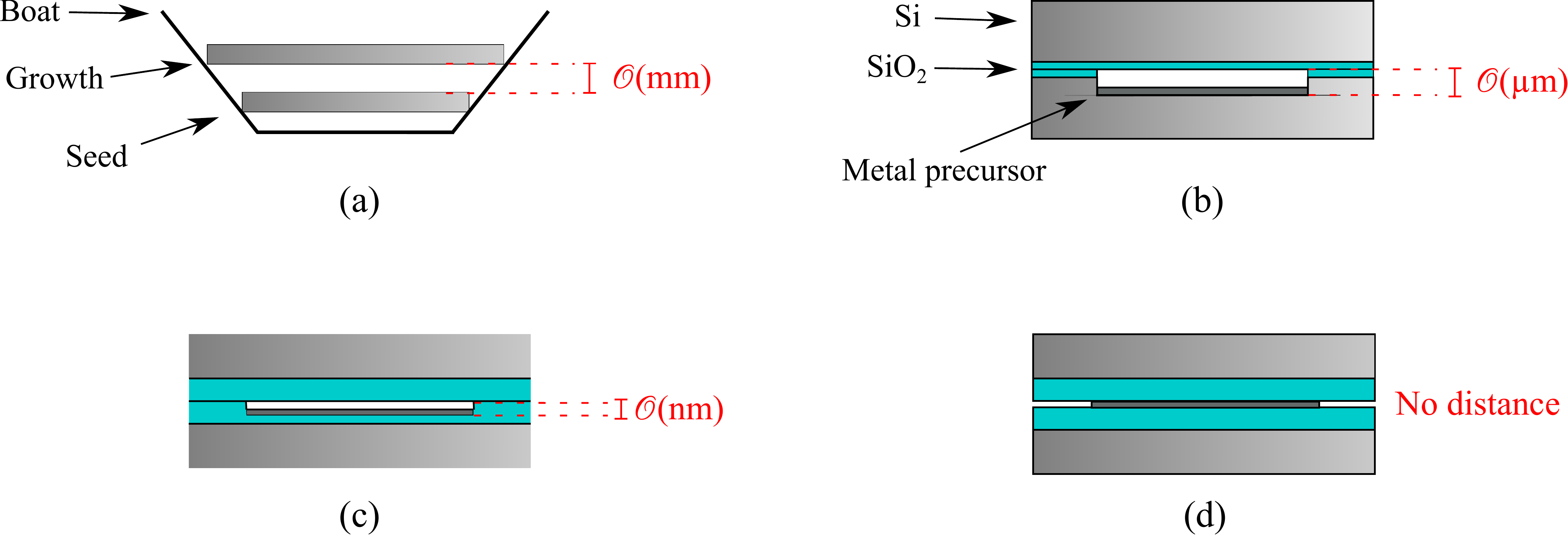}
\caption[Close-proximity CVD distance control]{Illustration of the gaps, introduced into the microreactor with (a) mm-, (b) µm-, (c) nm-distance, and (d) no distance.}
\label{fig:4.1_DistanceControl}
\end{figure}

An exciting parameter in the close-proximity CVD approach, which has been very sparsely studied in the literature so far, is the distance between the two substrates of the microreactor. In reality, impurities are ever present, even if only nanometre-scaled. These particles can settle between the two substrates and thus vary the distance and influence growth. In order to investigate as many different orders of magnitude as possible, three different gap sizes in the range of millimeters, micrometers, and nanometers were realized using different techniques. The approaches are explained in the following and illustrated in figure \ref{fig:4.1_DistanceControl}.

The simplest way to create a gap is to use a top substrate with a different width than the underlying bottom sample. The substrate above is kept at a different height than the substrate below due to the boat shape, as it is depicted in figure \ref{fig:4.1_DistanceControl}. Unfortunately, it is not easy to measure the emerging space precisely since a slight inclination of either sample can already vary the distance. Nevertheless, in this case, the order of magnitude lies in the mm range. The width of the mm-distance growth samples used in this work were $10.5$, $11$, and $11.5\,mm$.

To ensure a distance of several ten to hundred micrometers, a diamond saw was used to cut into the wafer in small sections with micrometer precision. A space of $1\,mm$ was kept at the left and right border to allow the growth substrate to rest on the seed sample. The width of the blade amounts to 28\textit{\,µm} and the depths that were investigated here are $25$, $50$, $100$, and 200\textit{\,µm}.

Lastly, to realize an nm-range gap, the SiO\textsubscript{2} layer was partly etched with the help of an ammonium fluoride etching mixture. First, the sample was covered on the left and right with polyimide tape from Kapton, which is resistant to the etchant, and then placed inside the acid. The etch rate equals approximately $75\,nm$ per minute, and thus $20$, $40$, and $60\,nm$ dips were produced with waiting times of $16$, $32$, and $48\,s$. Eventually, the metal precursor was deposited onto the seed substrates, as it is described in the next section.

\subsection{Precursor Synthesis}
\label{sec:PrecursorSynthesis}

This chapter explores two different variants of applying the transition metal precursor to the seed substrate. Within one, the precursor was dispersed in a solution and refined by subsequent sonication and centrifugation. The resulting supernatant was then coated onto the seed substrate by a deposition method.

\subsubsection{Liquid Precursor}
\label{sec:LiquidPrecursor}

The different deposition methods\cite{Alzakia.2021} include for one drop-casting, which involves casting droplets of 2D material dispersions onto a target substrate with subsequent evaporation of the solvent by heat. A further technique is dip coating, where the sample is immersed for a short time in a container containing the dispersion of 2D material and then pulled back to dry. While these two approaches are very straightforward, they suffer from difficulties in accurately controlling thickness and uniformity. A slightly more sophisticated way to achieve homogeneous films is spin-coating. It involves spinning the dispersion at a very high speed on the target substrate.

Moreover, there are a couple of published studies on the LPE of 2D materials.\textsuperscript{\cite{Sahoo.2020,Adilbekova.2020,Jha.2017,Coleman.2011}} They investigate different solvents, such as IPA, acetone, or more sophisticated ones, such as N-methyl-2-pyrrolidone, and various sonication and centrifugation procedures. Actually, in \cite{OBrien.2014}, a close-proximity CVD setup with LPE precursors has been attempted. The recipe includes $30\,g$ of WO\textsubscript{3} powder, added to $100\,mL$ of IPA with subsequent sonication for $5\,h$ and $120\,min$ of centrifugation at $5\,krpm$. Finally, the dispersion was drop-casted on the surface and heated on a hot plate until the IPA evaporated.

In terms of this thesis, the first trial was thus attempted with an equal ratio of WO\textsubscript{3} powder from Alfa Aesar with $99.99\,\%$ purity, dissolved in IPA, followed by sonication for several hours, subsequent drop-casting and heating on a hotplate. However, this method resulted in very inhomogeneous films, including evaporation rings, formed during bakeout. Therefore, the next step was to change the strategy to spin coating and intent at different times and rotation speeds. However, most of the dispersion was spun off, and hence the WO\textsubscript{3} concentration was increased to $1.5\,g$ WO\textsubscript{3} powder per milliliter IPA.

\begin{figure}[h!]
\centering
\includegraphics[width=13cm]{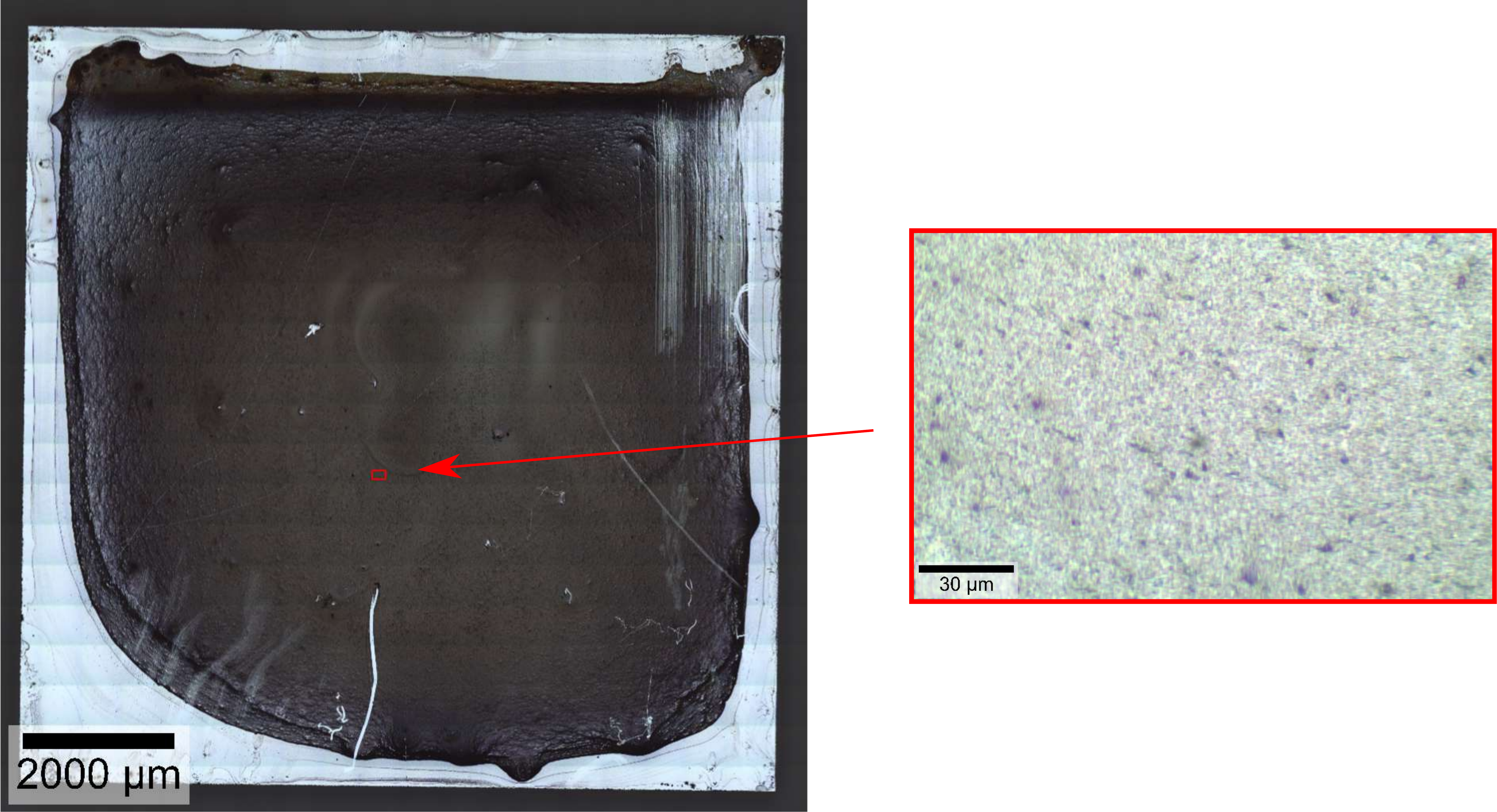}
\caption[Seed substrate with spin-coated liquid WO\textsubscript{3} precursor]{The seed substrate in light blue, with the spin-coated, dispersed WO\textsubscript{3} precursor on top in black. The red crop reveals a magnified section with dispersed WO\textsubscript{3}, in the form of black dots.}
\label{fig:4.2_LiquidPrecursor}
\end{figure}

A way to improve wettability and achieve homogeneous films is by increasing the hydrophilicity of the surface. SiO\textsubscript{2} can be made more hydrophilic by an oxygen plasma treatment. A final sample can be seen in figure \ref{fig:4.2_LiquidPrecursor}. It was oxygen-plasma treated for $5\,mins$ at $100\,\%$ power with subsequent spin coating for $10\,s@500\,rpm$ plus $60\,s@1000\,rpm$ in a $Atto$ $Plasmacleaner$ from $Diener$.

\subsubsection{Solid Precursor}
\label{sec:4.2_SolidPrecursor}

The alternative way of applying tungsten oxide involves sputtering of tungsten and subsequent oxidation in a furnace to WO\textsubscript{3}. Sputtering is a technique in which a surface of a metal target is bombarded with argon-ions, resulting in the ejection of target material onto the substrate. The different deposition thicknesses that have been examined were $5\,nm$, $10\,nm$, and $20\,nm$. Oxidation was then carried out in an oven at $500\,$°$C$, at one millibar. The gas flow consisted of $200\,sccm$ Ar and $200\,sccm$ O\textsubscript{2}, with the process lasting twelve hours, including a ramp-up of one hour and natural cooling after a total of eleven hours.

As it is illustrated later in section \ref{sec:InitialParameterStudies}, it was observed that the TMD on the growth substrate tends to grow on the border of the WO\textsubscript{x} precursor of the seed substrate. Therefore, to enable a high surface coverage of 2D material, the sputtered tungsten was patterned by a lithography process. Photolithography (PLT) and electron-beam lithography (EBL) are two methods used to create patterns in thin-film fabrication. Today's EBL standards achieve a lateral resolution of down to $20\,nm$. \textsuperscript{\cite{PatrickP.Naulleau.NovDec2009}}

\begin{figure}[h!]
\centering
\includegraphics[width=15cm]{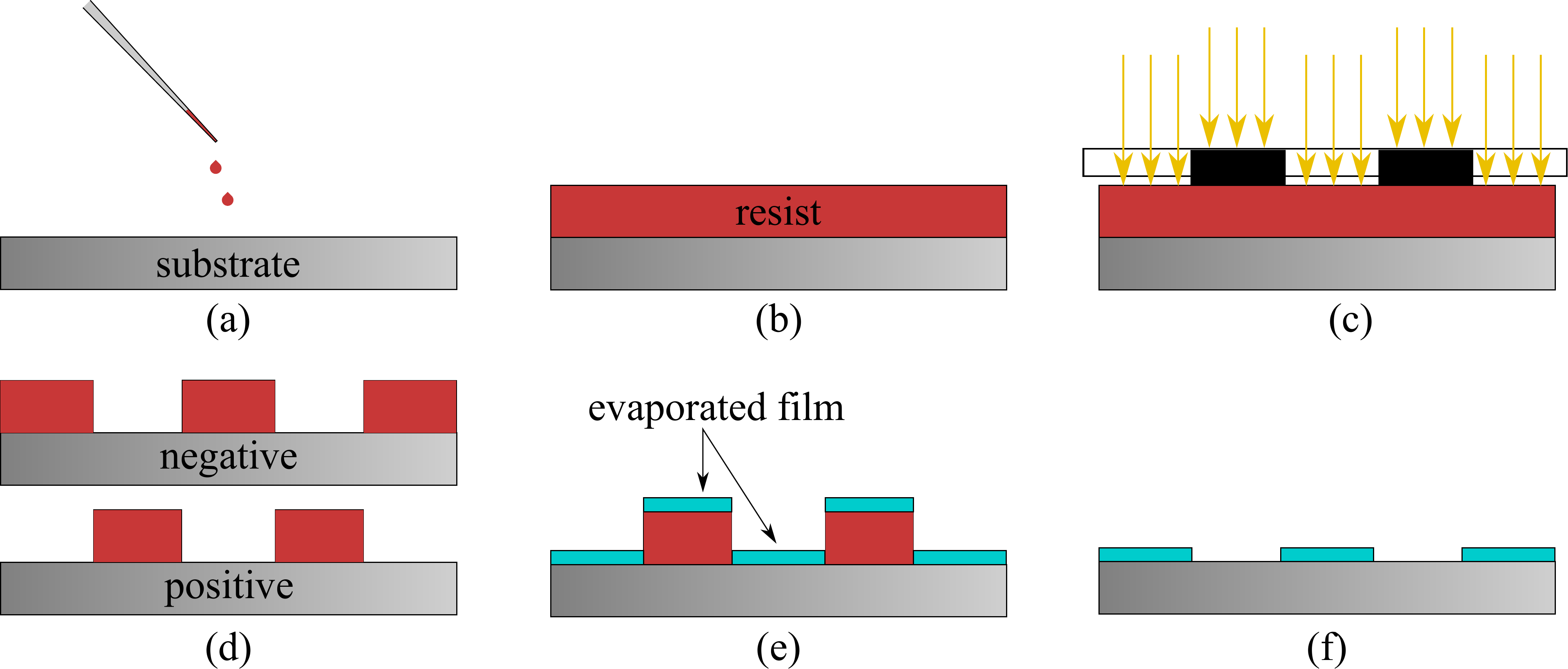}
\caption[Steps in a PLT process]{PLT process steps: (a) \& (b) Spincoating photoresist onto the sample. (c) Exposure of the sample to UV light through a mask to transfer the pattern to the resist. (d) Remaining patterns for negative and positive resists after development. (e) \& (f) Material deposition and lift-off.}
\label{fig:4.2_Litho}
\end{figure}

The process involves spin-coating of resist onto a substrate, with the resist being sensitive to UV light in the case of PLT or to focused electron beams in the case of EBL. Depending on whether a positive or negative photoresist has been applied, the exposed or unexposed area changes its solubility. A solvent, also called a developer, can eventually remove the loosened area. A visual illustration of the process steps involved in PLT is provided in figure \ref{fig:4.2_Litho}. The operating principle of EBL is discussed in more detail in section \ref{sec:SEM}, in which the Scanning Electron Microscope (SEM) is introduced.

In this work, a Hg lamp was used in the case of PLT. It was aligned onto a mask of either printer ink or aluminum to shield portions of the wafer from radiation. In EBL, the exposure was carried out by a focused electron beam that scans over the sample in a predefined pattern.

\begin{figure}[h!]
 \centering
 \subfigure[]{\includegraphics[width=0.45\textwidth]{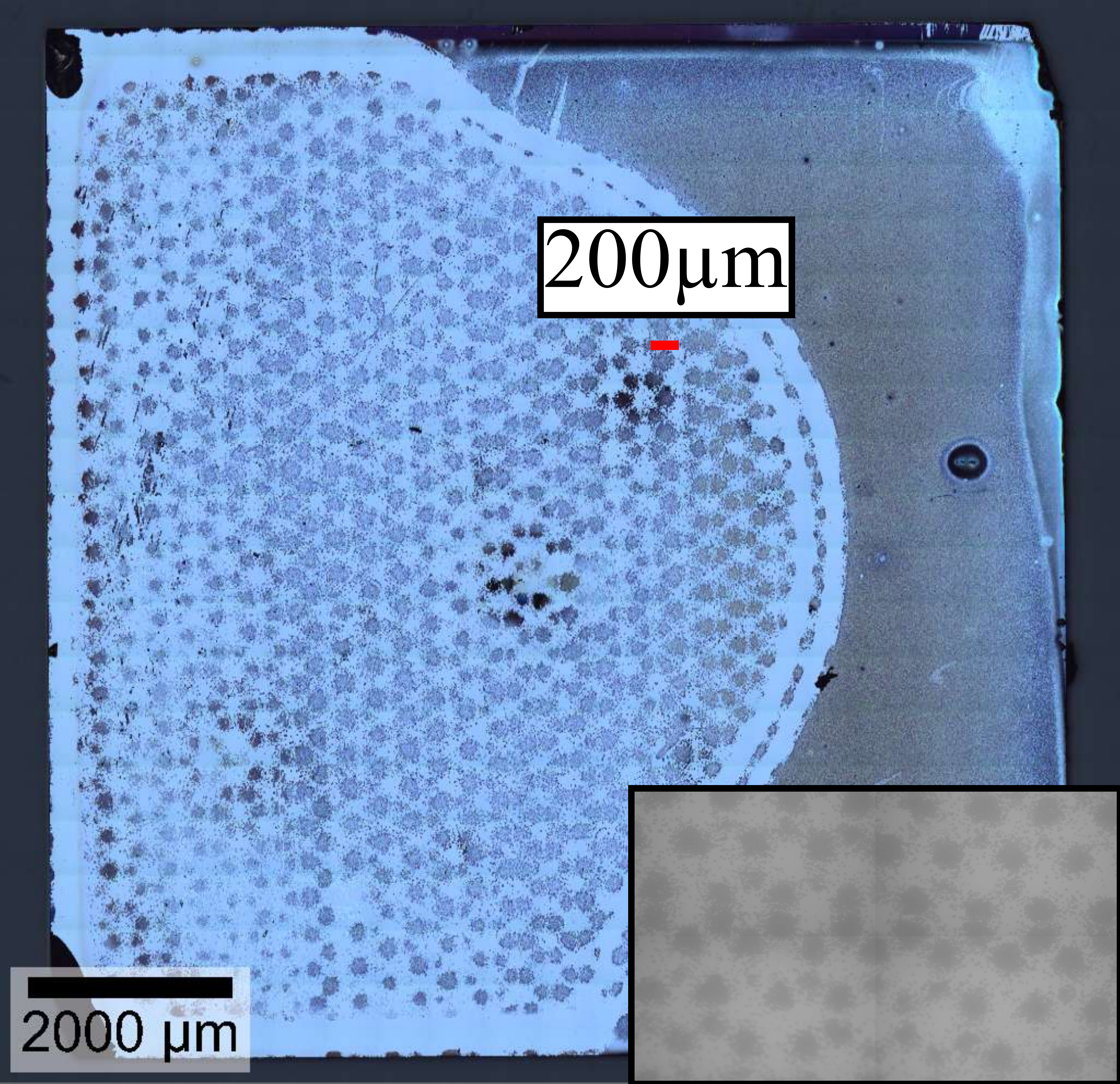}}\quad
 \subfigure[]{\includegraphics[width=0.45\textwidth]{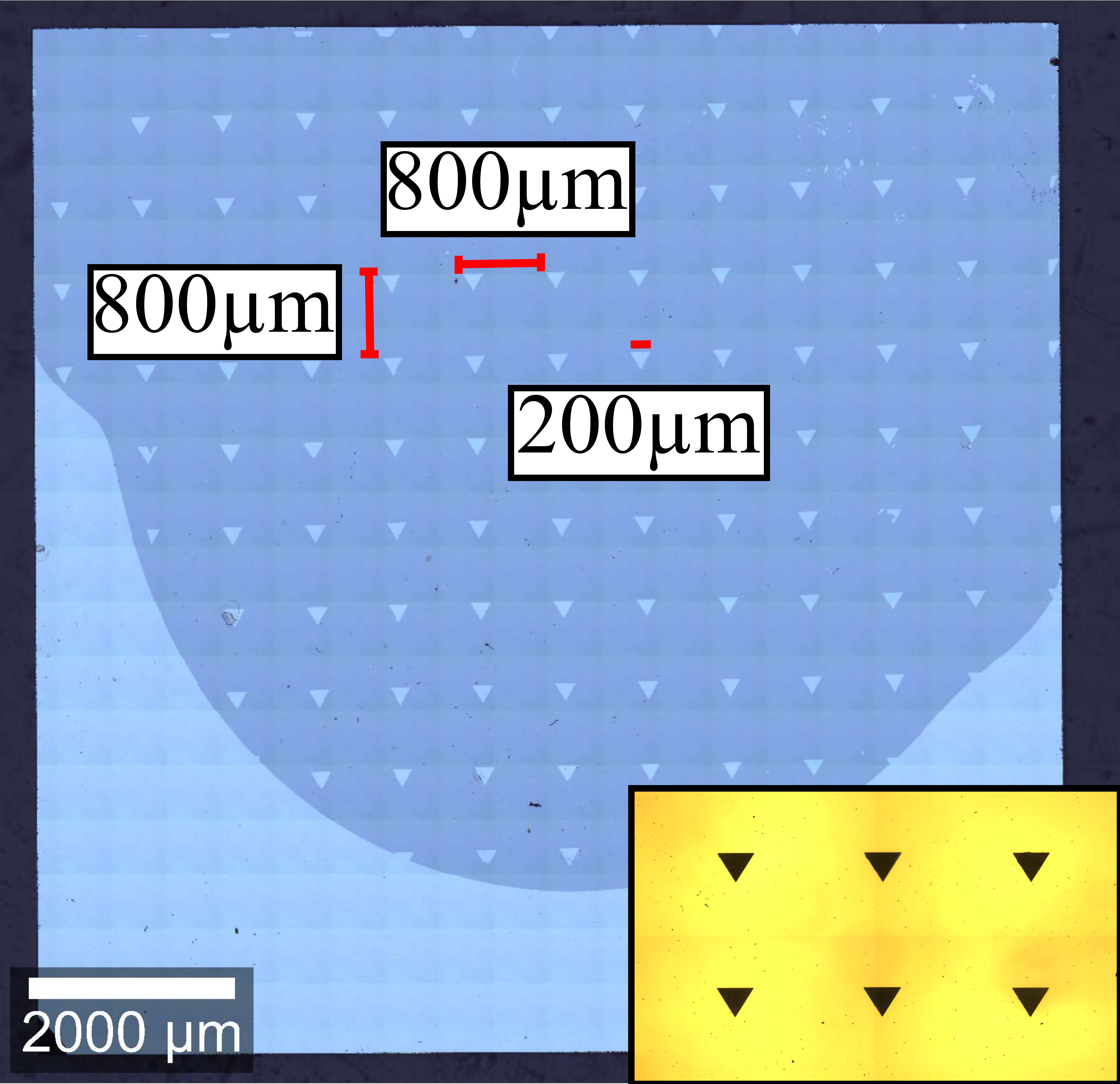}}
 \caption[Patterned seed substrates and masks]{Patterned seed substrates, including the printed mask in (a) and the EBL mask in (b).}
 \label{fig:4.2_MasksResults}
\end{figure}

Two types of masks have thus been made to enable the precursor pattern. The initial, very disordered mask was fabricated using an office printer that prints ink onto a foil. This foil was then taped onto a translucent glass that acted as a mask holder. The recipe to realize the PLT for the pattern on a seed substrate is shown in Table \ref{tab:LithoMasksRecipes}. After the individual steps were executed, tungsten was sputtered onto the wafer, and finally, a lift-off was performed. For this, the sample was immersed in acetone for several minutes in an ultrasonic bath until the remaining resist was dissolved and the metal on top lifted off. An image of the resulting seed substrate, along with a crop of the photomask, is provided in figure \ref{fig:4.2_MasksResults} (a).

A further mask was fabricated to sharpen the resulting structures, using the much more accurate EBL technique. Initially, 500\textit{\,nm} of aluminum was evaporated onto a glass substrate. After that, the steps for EBL, which are also enumerated in Table \ref{tab:LithoMasksRecipes}, were performed. The excess aluminum was then removed to release the pattern by etching the aluminum for seven minutes in an ultrasonic bath with a phosphoric acid etchant. To take off the photoresist, the sample was immersed in acetone and IPA for $10\,mins$ each and rinsed with DI water. This mask was then applied to another glass substrate with vapor-deposited aluminum to change from a dark field to a bright field mask.

The recipe for the seed substrate of the EBL mask can also be observed in Table \ref{tab:LithoMasksRecipes}. The rest of the process was similar to the PLT one explained above. An optical image can be found in figure \ref{fig:4.2_MasksResults} (b).

\begin{table}[h!]
\begin{center}
\begin{tabular}{cccccc} \toprule
 & Fotoresist & Spin-coating & Post bake & Exposure & Development \\ \midrule
\makecell{Printed \\ mask seed} & \makecell{ma-N 1420 \\ (negative)} & \makecell{$10\,s@500\,rpm$ \\ $30\,s@3000\,rpm$} & $120\,s@120\,$°$C$ & $15\,s$ & \makecell{ma-D 533/S \\ for $90\,s$} \\
\rule{0pt}{5ex}
EBL mask & \makecell{ARP 679.04 \\ (negative)} & \makecell{$5\,s@500\,rpm$ \\ $60\,s@4000\,rpm$} & $180\,s@150\,°C$ & / & \makecell{AR600-55 \\ for $60\,s$} \\
\rule{0pt}{5ex}
\makecell{EBL mask \\ seed} & ma-N 1420 & \makecell{$10\,s@500\,rpm$ \\ $30\,s@3000\,rpm$} & $120\,s@120\,$°$C$ & $15\,s$ & \makecell{ma-D 533/S \\ for $90\,s$} \\ \bottomrule
\end{tabular}
\caption[Recipes for the pattern of the seed substrates]{Recipes for the pattern of the seed substrates.}
\label{tab:LithoMasksRecipes}
\end{center}
\end{table}

\subsection{Design of Experiments}
\label{sec:DesignOfExperiment}

In order to achieve large-scale growth, the first step was to conduct a parameter study to determine the intervals at which solid growth occurs. Once the variables were narrowed down, their optimum was sought in the encountered subspace. A DoE was implemented for this purpose.

The practical implementation was carried out using the commercially available software Cornerstone from camLine. The fitted model includes a constant term, linear and quadratic correlations to the input variables, and first-order interactions. This model corresponds exactly to the formula from equation \ref{equ:DoE}.

In order to minimize the number of necessary experiments and extract the maximum amount of information, an optimal experimental design must be chosen. When designing experiments to estimate statistical models, optimal experimental designs allow parameters to be estimated without bias and minimal variance. Here, a D-optimal design is employed. In the case of D-optimality, the determinant of the information matrix of the design is maximized. This results in maximizing the differential Shannon information content of the parameter estimates.\textsuperscript{\cite{Shahmohammadi.2020}}

Nevertheless, the evaluation of the growth poses the biggest challenge. Output variables must be encountered to accurately, quantitatively, and efficiently evaluate the quality and scale of the resulting 2D material. The properties to be estimated by these output parameters are the growth size, the interconnectivity of the individual flakes, and their thickness.

\begin{table}[b!]
\begin{center}
\begin{tabular}{c m{10.0cm}} \toprule
Parameter & Description \\ \midrule
\makecell{Surface coverage \\ (Growth size)} & The percentage of the growth substrate's area, that is covered by 2D material. \\
\rule{0pt}{5ex}
\makecell{Flake size \\ (Growth size)} & The biggest side length of the second to largest individual flake. It is expected to estimate the VW growth type. \\
\rule{0pt}{7ex}
\makecell{PUF square \\ (Interconnectivity)} & The largest square, sufficiently covered by 2D material to allow current to flow between any arbitrarily placed electrodes. It represents the maximum size that the PUF can take. \\
\rule{0pt}{5ex}
\makecell{Layer number \\ (Thickness)} & The average number of layers that the 2D material possesses inside the biggest PUF square. \\ \bottomrule
\end{tabular}
\caption[Evaluation parameters that were chosen in the DoE]{Description of the evaluation parameters that were chosen in the DoE.}
\label{tab:EvaluationParameters}
\end{center}
\end{table}

In order to achieve high efficiency, the evaluation is mainly performed optically. The adopted output parameters are detailed in Table \ref{tab:EvaluationParameters}, including the property that they are expected to represent in brackets.


\subsection{PUF Fabrication Steps}
\label{sec:PUFFabricationSteps}

\begin{figure}[h!]
\centering
\includegraphics[width=15cm]{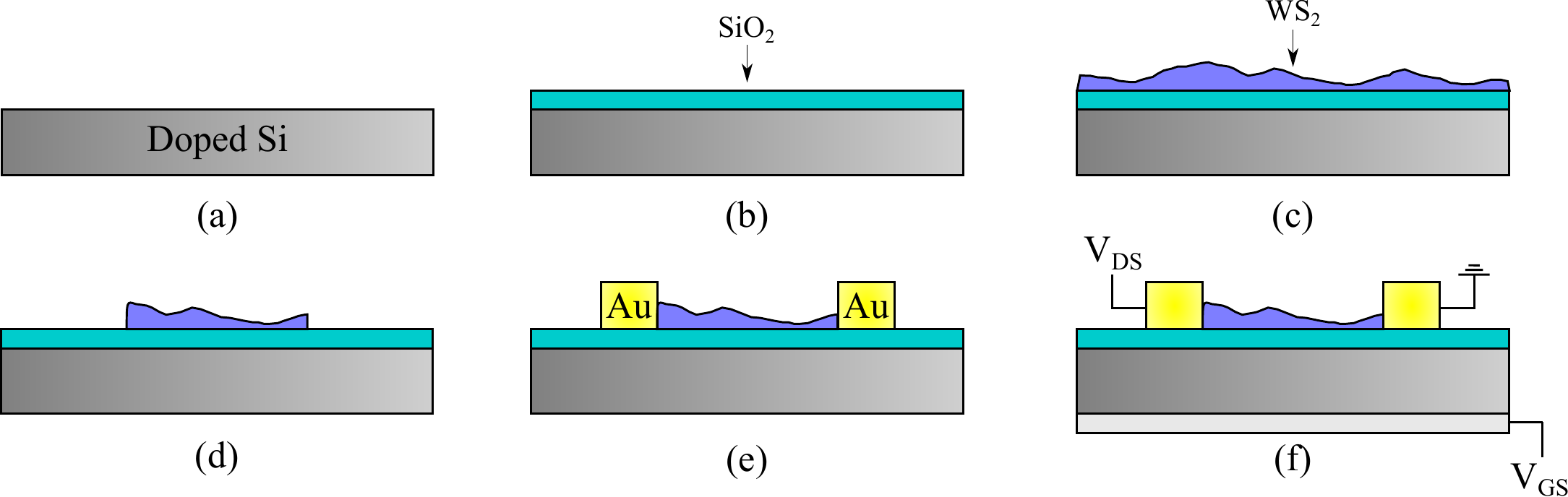}
\caption[Steps involved in PUF fabrication]{PUF fabrication process: (a) Starting with the empty Si substrate (b) Thermal oxidation of $300\,nm$ SiO\textsubscript{2} (c) WS\textsubscript{2} CVD growth (d) Pattern of the 2D material layer (e) Contacting the PUF (f) Backgating of the sample}
\label{fig:4.4_ProcessSteps}
\end{figure}

The fabrication of a PUF prototype starts with an empty Si substrate as sketched in figure \ref{fig:4.4_ProcessSteps} (a). In this thesis, samples with a thickness of 500\textit{\,µm}, and lateral dimensions of $10x10\,mm^2$ are employed. It should be doped in order to be able to manipulate the channel current by a gate contact and thus exponentially increase the number of CRPs.

The substrate then undergoes thermal oxidation until a 300nm thick insulating layer of SiO\textsubscript{2} is formed, as illustrated in (b). Then, in (c), the 2D material is grown on top via CVD. The optimized growth parameters found in this work can be viewed in table \ref{tab:DoE_Output}. Finally, the TMD is structured and contacted, shown in (d) and (e). The details are outlined in section \ref{sec:5.4_PatterningAndContacting}. To obtain a more sophisticated PUF design, the wafer can be additionally equipped with a back gate, as demonstrated in (f).

\subsection{Characterization}
\subsubsection{Optical Microscopy}

Characterization methods must be utilized to detect 2D material and determine its quality. The simplest and most efficient one is optical microscopy. Here, the sample is placed under a lens and is subsequently illuminated, resulting in magnification. As previously mentioned, TMDs on SiO\textsubscript{2} can be easily identified since they exhibit a strong contrast. Furthermore, this contrast changes when the TMDs differ in even only a single layer. In other words, one can determine if there is growth on the sample, identify its location and size, and estimate the layer number.

Nevertheless, there are also a couple of downsides to optical microscopy. It is difficult, if not impossible, to identify defects, such as grain boundaries in the 2D material. In addition, the optical method is limited by the diffraction limit of light. This poses a constraint since the WS\textsubscript{2} flakes sometimes go below this limit. It can be calculated by the following formula:

\begin{equation}
d=\frac{\lambda}{2NA}
\label{equ:OpticalMicroscopy}
\end{equation}

where $\lambda$ stands for the wavelength of the light and $NA$ for the numerical aperture of the objective. Thus, the detection limit usually lies within $0.5$ to 0.2\textit{\,µm}. The optical microscopes used in this work are the $LEICA$ $DM6$ $M$ and the $Witec$ $Alpha$ $300R$ spectrometer.

\subsubsection{Raman Spectroscopy}
\label{sec:RamanSpectroscopy}

A more advanced method to characterize TMDs is Raman spectroscopy. It examines the vibrational, translational, and various other modes in a system. When light interacts with a surface, most of it is reflected elastically. However, one photon in about $10^5$ scatters inelastically as it interferes with the molecular or lattice vibrations in the sample. This leads to characteristic peaks that are shifted from the laser wavelength by a certain magnitude. This distance is usually measured and displayed in the unit of wavenumbers, namely $cm^{-1}$. The frequency shift of the light depends on the bonds and symmetries in the material, which leads to characteristic peaks for each Raman active vibrational mode and is thus a unique fingerprint for the individual compounds.

Raman spectroscopy is a very prominent method for the characterization of TMDs as it brings several merits with it. One can identify their structures, detect the defects and even quantify the number of layers very precisely. Moreover, similar to optical microscopy, it is a non-destructive method.

Most effects that occur in Raman spectroscopy can be described by classical Raman theory, which is explained by the following calculations. The electric field strength of a monochromatic laser beam, incident on the sample, fluctuates as a function of time:

\begin{equation}
E=E_0 \cdot cos(\omega_0 \cdot t)
\label{equ:RamanClassic1}
\end{equation}

Here, $E_0$ is the amplitude of the vibration, and $\nu_0$ the laser frequency. If the laser irradiates a TMD, an electric dipole moment $P$ is induced:

\begin{equation}
P = \alpha E = \alpha E_0 \cdot cos(\omega_0 \cdot t)
\label{equ:RamanClassic2}
\end{equation}

with $\alpha$ being a proportionality constant, called polarizability. Further, the nuclear displacement $q$ of the molecule vibrating at a frequency $v_m$ amounts:

\begin{equation}
q=q_0 \cdot cos(\omega_m \cdot t)
\label{equ:RamanClassic3}
\end{equation}

where $q_0$ stands for the vibrational amplitude. One can now approximate the polarizability for a small vibrational amplitude $q_0$ by a Taylor series:

\begin{equation}
\alpha = \alpha_0 + \frac{\partial \alpha}{\partial q}\big|_0 \cdot q_0 + ...
\label{equ:RamanClassic4}
\end{equation}

Here, $\alpha_0$ indicates the polarizability at the equilibrium position, and $\frac{\partial \alpha}{\partial q}\big|_0$ the rate of change of the polarizability with respect to a change in $q$, evaluated at the equilibrium position. If one now combines equations \ref{equ:RamanClassic2}, \ref{equ:RamanClassic3} and \ref{equ:RamanClassic4}, one obtains the following:

\begin{flalign}
\begin{aligned}
P& = \alpha E_0 \cdot cos(\omega_0 \cdot t)= && \\
&=\alpha_0 E_0 \cdot cos(\omega_0 \cdot t) + \frac{\partial \alpha}{\partial q}\big|_0 \cdot q_0 E_0 \cdot cos(\omega_0 \cdot t) \cdot cos(\omega_m \cdot t)&&
\label{equ:RamanClassic5}
\end{aligned}
\end{flalign}

which can be finally rewritten to:

\begin{flalign}
\begin{aligned}
P &= \alpha_0 E_0 \cdot cos(\omega_0 \cdot t) + &&\\
&+\frac{1}{2} \frac{\partial \alpha}{\partial q}\big|_0 \cdot q_0 E_0 \cdot (cos((\omega_0+\omega_m) \cdot t)+cos((\omega_0-\omega_m) \cdot t))&&
\label{equ:RamanClassic6}
\end{aligned}
\end{flalign}

\begin{figure}[t!]
\centering
\includegraphics[width=10cm]{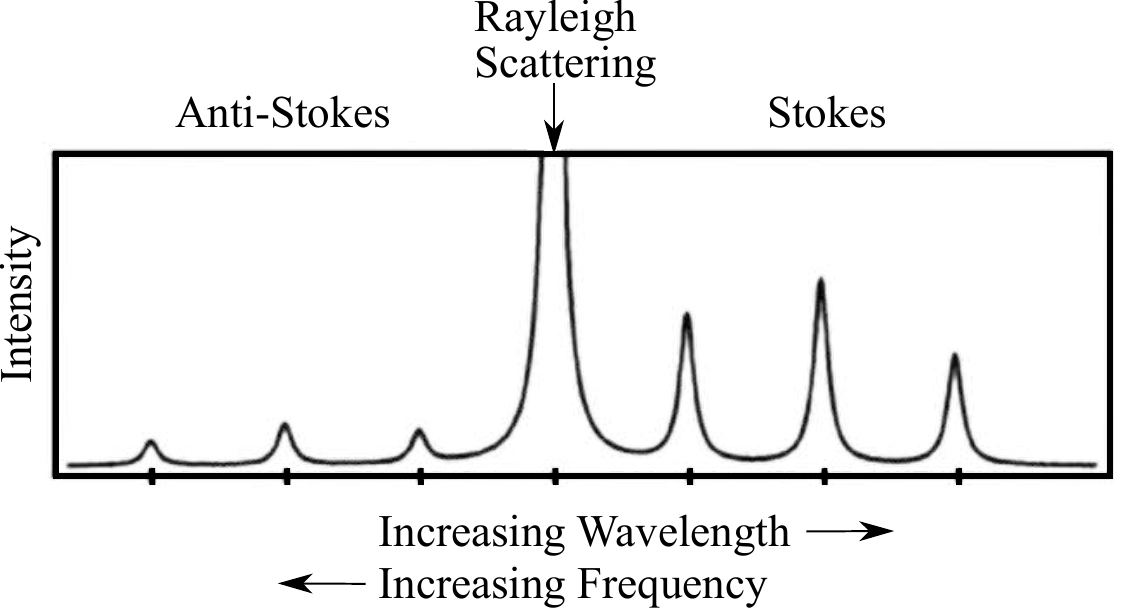}
\caption[Illustrative Raman spectrum]{An illustrative Raman spectrum showing the Anti-Stokes, the Rayleigh and the Stokes peaks, adapted from \cite{Ferraro.2003}.}
\label{fig:4.5_RamanSpectrum}
\end{figure}

According to classical theory, the first term in equation \ref{equ:RamanClassic6} describes the oscillating dipole that radiates light at frequency $\nu_0$, which is also called Rayleigh scattering. The second term results in the Anti-Stokes peaks, caused by inelastic Raman scattering along with a frequency shift of $-\nu_m$. And finally, the third term represents the Stokes scattering which exhibits a frequency shift of $\nu_m$. This means that a vibrational mode is Raman active when the rate of change of the polarizability with the vibration $\frac{\partial \alpha}{\partial q}\big|_0$ is non-zero. On the other hand, a vibration is infrared-active when the molecule's dipole moment changes with the vibration. An illustrative Raman spectrum is provided in figure \ref{fig:4.5_RamanSpectrum}. Note that the Anti-Stokes bands typically feature a lower intensity, since they represent the transition from the excited to the unexcited state.\textsuperscript{\cite{Ferraro.2003}}

\begin{figure}[b!]
\centering
\includegraphics[width=15cm]{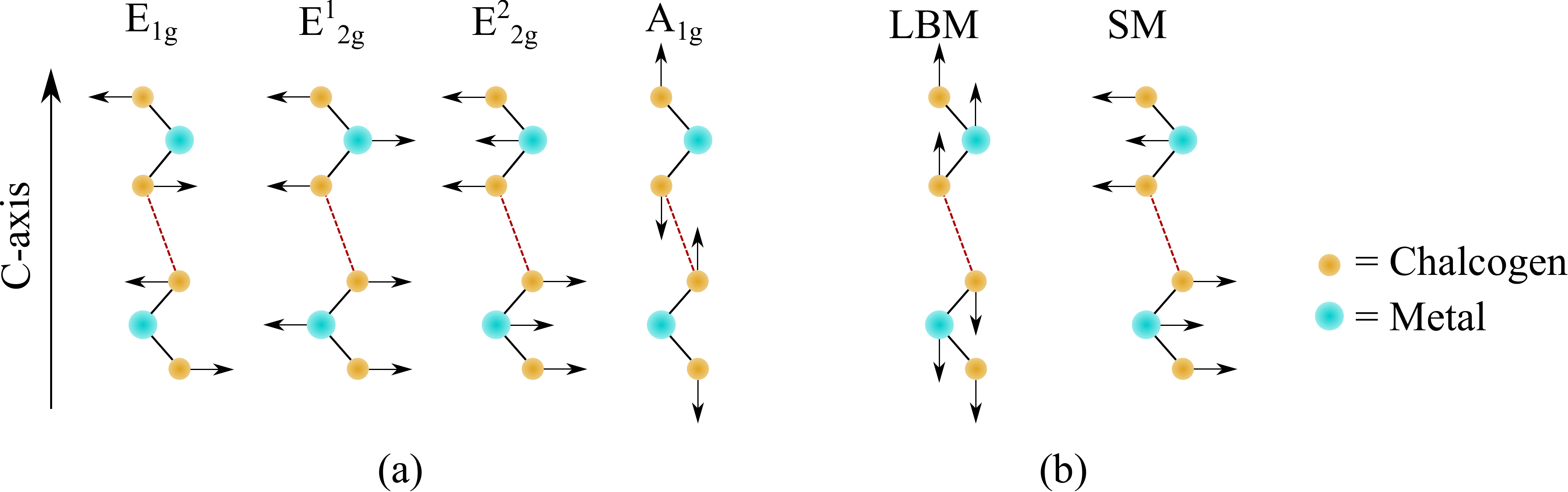}
\caption[Raman-active modes in TMDs]{Overview over the most important Raman-active modes in TMDs, adapted from \cite{Tonndorf:13}. LBM stands for layer breathing modes and SM for shearing modes.}
\label{fig:4.5_VibrationalModes}
\end{figure}

When it comes to TMDs, one can determine several characteristic modes that may be used to examine their properties. In bulk form, they possess the symmetry group $D_{6h}$ with the Raman active modes $A_{1g}$, $2E_{2g}$ and $E_{1g}$, where $2E_{2g}$ can be distinguished into $E^1_{2g}$ and $E^2_{2g}$. When narrowed down to a few layers, the symmetry group changes to $D_{3h}$ for odd-layer numbers with the Raman active modes $A_1'$, $E''$, and $2E'$. Even-layer TMDs own the symmetry group $D_{3d}$, with $3A_{1g}$ and $3E_g$.\textsuperscript{\cite{Li.2013,MolinaSanchez.2011}} A representative sketch of the most important Raman-active vibrational modes is given in figure \ref{fig:4.5_VibrationalModes} (a).

The two most significant modes, which were also mainly investigated for the analysis in this work, are the $E^1_{2g}$ and the $A_{1g}$ modes. In TMDs with few and odd layer numbers, these modes are called $E'$ and $A_1'$, and in the case of even layer numbers, $E_g$ and $A_{1g}$. These modes will henceforth all be called $E^1_{2g}$ and $A_{1g}$, regardless of the layer number. In reality, close to the $E^1_{2g}$ peak lies an acoustic mode, called $2LA(M)$. The two peaks merge, making it tricky to correctly evaluate the $E^1_{2g}$ mode, as it can often be distorted by a wrong fit. That is why $A_{1g}$ mode is more trustworthy. The intensity, position, and full width at half maximum (FWHM) of these two peaks provide information about the quality and number of the layers. An assignment of the peak characteristics to the layer numbers with atomic force microscopy (AFM) comparison measurement is provided in section \ref{sec:GrowthCharacterization}.

Lastly, in addition to the $E^1_{2g}$ and $A_{1g}$ modes, layer breathing modes (LBM) and shearing modes (SM) exist. They occur in the low-frequency region of the Raman spectrum and provide useful information about the number of layers. Their vibrational modes are sketched in figure \ref{fig:4.5_VibrationalModes} (b).

The Raman spectroscope used in this work is the $Witec$ $Alpha$ $300R$ spectrometer with a laser wavelength of either $405$ or $532\,nm$.

\subsubsection{Photoluminescence Spectroscopy}

In photoluminescence (PL), the sample is irradiated with light of a particular wavelength. When the energy of the light is higher than that of the bandgap, photons are absorbed, and electrons are excited into the conduction band. At the same time, holes are created in the valence band. The excited electrons now undergo various energy relaxation events due to Coulomb scattering and the excitation of phonon modes. At some point, they will get close to the conduction band minimum, forming excitons with the positrons at the valence band maximum, which recombine under the emission of a photon. It can be measured to determine the characteristic bandgap of the material.

Equivalent to Raman, there is a PL peak shift in TMDs when the layer number is varied. This stems from the strong interlayer coupling of the $d$-orbitals of the transition metal and the anti-bonding $p_z$-orbitals of the chalcogen atoms at the $\Gamma$ and $K$ points.\textsuperscript{\cite{L.M.J.J.Peters.June2021}} Additionally, the intensity of the peaks increases exponentially in monolayer material.

Similarly to Raman, the intensity, the FWHM, and the position of the peaks, previously enumerated in figure \ref{fig:3.1_WS2CrystalStructureAndBandDiagram}, will be assigned to the layer numbers by an AFM comparison measurement in section \ref{sec:GrowthCharacterization}. To record PL, the same experimental setup was employed as for the Raman analysis.

\subsubsection{Atomic Force Microscopy}

\begin{figure}[h!]
\centering
\includegraphics[width=6cm]{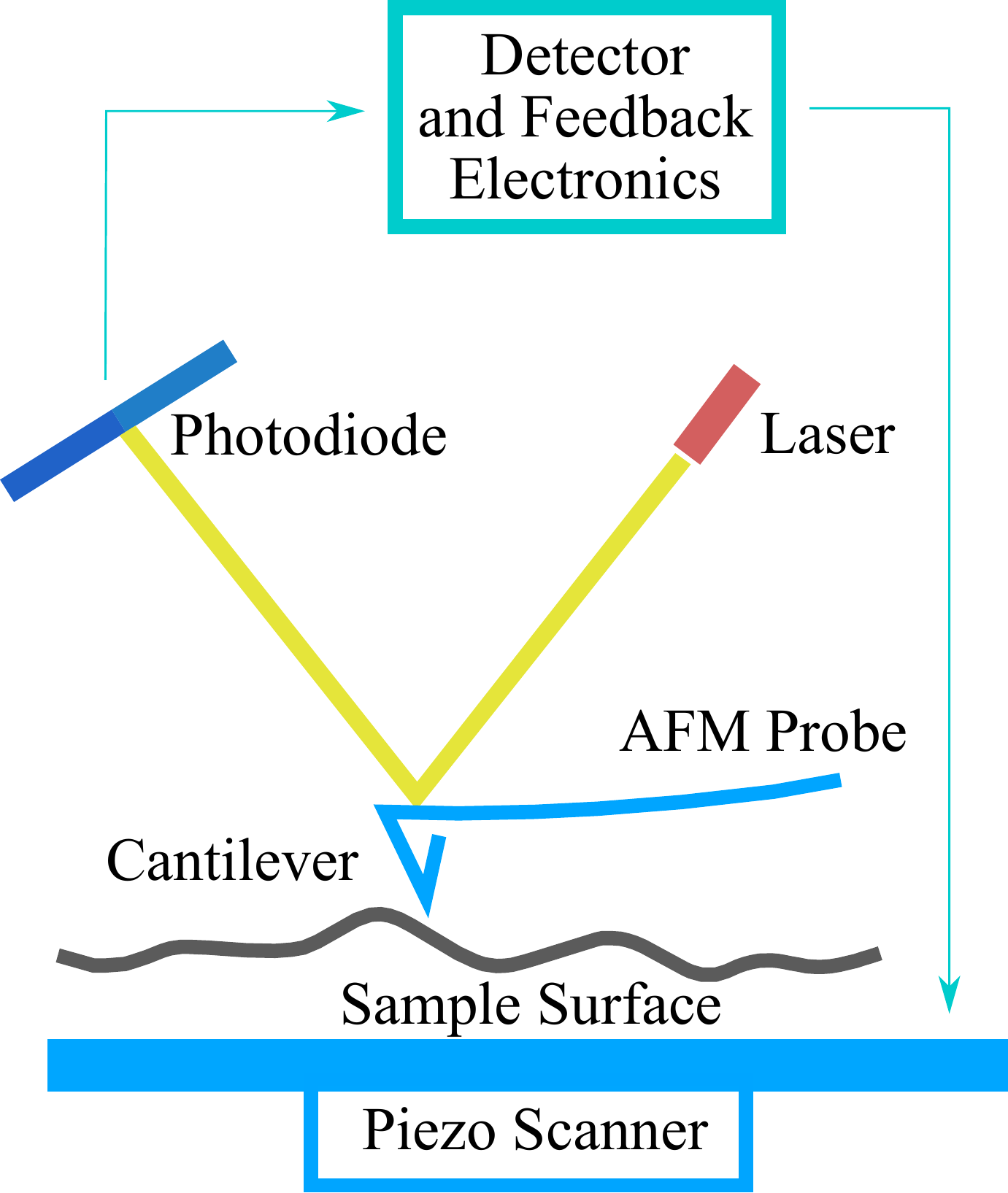}
\caption[Working principle of the AFM]{The working principle of the AFM, adapted from \cite{NanoAndMoreGmbH.16.05.2022}. The sample is scanned by a cantilever, whose deflection changes as it moves over structures of different heights. This displacement is detected by a laser beam and finally measured by a position-sensitive photodetector. Thereby, 3D near-atomic-resolution images of the surface can be generated.}
\label{fig:4.5_AFM}
\end{figure}

AFM is a high-resolution, non-optical imaging technique that is capable of examining surfaces at the nanoscale.\textsuperscript{\cite{NanoAndMoreGmbH.16.05.2022}} The principle of operation is illustrated in figure \ref{fig:4.5_AFM}. The sample is scanned by a cantilever attached to an AFM probe in a raster pattern. A piezoelectric sensor controls the lateral and vertical position of the probe. The deflection of the cantilever changes as it moves over structures of different heights. A laser beam measures this displacement. It is reflected from the back of the probe and directed into a position-sensitive photodetector. A feedback loop controls the vertical extension of the scanner to achieve a nearly constant deflection and thus interaction force. Hence, three-dimensional topographic images, with atomic or near-atomic resolutions, of the surface can be generated.\textsuperscript{\cite{Gross.2009}}

The topographies in this thesis were recorded in tapping mode throughout. This involves oscillating the cantilever using a piezoelectric actuator close to its resonant frequency. It is then brought close to the surface of the substrate, with the tip touching the sample at the lower end of the oscillation. This attenuates the amplitude, which then can be measured by the laser and photodiode.

In the case of TMDs, AFM can be employed to obtain very accurate images of individual flakes. Even grain boundaries can be resolved. Furthermore, it is possible to determine the height of the 2D material with $<\,nm$ resolution and thus assign the number of layers trustfully.\textsuperscript{\cite{Hajiyev.2013}} The AFM measurements in this work were performed on a $Jupiter$ $XR$ $AFM$ from $Oxford$ $Instruments$ with a $RTMSEP$ tip from $VECCO$ in tapping mode.

\subsubsection{Scanning Electron Microscopy}
\label{sec:SEM}

In the case of SEM, a focused electron beam scans across a sample in a raster-wise pattern. It then detects the interactions of the electrons with the surface by a plethora of procedures.\textsuperscript{\cite{.2003,Zhou.2006}} Various types of electrons can be detected by this method. The most frequently exploited ones are secondary electrons and backscattered electrons. The former are created by the inelastic scattering of electrons in the beam volume at the sample, which leads to the ejection of low-energy secondary electrons. Their inherent low energy results in a smaller mean free path, and thus they can only be detected if they are generated close to the surface. This type of electron is mainly used to record high-resolution images of the surfaces.

Backscattered electrons, on the other hand, hold a high energy. They are elastically scattered within the beam volume. Since the amount of backscattered electrons is proportional to the atomic number of the atoms, they are used to distinguish between materials with different atomic numbers.

The SEM column consists of an electron source and a set of lenses. The electrons are condensed into a beam and finally focused on the sample through a final lens, also called objective lens. Coils direct the beam at the source and the final lens. The wavelength of the electron beam can be determined by the following formula:

\begin{equation}
\lambda = \frac{h}{\sqrt{2 e V_a m_e}}
\label{equ:SEM}
\end{equation}

where $V_a$ stands for the acceleration voltage and $m_e$ for the free electron mass. Consequently, the wavelengths achieved can adopt minimal values in the range of $1-40\,pm$ at accelerating voltages of $30-1\,kV$, leading to high-resolution images.

\subsubsection{Electrical Measurements}

\begin{figure}[b!]
\centering
\includegraphics[width=13cm]{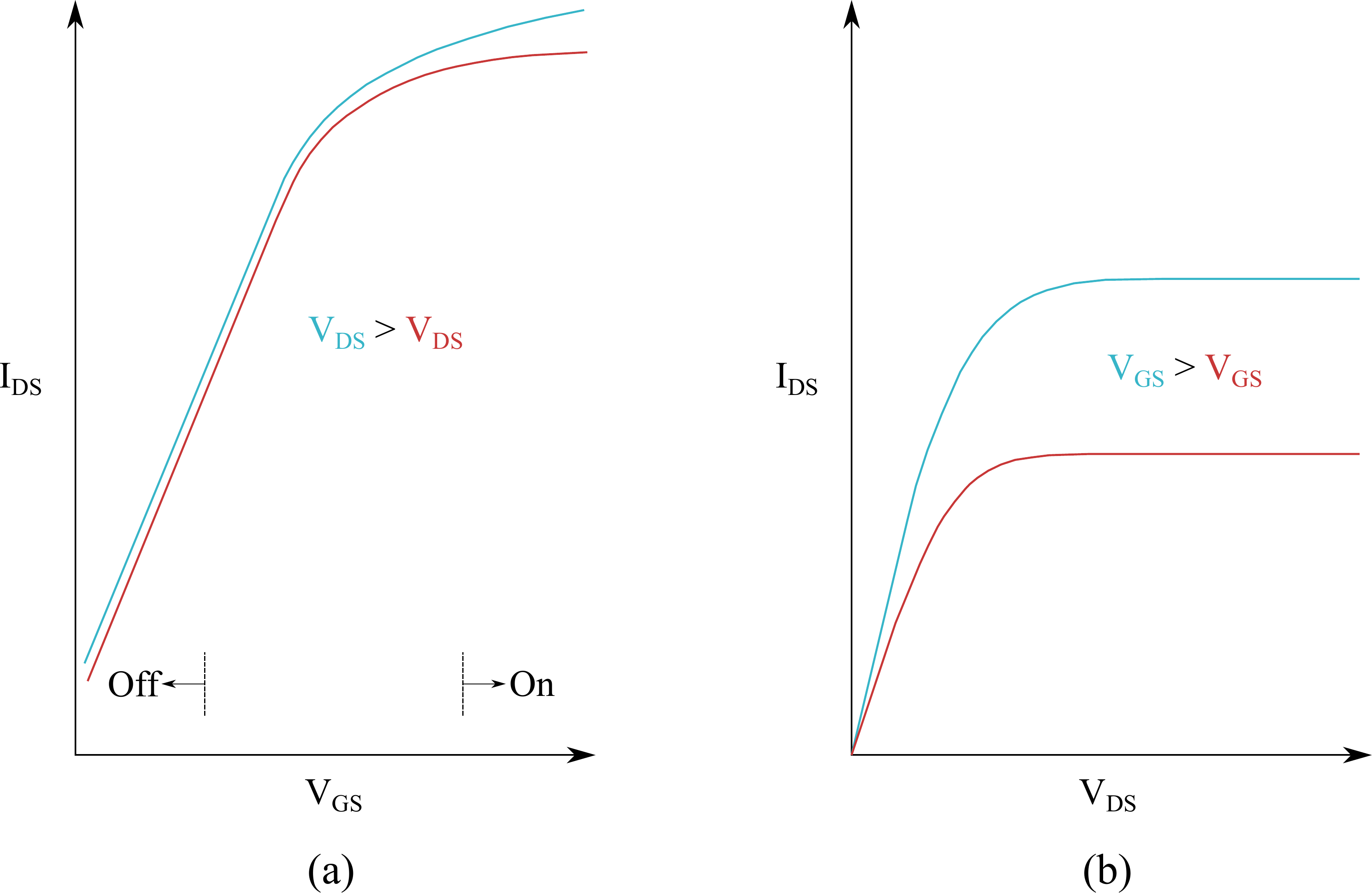}
\caption[A typical electrical signal of an n-type transistor]{A typical electrical signal of an n-type transistor with a voltage sweep of (a) the gate-source voltage $V_{GS}$ and (b) the drain-source voltage $V_{DS}$.\textsuperscript{\cite{ManishChhowallaDebdeepJenaandHuaZhang.2016}}}
\label{fig:4.5_Electrical}
\end{figure}

The device structure of the PUFs and their contacts can be observed in figure \ref{fig:4.4_ProcessSteps}. The electrical signals are measured by holding either the gate voltage ($V_{GS}$) or the drain-source voltage ($V_{DS}$) constant and sweeping the other. Characteristic curves for similar device structures can be found in figure \ref{fig:4.5_Electrical}. The CRPs of the PUFs are ultimately generated by the electrical signals flowing through the random percolation network made of WS\textsubscript{2}.

There are some quantities that represent the quality of the transistor. One of them is the on/off ratio, which indicates the current ratio between the on and off states of the device. A small ratio indicates that the device is not able to turn off properly, which leads to unwanted leakage current. Another factor is the field-effect mobility, which is defined by the following formula:

\begin{equation}
\mu_{FE}=\frac{1}{C_{OX}} \frac{\partial I_{DS}}{\partial V_{GS}} \frac{L}{W} \frac{1}{V_{DS}}
\label{equ:Electrical}
\end{equation}

where $C_{OX}$ stands for the oxide capacitance, $\frac{\partial I_{DS}}{\partial V_{GS}}$ for the transconductance, $L$ for the channel length, and $W$ for the channel width. It is essential to mention here that when fabricating a PUF, one does not strive to maximize the field-effect mobility as usual but to achieve as random an output result as possible.\textsuperscript{\cite{ManishChhowallaDebdeepJenaandHuaZhang.2016}}

\clearpage
\newpage
\thispagestyle{plain}

\section{Results \& Discussion}
\label{sec:ResultsDiscussion}
\subsection{WS\textsubscript{2} Growth Characterization}
\label{sec:GrowthCharacterization}

In the first section of this chapter, the PUF material WS\textsubscript{2} is characterized by optical analysis, AFM, Raman, and PL. This is followed by a discussion about the preliminary optimization of the CVD growth via trial and error. Next, the results of the systematic optimization using a DoE approach are presented. Finally, the material is successfully patterned, and the first contacting trials are executed.

\subsubsection{Optical Analysis}

\begin{figure}[t!]
\centering
\includegraphics[width=12cm]{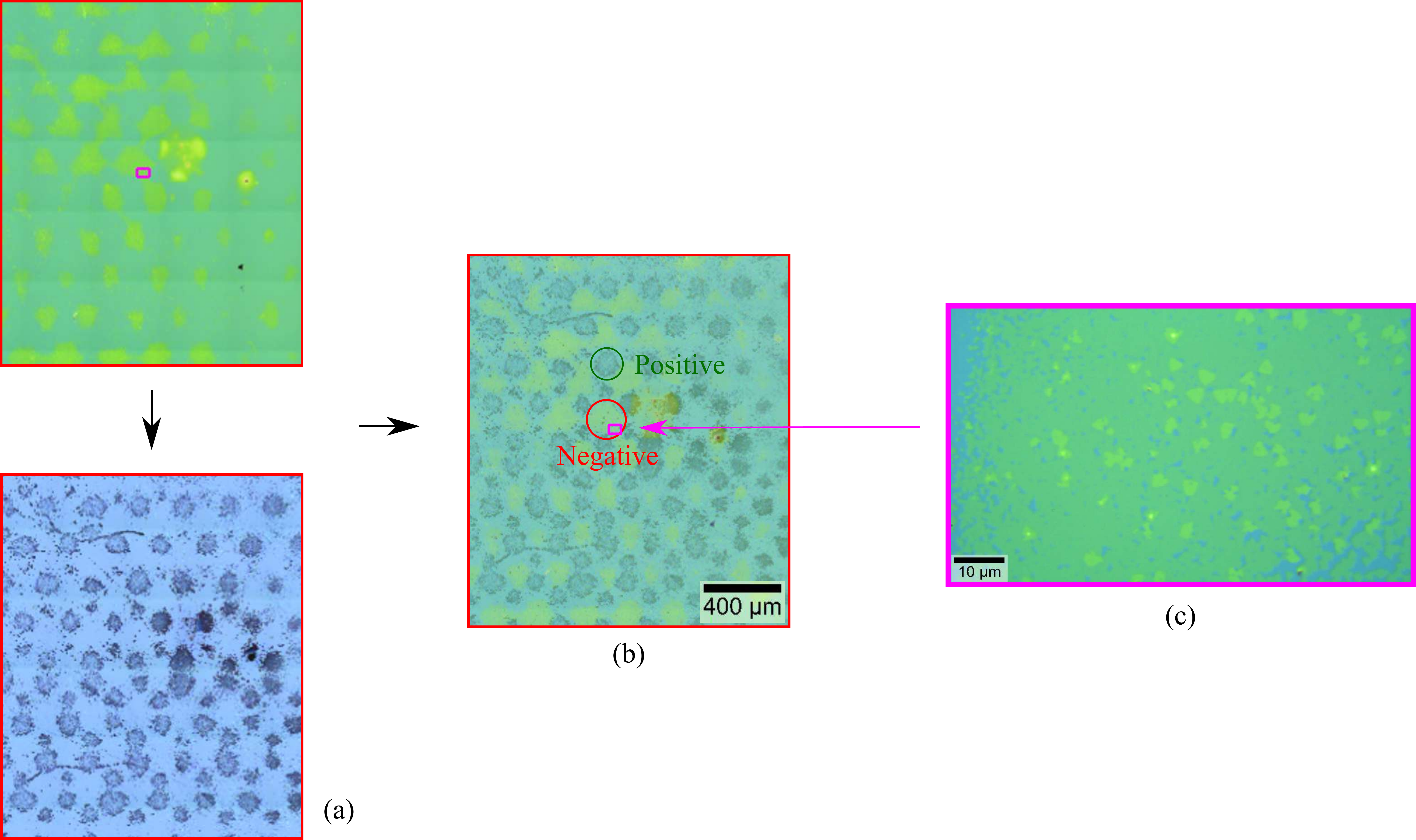}
\caption[An optical image with negative and positive growth]{An optical image of the growth substrate with the light green WS\textsubscript{2} on the top and the seed substrate with the black precursor on the bottom in (a), stacked as in the microreactor, to illustrate the negative and positive growth regimes in (b). A supplementary magnified image in (c), zooming into the pink crop in (b). The printed mask from figure \ref{fig:4.2_MasksResults} (a) was applied to create the pattern.}
\label{fig:5.1_OpticalImage}
\end{figure}

An optical image of the growth can be found in figure \ref{fig:5.1_OpticalImage}. The growth substrate featuring the light green 2D material on the slightly darker SiO\textsubscript{2} can be seen in the top image in (a). Underneath, one can find the seed substrate containing the patterned WO\textsubscript{3} exhibiting a black color and the Si underground in blue. As previously stated, the wafers differ in color depending on the oxide thickness. Hence, growth experiments were conducted in the context of this work, which show that the oxide thickness and thus the surface color does not significantly influence the growth of WS\textsubscript{2}.

To obtain (b), the transparency of the image of the growth substrate was lowered and it was mirrored. Subsequently, it was placed over the image of the seed substrate, exactly as in the real microreactor, in order to estimate the spatial component of the growth as a function of the position of the precursor.

Now, one can generally define three different growth regimes: positive, negative, and flat. Positive growth occurs just beneath the structured WO\textsubscript{3} precursor, while negative growth emerges in the vacant gaps. The latter appears as the metal precursor diffuses into the empty spots throughout the process. Two circles in the middle image of \ref{fig:5.1_OpticalImage} highlight the areas respectively. When the precursor is not structured but applied as a continuous, large-area homogeneous film, it is in the following referred to as flat regime.

A magnified image of the pink crop is provided in figure \ref{fig:5.1_OpticalImage} (c). The light green TMD appears to exhibit an SK growth mode, i.e. layer plus island growth. It shows darker monolayer as the base and brighter, few-layered triangular features atop. The monolayer commences coalescing into a continuous film, a few 100\textit{\,µm} in size. Note that the variation of the brightness allows the monolayer to be distinguished from the multilayered structures on top.

\subsubsection{Atomic Force Microscopy}

\begin{figure}[t!]
\centering
\includegraphics[width=12cm]{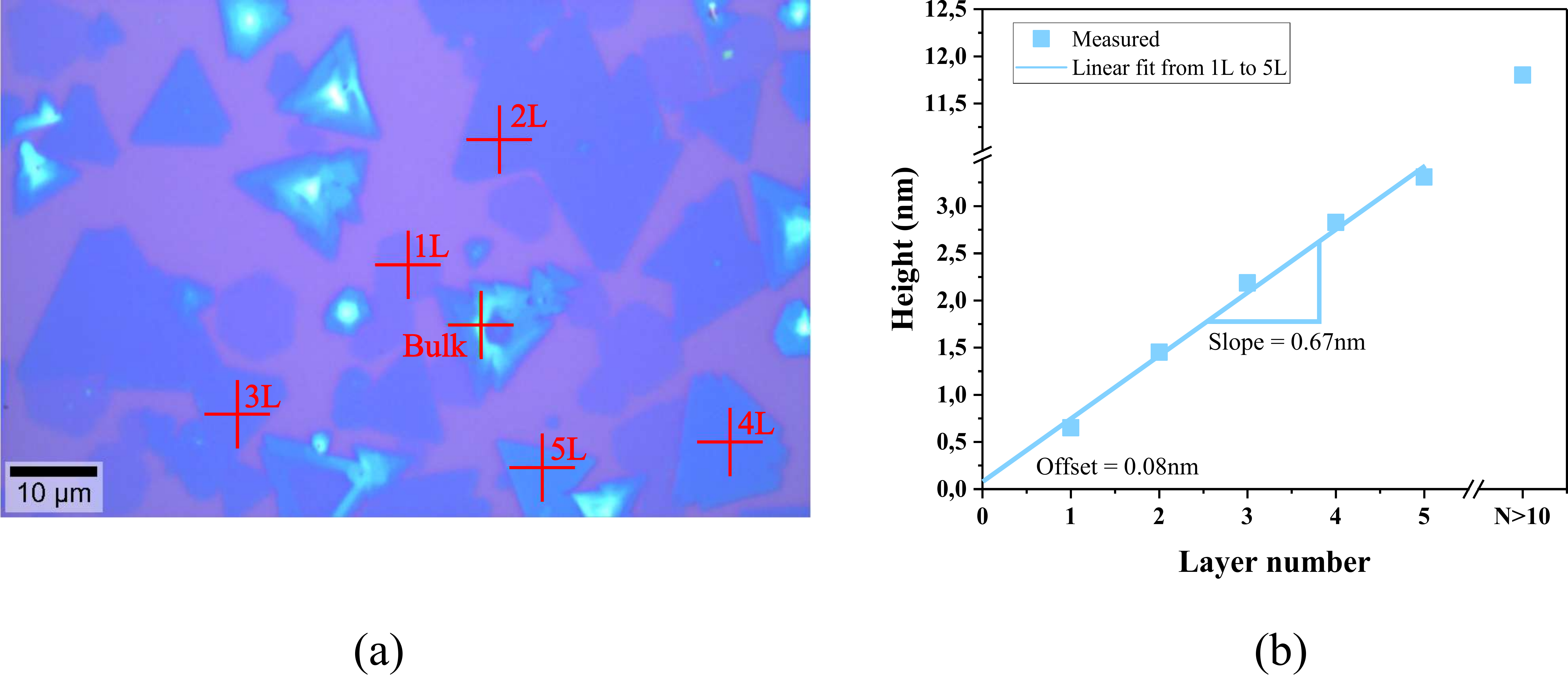}
\caption[Assigning the AFM height to the layer number]{An optical image of WS\textsubscript{2} flakes used to assign the Raman and PL characteristics to the layer numbers, labeled with the spot at which the spectra from the next section were acquired in (a). The fit of the height of the flakes, measured via AFM in (b).}
\label{fig:5.1_AFMLayers}
\end{figure}

In the following, several flakes with ascending brightness were investigated. Their thickness was measured with AFM and assigned to a layer number. An optical image of the flakes along with the measured thicknesses can be viewed in figure \ref{fig:5.1_AFMLayers}. In (a), the respective flakes are labeled with the deduced layer numbers. The crosses indicate the sites at which the Raman and PL spectra from the following subsection were acquired. Based on these measurements, the characteristic properties of the spectra, such as peak position or intensity, can be derived as a function of the layer number. The motivation for performing such an analysis is to efficiently determine the flakes' thickness to analyze the layer number of a large amount of samples in the DoE in section \ref{sec:5DesignOfExperiment}.

The heights of layers one through five were fitted by a straight line. The slope, and thus the thickness of each layer amounts to approximately $0.67\,nm$, which is in agreement with the literature.\textsuperscript{\cite{Godin.2017}} Additionally, note that an offset of $0.08\,nm$ is present, which can be attributed to the different tip-surface interactions when measuring on SiO\textsubscript{2} and WS\textsubscript{2}. Additionally, when examining subfigure (a), the increment in the flake's brightness can now be roughly assigned to a layer number. Finally, note that the WS\textsubscript{2} occurs in hexagonal and triangular shapes, which, as mentioned above in section \ref{sec:ChemicalVapourDepositionInDetail}, depends on the ratio of sulfur and tungsten provided during the process.

\begin{figure}[b!]
\centering
\includegraphics[width=12cm]{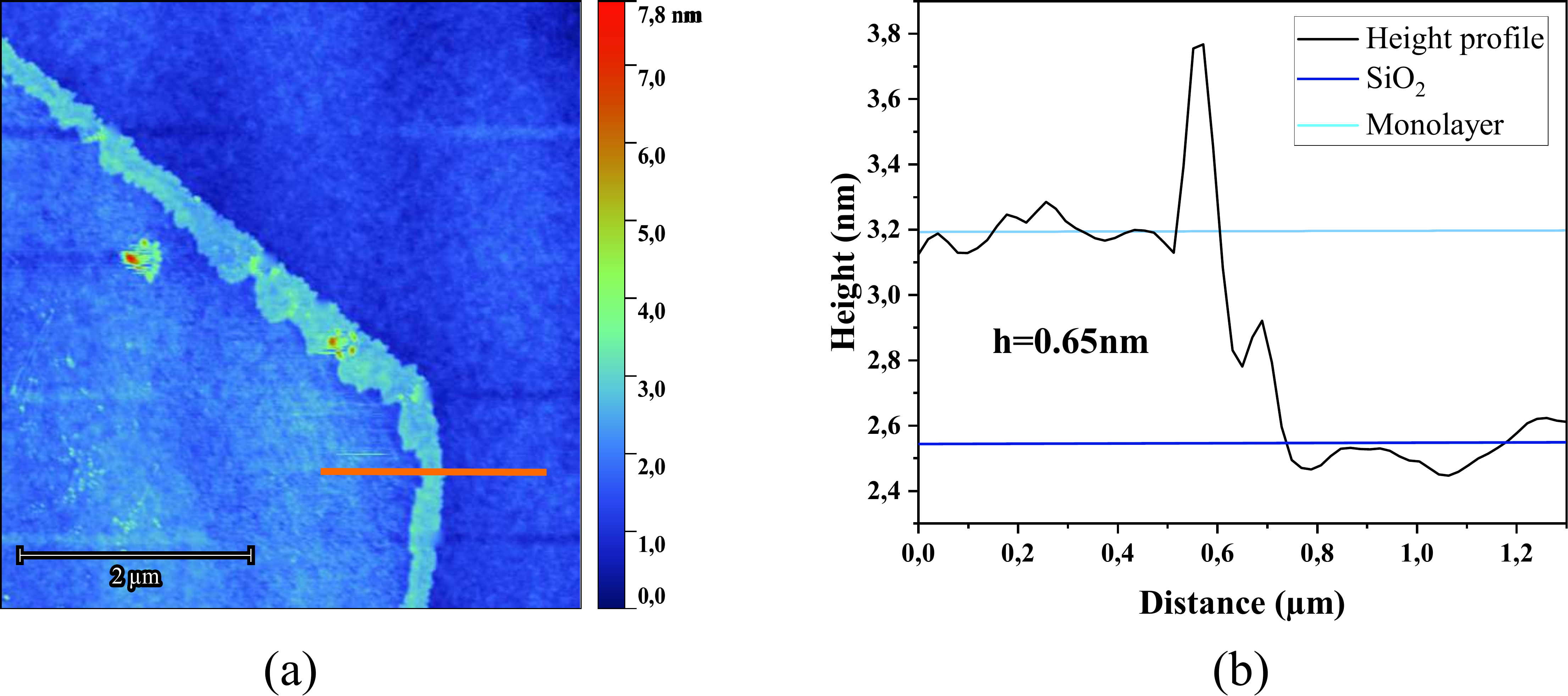}
\caption[AFM monolayer step height evaluation]{AFM image of a monolayer flake in (a) with the corresponding height profile in (b), measured along the orange line. The SiO\textsubscript{2} and WS\textsubscript{2} plateaus were fitted by constant functions and the height was estimated to be $0.65\,nm$.}
\label{fig:5.1_AFM_ML}
\end{figure}

Two example evaluations are provided below to illustrate the approach to determining the thickness of a flake. Figure \ref{fig:5.1_AFM_ML} (a) shows an AFM image of a fragment of a single-layer flake in hexagonal form. The respective height profile in (b) was recorded along the orange line. To this end, the SiO\textsubscript{2} background was smoothed using the software Gwyddion. Subsequently, the two plateaus and thus the height of the flake was fitted by a constant function. The extracted thickness of the single layer amounts to $0.65\,nm$.

\begin{figure}[t!]
\centering
\includegraphics[width=12cm]{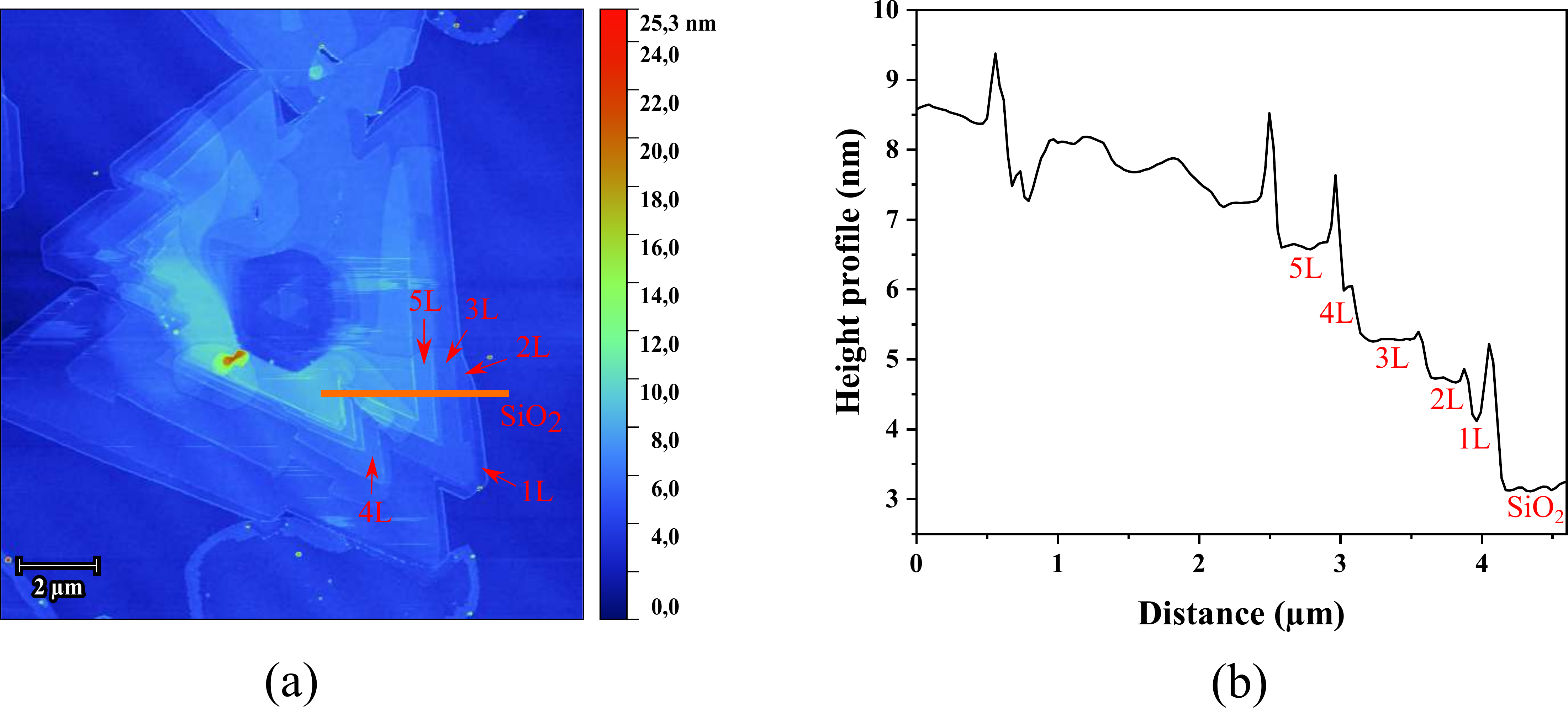}
\caption[AFM bulk step height evaluation]{AFM measurement of a bulk flake in (a) with the corresponding height profile in (b), measured along the orange line. The plateaus of the different layers are indicated in both images.}
\label{fig:5.1_AFM_Bulk}
\end{figure}

Figure \ref{fig:5.1_AFM_Bulk} depicts a thicker flake consisting of superimposed layers. This type of flake is referred to as bulk as the number of layers increases when approaching the center. Layers one to five, measured along the orange line were indicated in the figure.

\subsubsection{Raman \& Photoluminescence spectroscopy}

\begin{figure}[b!]
\centering
\includegraphics[width=12cm]{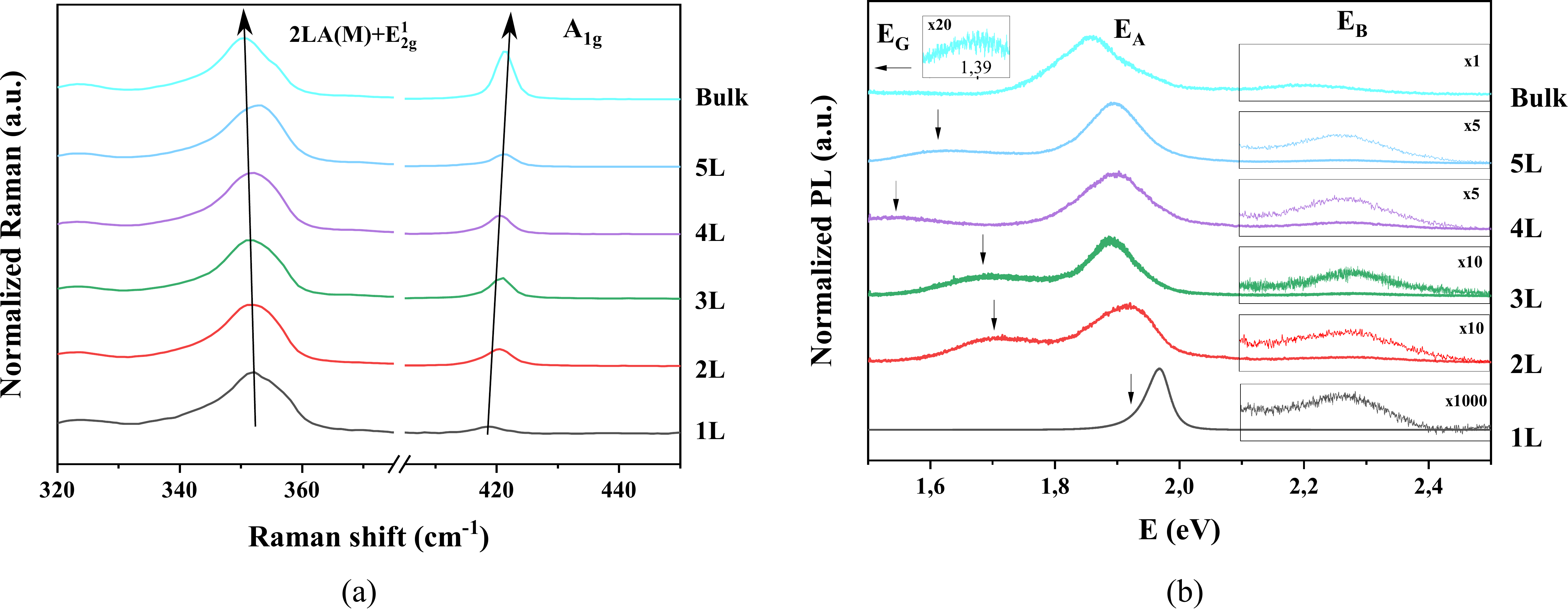}
\caption[Raman and PL layer dependent spectra]{The normalized spectra of various layer numbers of the WS\textsubscript{2} flakes in figure \ref{fig:5.1_AFMLayers}, with Raman in (a) and PL in (b). The two arrows in (a), which serve as guide-to-the-eye, imply that the $2LA(M)+E^1_{2g}$ mode shifts blue and the $A_{1g}$ mode shifts red, which is consistent with theory. Moreover, the PL-peaks behave as expected, from the band diagram introduced in figure \ref{fig:3.1_WS2CrystalStructureAndBandDiagram}.}
\label{fig:5.1_AllPeaks}
\end{figure}

The spectra, acquired at the points indicated with red crosses in figure \ref{fig:5.1_AFMLayers} can be found in figure \ref{fig:5.1_AllPeaks}. Raman was recorded with a green laser at $532\,nm$ wavelength, whereas for PL, blue laser excitation at $405\,nm$ was used. The spectra were captured in wavelength units To fit the peaks and extract their parameters, the background was first removed using a shape subtraction in the software \textit{Project Five}. A Jakobi transformation was then performed to correctly convert the units to $cm^{-1}$ and $eV$, respectively. Last but not least, the spectra were normalized.

In (a), the section of the two critical Raman peaks $2LA(M)+E^1_{2g}$ and $A_{1g}$ is depicted. The $E^1_{2g}$-mode stiffens together with the $2LA(M)$ peak and the $A_{1g}$ peak blueshifts. This is consistent with the theory described in section \ref{sec:RamanSpectroscopy}. Note that the arrows do not coincide perfectly with the peak maxima and serve merely as a guide to the eye. Furthermore, the intensity ratio of the two peaks $A_{1g}/[{E^1_{2g}+2LA(M)}]$ seems to grow with increasing layer number, since there are more out-of-plane modes.

In (b), the peaks of the excitons, $E_A$, and $E_B$, and the indirect transition $E_G$ to be expected from figure \ref{fig:3.1_WS2CrystalStructureAndBandDiagram} can be observed. The $E_B$ peaks are magnified as indicated, and the $E_{G}$ peaks are marked with arrows. The optical bandgap $E_G$ appears to shift red with increasing layer number, except for layer five. This could be caused by the fact that the Gaussian laser beam has a spatial expansion and thus may have picked up the signal of lower layer numbers at the borders of the flake. In the following, the peaks are fitted, and their characteristic properties are extracted.

\begin{figure}[b!]
\centering
\includegraphics[width=12cm]{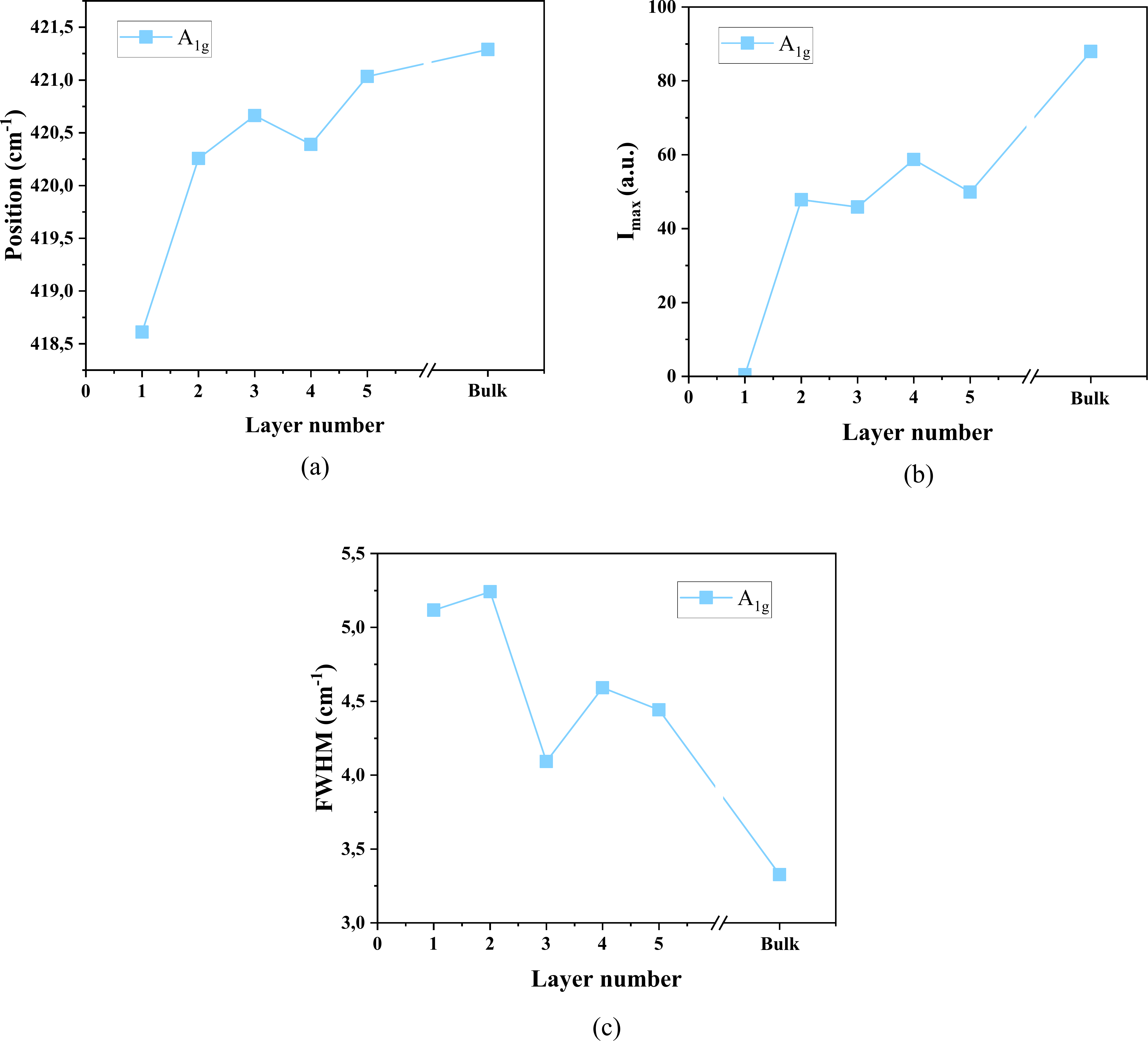}
\caption[Shift of the characteristic properties of the Raman mode A\textsubscript{1g} in dependence of the layer number]{The shift of (a) the peak position, (b) the intensity, and (c) the FWHM of the A\textsubscript{1g}-Raman mode with the layer number. Since the position depends least on the fit, it is used in the following as a reference for the layer number.}
\label{fig:5.1_A1g}
\end{figure}

In the case of Raman, only the properties of the $A_{1g}$ mode are presented and exploited since it is more reliable than the $E^1_{2g}$ peak. The latter merges with the $2LA(M)$ acoustic mode and thus must be extracted by a multi-peak fit. This is laborious in the analysis and leads to increasingly untrustworthy results. 

In figure \ref{fig:5.1_A1g}, it can be seen that the peak position (a) increases with thickness. It reads approximately $418.5cm^{-1}$ for monolayer and reaches $421.3cm^{-1}$ in the case of bulk WS\textsubscript{2}. The intensity, illustrated in (b), of monolayer WS\textsubscript{2} seems very weak. It remains reasonably constant in the range of two to five layers and then increases again in the case of bulk. Finally, one can consider the FWHM, which, however, is strongly dependent on the fit. It is plotted in (c) and indicates a descending trend with an increasing layer number. Since the position depends least on the fit, it is used in the following as a reference for the layer number.

\begin{figure}[b!]
\centering
\includegraphics[width=12cm]{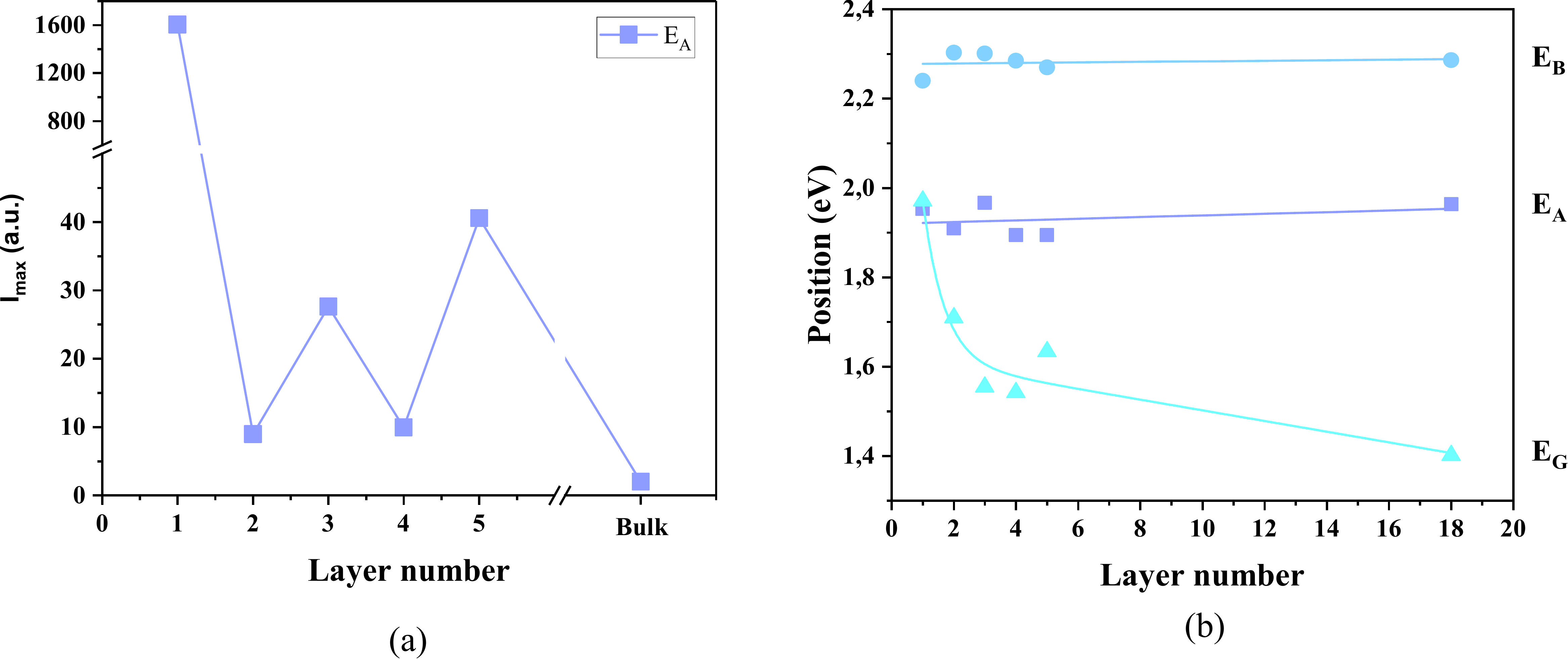}
\caption[Shift of the characteristic properties of the PL signal in dependence of the layer number]{The PL peaks' shift of the (a) intensity and (b) peak position. The intensity of the $E_A$ peak can be used with confidence to distinguish monolayers from multilayers. In addition, the strong blueshift of the $E_G$ peak from around $2$ to $1.4\,eV$ can be measured to identify bulk-WS\textsubscript{2}.}
\label{fig:5.1_PL}
\end{figure}

The intensity of the $E_A$ peak and the positions of all three PL peaks can be found in figure \ref{fig:5.1_PL}. As it can be observed in (a), the intensity of the $E_A$ peak in monolayer WS\textsubscript{2} increases by two orders of magnitude as the band transition shifts from an indirect to a direct gap. This offers a simple way to distinguish monolayer from few-layer WS\textsubscript{2}.

In addition, it can be seen in (b) that the positions of the peaks $E_A$ and $E_B$ remain more or less constant, whereas the indirect band transition $E_G$ shifts blue with increasing layer number. This is mostly consistent with the DFT calculation presented in figure \ref{fig:3.1_WS2CrystalStructureAndBandDiagram}. The only difference is, that the spin-orbit coupling and thus the distinction into A and B peaks also occurs in monolayer WS\textsubscript{2} in reality.

\clearpage
\newpage

\subsection{Growth Behaviour}
\label{sec:GrowthBehaviour}

Each experimental setup is unique, and the CVD process can depend on even minor details such as the tube diameter or the PID controller values of the furnaces. Therefore, it is important not only to enumerate the final parameters but also to document the optimization process. In the following, this is accomplished for the solid precursor.

\subsubsection{Initial Parameter Studies}
\label{sec:InitialParameterStudies}

In order to narrow down the parameters to the subspace where the growth occurs, the variables were initially studied without a systematic DoE process. 

First, $20\,nm$ tungsten was sputtered without any pattern onto the seed substrate and oxidized for two hours at $500$°$C$ at $1\,mbar$ involving $200\,sccm$ O\textsubscript{2} and $200\,sccm$ Ar flow. During the CVD process, the metal temperature was kept at $900$°$C$ and the sulfur temperature at $200$°$C$. The pressure was $1\,mbar$ and the flow consisted of $100\,sccm$ Ar. The process times of the experiments were $15$ and $30\,min$. The first attempts yielded structures that lay below the optical resolution limit but nevertheless produced a characteristic bulk-WS\textsubscript{2} Raman signal.

In the next step, the sulfur temperature was dropped to $120$°$C$, as about $1\,g$ of sulfur was consumed during each process. Furthermore, a hydrogen component of $16\,sccm$ was added to the argon stream. As a result, the first µm-scale structures were obtained, which were made of monolayers with higher-layered islands on top. However, the crystal layer remained highly polycrystalline. It was further observed that the WS\textsubscript{2} on the growth substrate tended to grow at the edge of the sputtered precursor on the seed substrate. It appeared that all the sulfur was trapped by the borders of the metal precursor. The idea was to introduce a gap between the microreactor to promote growth in the center of the sample.

Processes with mm, µm, and nm separation were performed. Nevertheless, the growth did not improve perceptibly. An alternative idea was to pattern the precursor on the seed and aim for negative growth. Therefore, the strategy was changed to structure the sputtered tungsten using the PLT method detailed in section \ref{sec:4.2_SolidPrecursor}. Sputter thicknesses of $5$, $10$, $20$ and $140\,nm$ were attempted with oxidation times of $2$, $5$ and $10\,h$. It was found that the optimal sputter height of $10\,nm$ and an oxidation time of $10\,h$ yielded a process with high surface coverage and well-defined large-area monolayers, mainly in the negative growth regime.

\begin{figure}[t!]
\centering
\includegraphics[width=9cm]{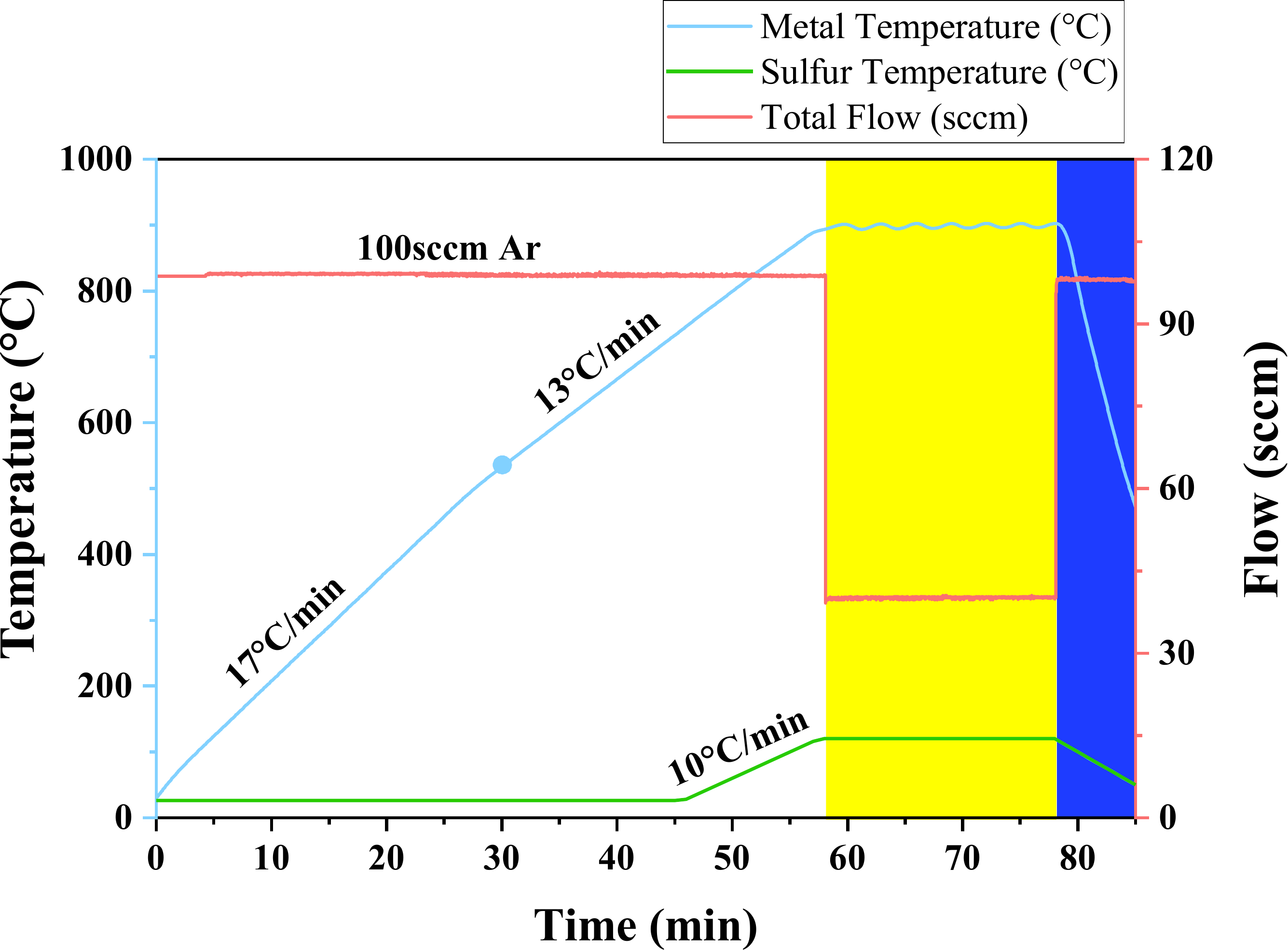}
\caption[The preliminary optimized process without a systematic DoE study]{The preliminary optimized process without a systematic DoE study. The cooling down was accelerated by air-cooling and the pressure amounted to $10\,mbar$ throughout. The boats were placed in the center of the furnaces and the Ar/H\textsubscript{2} ratio was kept at $50\,\%$ during the process.}
\label{fig:5.2_WholeProcess}
\end{figure}

The final significant improvement occurred by increasing the pressure to $10\,mbar$. The optimized process, including heating-up and cooling-down phases, is shown in figure \ref{fig:5.2_WholeProcess}.

To maintain a clean process, the sulfur that had accumulated in the tube during the optimization processes was baked out. Unfortunately, growth disappeared with it, although the parameters were maintained and the substrates originated from the same batch. This was the case for the following seven processes until the sulfur atmosphere in the tube was restored.

The liquid precursor described in section \ref{sec:LiquidPrecursor} was similarly optimized. The improved process includes the following parameters: A metal furnace temperature of $850$°$C$, a sulphur furnace temperature of $120$°$C$, $40\,min$ process time, $25\,sccm$ Ar and $25\,sccm$ H\textsubscript{2}. The process was carried out at $1.7\,mbar$, and air cooling was included.

The growth produced a high surface coverage with continuous monolayers of up to several millimeters. However, the structures appeared not always well defined and were often distorted. Moreover, it is not easy to control the thickness of the precursor film during spin coating, see figure \ref{fig:4.2_LiquidPrecursor}. In addition, the 2D material seems to grow only in the areas with the thickest precursor. Such factors increasingly make the growth unreliable. Therefore, the DoE was carried out using the solid precursor.

\subsubsection{Distance Behaviour}
\label{sec:DistanceBehaviour}

Prior to beginning the systematic DoE optimization of the growth, the ideal construction was determined by applying the encountered process to the different distances.

The spacings were described in section \ref{sec:MicroreactorDistanceControl} and involved 20, 40, and $60\,nm$ for the HF dip, 25, 50, 100, and 200\textit{\,µm} for the saw-cut samples, and a few $mm$ for the wider substrates. All processes were carried out with a sputtered, flat metal precursor layer. Therefore, generally little growth is expected as no patterning has taken place. However, this should not be a problem as the following experiments are only performed to determine the best spacing.

The narrowest separation of $20\,nm$ showed moderate surface coverage in the center of the sample with a clear WS\textsubscript{2} Raman signal. However, the structures could not be resolved optically. The other two substrates at 40 and $60\,nm$ did not produce any growth in the distant part.

In the micrometer scale, the samples with 25, 50, and 100\textit{\,µm} show a slight light green tinge in the center of the sample. It features a weak WS\textsubscript{2} Raman signal and decreases rapidly with increasing distance. The substrate with a distance of 200\textit{\,µm} was still vacant after the process.

Surprisingly, the substrates with a spacing of a few $mm$ showed moderate growth. The samples with a width of $11\,mm$ resulted in very high surface coverage, yet the flake structures turned out to be less than 0.1\textit{\,µm} in size. In addition, the $11.5\,mm$ wide sample repeatedly resulted in voids.

In summary, growth was best at the shortest distance of $20\,nm$. However, the processes that did not include vertical space still yielded the highest surface coverage and the best growth quality. Therefore, the DoE was performed in the next chapter using samples with no spacing. It should be noted that these results indicate low reproducibility of the CVD process, as even small impurities in the nanometer range can affect growth negatively.

\clearpage
\newpage

\subsection{Design of Experiments}
\label{sec:5DesignOfExperiment}

\subsubsection{Reproducibility and Consistency Studies}

\begin{figure}[b!]
\centering
\includegraphics[width=12cm]{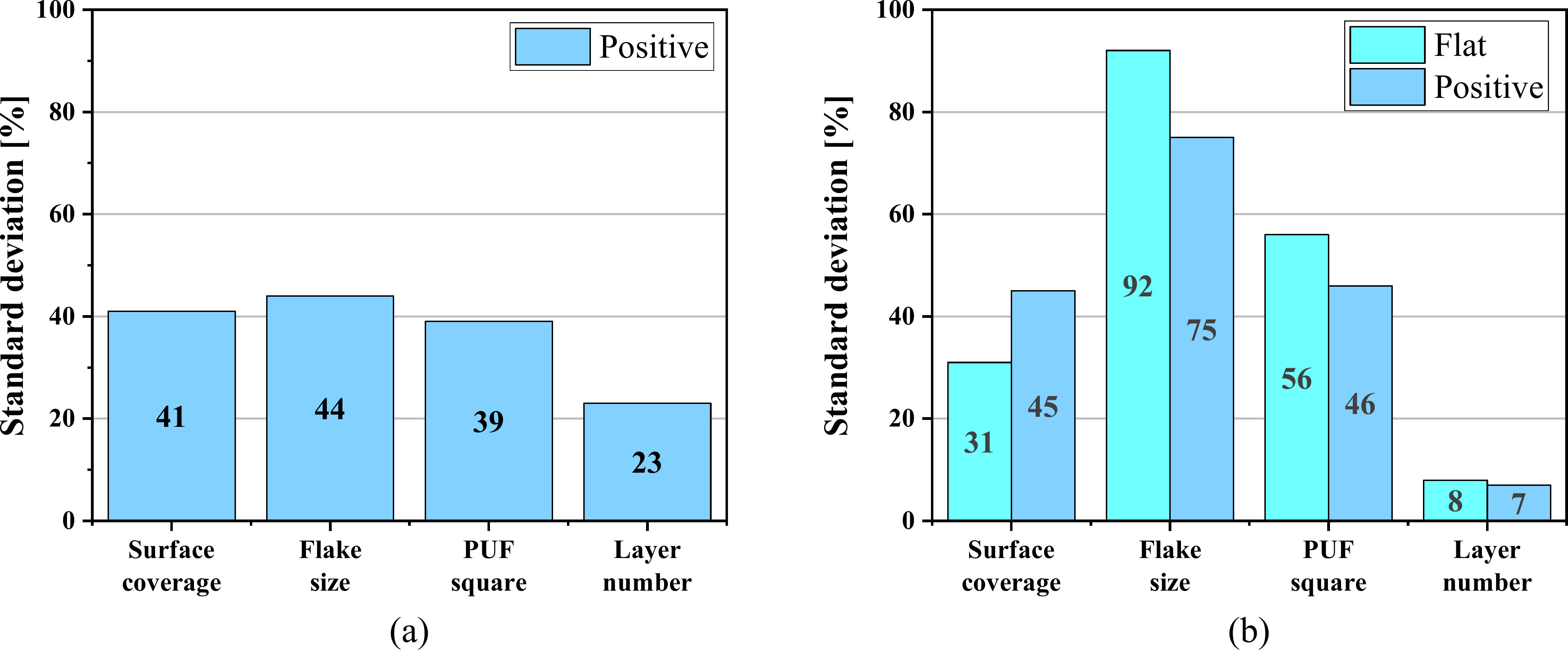}
\caption[DoE reproducibility and consistency studies]{The standard deviation of the four evaluation parameters for (a) the four reproducibility and (b) the five consistency processes. In both cases, the flake size seems to be the most unreliable quantity and the layer number shows the smallest standard deviation.}
\label{fig:5.2_ReproducibilityConsistency}
\end{figure}

To perform the systematic optimization, a reproducibility study was first conducted to estimate the reliability of the growth and evaluation parameters introduced in table \ref{tab:EvaluationParameters}. The process was repeated four times in a row. The parameters of the reproducibility process differed slightly from those of the pre-optimized process. The differences involved using $1\,mbar$ instead of $10\,mbar$ and growth lasting $30\,min$ instead of $20$. These modifications resulted in average surface coverage and medium growth quality.

Here and in the following, the processes comprise two samples in order to average out the growth fluctuations. The mean of the two is taken in the evaluation. In the end, the standard deviation was calculated for each evaluation parameter. The results are presented in figure \ref{fig:5.2_ReproducibilityConsistency} (a). Note that only positive growth was considered, as the samples did not contain any flat parts, and unfortunately, no negative growth occurred. The reason for the latter is that the new mask design from figure \ref{fig:4.2_MasksResults} (b) was used for the first time and the growth was not yet optimized for it. However, the pattern was further used in the expectation that the DoE would encounter new optimization parameters that yield enhanced negative growth. Considering the standard deviations, it can be concluded that the number of layers appears to be the most reliable parameter and that the flake size has the largest deviations, closely followed by the surface coverage and the PUF square.

A further validation process, which served to verify the consistency of the experimental design and thus to determine whether the processes at the start and end of the DoE were equivalent, was carried out five times. It was performed once before, again in the middle, and finally three times after the DoE. The same parameters as in the pre-optimized process were adopted, resulting in enhanced growth in comparison to the reproducibility study. Here, the flat part was included, but anew hardly any negative growth occurred.

The results are shown in figure \ref{fig:5.2_ReproducibilityConsistency} (b). Repeatedly, the number of layers is the most consistent parameter, and the PUF size and surface coverage remain in the same order of magnitude as in the reproducibility study. The standard deviation of the flake size reaches a maximum value of $92\,\%$ for the flat part. An issue regarding the flake size is that, on the one hand, it is infeasible to locate the largest flake in a huge sample with tens of thousands of flakes. On the other hand, the flakes coalesce to form a layer, such that sometimes only tiny flakes can be detected in a sample with decent growth.

It should be noted here that the uncertainties that arise during fabrication endorse the approach of a PUF based on 2D materials. Nevertheless, the question remains to be answered whether the electrical signal will turn out just as randomly. In summary, growth has moderate reproducibility with respect to the proper evaluation parameters. This must be taken into account when interpreting the results of the DoE.

\begin{figure}[b!]
\centering
\includegraphics[width=12cm]{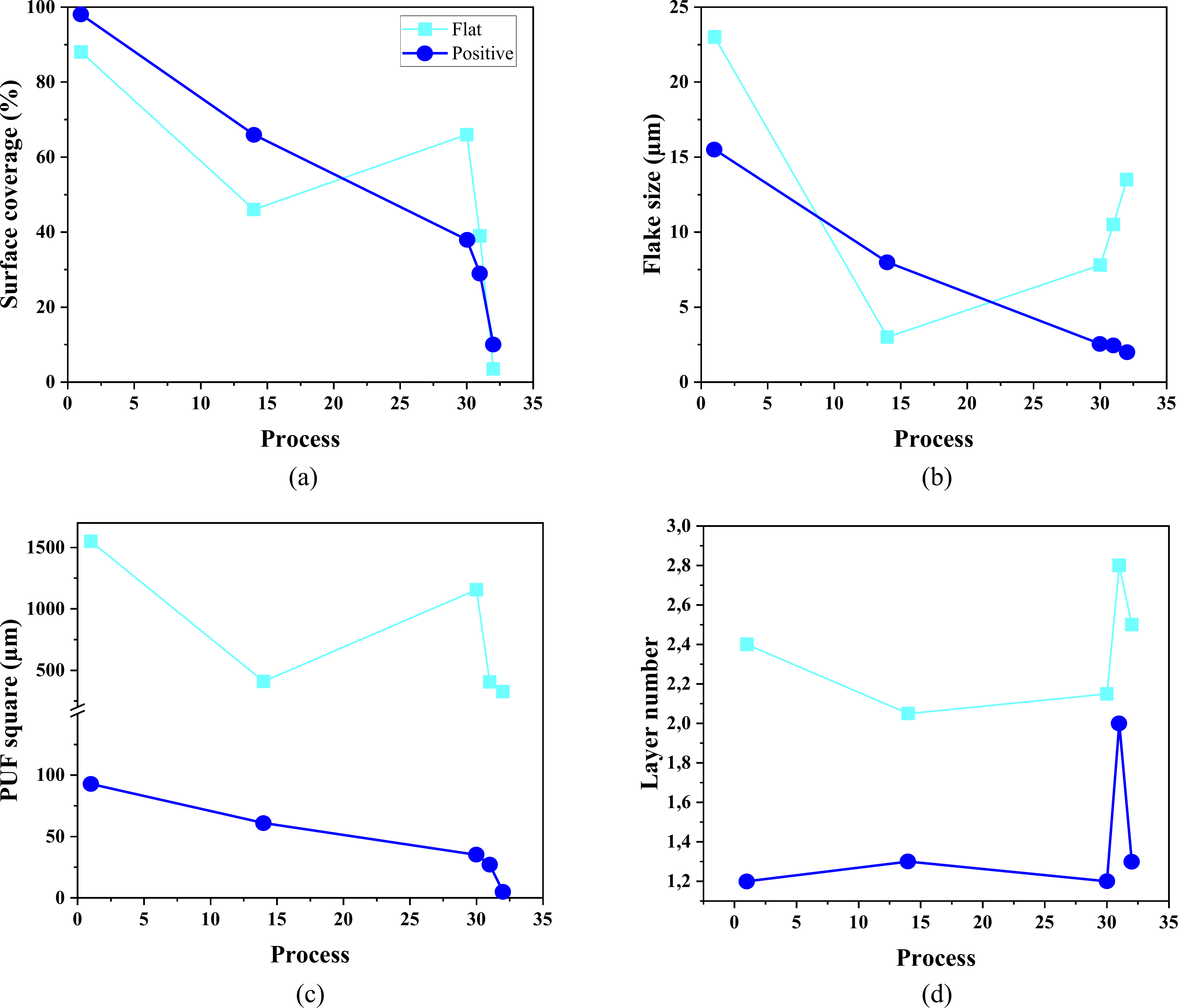}
\caption[DoE consistency studies over time]{The results of the consistency studies over time in the form of process numbers. Surface coverage in (a) and PUF size in (c) decrease with increasing process number, indicating degradation of the furnace. The triangle size in (b) fluctuates strongly and the layer number in (d) remains mostly constant with the exception of process 32.}
\label{fig:5.2_ConsistencyOverTime}
\end{figure}

Finally, figure \ref{fig:5.2_ConsistencyOverTime} shows the four evaluation parameters of the consistency processes as a function of the chronological process number to investigate the time dependence of the degradation. The parameters surface coverage, flake size, and PUF square indicate that the experimental setup has degraded after 33 DoE processes. Furthermore, it can be observed that the number of layers remains reasonably stable with an exception in the fourth consistency process. Furthermore, the thicknesses of the two different growth types, positive and negative, seem to differ significantly.

\subsubsection{Model and Example Analysis}
\label{sec:ModelAndExampleAnalysis}

\begin{table}[b!]
\begin{center}
\begin{tabular}{cccccc} \toprule
Metal temperature & H\textsubscript{2}/Ar ratio & Sulfur temperature & Time & Pressure & Air cooling \\ \midrule
800°$C$ & $10\,\%$ & 100°$C$ & $15\,min$ & $2\,mbar$ & $Yes$ \\
850°$C$ & $23\,\%$ & 110°$C$ & $25\,min$ & $5\,mbar$ & $No$ \\
900°$C$ & $37\,\%$ & 120°$C$ & $35\,min$ & $10\,mbar$ &  \\
 & $50\,\%$ &  & $45\,min$ & $20\,mbar$ &  \\
 &  &  &  & $50\,mbar$ &  \\ \bottomrule
\end{tabular}
\caption[DoE input model]{DoE Input Model.}
\label{tab:DoE_InputModel}
\end{center}
\end{table}

The chosen input parameters for the DoE model are listed in table \ref{tab:DoE_InputModel}. The metal temperatures were based on the parameters found to be optimal for the solid and liquid precursor in section \ref{sec:InitialParameterStudies}, and $800$°$C$ was appended. The total gas flow was kept fixed for all parameters, and the ratio between H\textsubscript{2} and Ar was varied in three equidistant steps from ten to $50\,sccm$.

The process times fall within a small range around the values of $20$ and $40\,mins$, encountered in the pre-optimized processes. The lower limit of the pressure is given by the lower bound of the measurement setup, i.e. by the minimum pressure at maximum gas flow, which amounts to $2\,mbar$. In order to obtain the upper limit, two further experiments were carried out at $50$ and $100\,mbar$, with the remaining parameters identical to those of the pre-optimized process. In the case of $100\,mbar$, hardly any growth occurred. The resulting structures at $50\,mbar$ were inferior to those at $10\,mbar$, but could theoretically be enhanced by alternative combinations. Therefore, the upper limit was chosen to be $50\,mbar$. Finally, it was checked whether the fast cool-down with air cooling at the end of the process has any effect. When multiplying the number of values of the individual parameters, 1152 processes would be obtained in a factorial experimental design, which could be reduced to 27 using a DoE approach.

The parameters of the individual processes were determined and randomized using the Cornerstone software. All samples were prepared in advance and consecutively oxidized. Hence, the samples all originated from the same batch and the blocking property was ensured. In addition, the processes of reproducibility and consistency studies were included in the evaluation, which implies that repetitions of the same process could be involved. Further, the growth seems sufficiently reproducible when regarding the proper evaluation parameters. Thus, all the fundamental criteria of the DoE have been fulfilled.

\begin{figure}[t!]
\centering
\includegraphics[width=15cm]{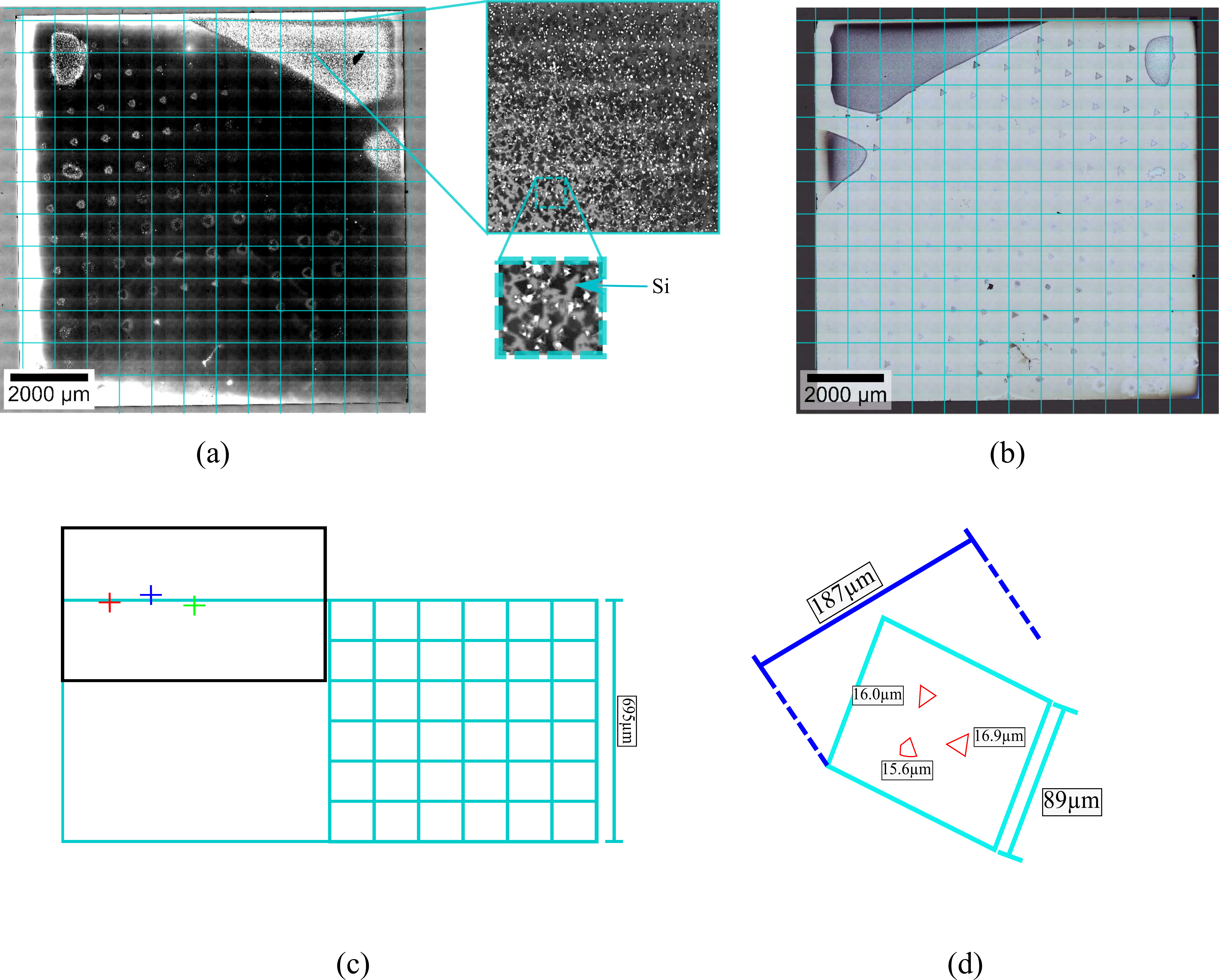}
\caption[An example analysis for the DoE]{An example analysis for the DoE. (a) The growth substrate with a black and white filter and a grid superimposed to estimate surface coverage in the flat and negative regions. (b) The seed substrate overlaid with a grid to estimate the proportion of area covered with precursor and thus determine the proportions of the growth zones. (c) The biggest PUF square from (a) with a grid to determine the average layer number. The upper left crop shows three flakes that occur most frequently in the PUF square, based on the color, which is equivalent to the layer number. Their thickness is determined by analyzing the Raman and PL characteristics to simplify the estimation of the average layer number. (d) Finally, a crop of a positive growth zone with the indicated determined evaluation parameters.}
\label{fig:5.2_ExampleEvaluation}
\end{figure}

In the following, an exemplary analysis is presented where the evaluation parameters are obtained from the optical images of a specific sample, provided in the figure \ref{fig:5.2_ExampleEvaluation}. First, the determination of the surface coverage is discussed. In the case of flat and negative growth, a high-resolution stitching image of the growth substrate was acquired. A black and white filter was then applied and the contrast was increased to visualize all growth. The respective image is shown in sub-figure (a). A stitching image of the seed substrate was then recorded, which can be seen in (b). A grid was placed on top of both images to better estimate the percentage of covered area for each square individually. Finally, the proportions of the flat and negative parts on the seed substrate were determined and combined with the proportions covered by WS\textsubscript{2} on the growth substrate. Since the image possesses a high resolution, the squares can be magnified and it is possible to estimate the surface coverage with reasonably high accuracy.

Sub-figure (a) indicates that a difficulty arises when the flat part is too small: growth can be bounded by its size. However, if the flat part is too large, hardly any growth occurs, as discussed previously in section \ref{sec:InitialParameterStudies}. A proper trade-off must therefore be made to encounter the optimal size of the flat part. In the case of positive growth, the second best-covered triangle of the pattern was selected and the proportion of WS\textsubscript{2} and SiO\textsubscript{2} was estimated. A section of the second best-covered triangle from the sample in (a) can be found in (d).

To determine the size of the biggest flake, those standing alone in the respective growth regimes were selected and the longest flanks were measured. Finally, the second largest value of all identified flakes was used. Figure (a) illustrates how challenging it can be to identify the largest flake under multiple thousands. In addition, as shown in (d), it may be difficult to determine whether a flake stands alone or not. These two factors lead to the high uncertainty in the standard deviation of the evaluation parameter.

To estimate the interconnectivity, i.e. the PUF square, the largest rectangle in each growth region was found that contained a large-scale, mostly continuous WS\textsubscript{2} film, as shown in figure (c) for the flat part and in figure (d) for the positive part. Then the smaller side was chosen for the acquisition of the square.

Finally, a further grid was positioned over the PUF square to determine the number of layers. Raman and PL signals were recorded for the flakes that appear to be optically most abundant. In the case of (c), the spectrum of the red cross can be assigned to a single-layered flake, since the PL intensity of the A peak was found to be 182 times higher than that of the green spectrum and 318 times higher than the intensity measured at the blue cross. Moreover, the $A_{1g}$ mode of the green-labeled flake was found to be $420.29\,cm^{-1}$ and thus seems to comprise two layers. A simple optical look at the high brightness of the flake with the blue cross shows that it consists of a high number of layers and can thus be called bulk. Finally, in each of the quadrants in the grid, the average number of layers was determined with the help of the brightness of the associated flakes. The mean value has then been calculated. The results of the example analysis are listed in table \ref{tab:ExampleEvaluationResults}.

\begin{table}[h!]
\begin{center}
\begin{tabular}{ccccc} \toprule
 & \makecell{Surface \\ coverage (\%)} & \makecell{2\textsuperscript{nd} biggest \\ flake ($nm$)} & PUF square (\textit{µm}) & Layer number \\ \midrule
Flat & $95$ & $20.8$ & $695$ & $3.2$ \\
\rule{0pt}{5ex}
Positive & $94$ & $16.0$ & $89$ & 1.3 \\
\rule{0pt}{5ex}
Negative & $<0.1$ & $1.2$ & $25$ & $2.0$ \\ \bottomrule
\end{tabular}
\caption[Example analysis results]{Results of the example analysis.}
\label{tab:ExampleEvaluationResults}
\end{center}
\end{table}

\subsubsection{From the Data to the Optimal Process}
\label{sec:5.3_Results}

The 27 experiments, and additionally the five consistency experiments, were carried out within three weeks. Subsequently, the samples were analyzed and the evaluation parameters extracted. The evaluation was accomplished again with the Cornerstone software.

\begin{table}[b!]
\begin{center}
\begin{tabular}{ccccccc} \toprule
Optimized model & Metal T & H\textsubscript{2}/Ar & Sulfur T & Time & Pressure & Cooling \\ \midrule
\makecell{Surface coverage \& \\ PUF square (Flat)} & 860°C & 27\% & 110°$C$ & 33$min$ & 27$mbar$ & / \\
\rule{0pt}{4ex}
+ Monolayer & 893°C & 11\% & 119°$C$ & 29$min$ & 36$mbar$ & $Yes$ \\
\rule{0pt}{4ex}
+ 2.5 Layer & 888°C & 20\% & 112°$C$ & 18$min$ & 43$mbar$ & $No$ \\
\rule{0pt}{4ex}
\makecell{Biggest flake (Flat)} & 857°C & 11\% & 111°$C$ & 38$min$ & 34$mbar$ & $Yes$ \\
\rule{0pt}{4ex}
\makecell{Surface coverage \& \\ PUF square (Positive)} & 852°C & 15\% & 113°$C$ & 30$min$ & 36$mbar$ & / \\ \bottomrule
\end{tabular}
\caption[DoE optimized parameters for different combinations of evaluation parameters]{DoE optimized parameters for different combinations of evaluation parameters.}
\label{tab:DoE_Output}
\end{center}
\end{table}

The obtained optimized parameters for different subspaces of the evaluation parameters can be found in table \ref{tab:DoE_Output}. Optimization was only applied to the flat and positive growth regimes, as the surface coverage in the negative region consistently remained below $0.5\%$. Furthermore, the negative PUF square never exceeded a value of 25\textit{\,µm}. Note here that this does not imply that negative growth does not work, as high-quality large-scale films have previously been obtained. However, the new EBL mask from \ref{fig:4.2_MasksResults} (b) would have to be adapted and optimized.

In the case of the flat, the surface coverage was optimized first, followed by the PUF square, and eventually their combination. Nevertheless, the resulting parameters differed only in the first decimal place. Furthermore, the two parameters were optimized together with a layer number of one and $2.5$. It is important to understand that a layer number of $2.5$ does not imply that a continuous film with a thickness of $2.5$ layers is produced, but that the average thickness may be $2.5$ layers. When a sample has an average layer count of $2.5$, it usually means that the thickness of the flakes varies widely.

Finally, the calculation was performed for the largest flake in the flat area and the surface coverage in combination with the PUF square in the positive part. If a parameter is expected to show no significant influence, it was marked with a slash. According to the DoE results in table \ref{tab:DoE_Output}, each parameter has an impact on growth, except for air cooling together with optimized surface coverage + PUF size in the flat and positive parts.

The experiments were then conducted successively. The optimization for surface coverage + PUF size for the flat part yielded the strongest growth and is presented below. The optimized processes with layer numbers and the largest flake, unfortunately, gave less than one percent surface coverage.

\begin{figure}[t!]
\centering
\includegraphics[width=15cm]{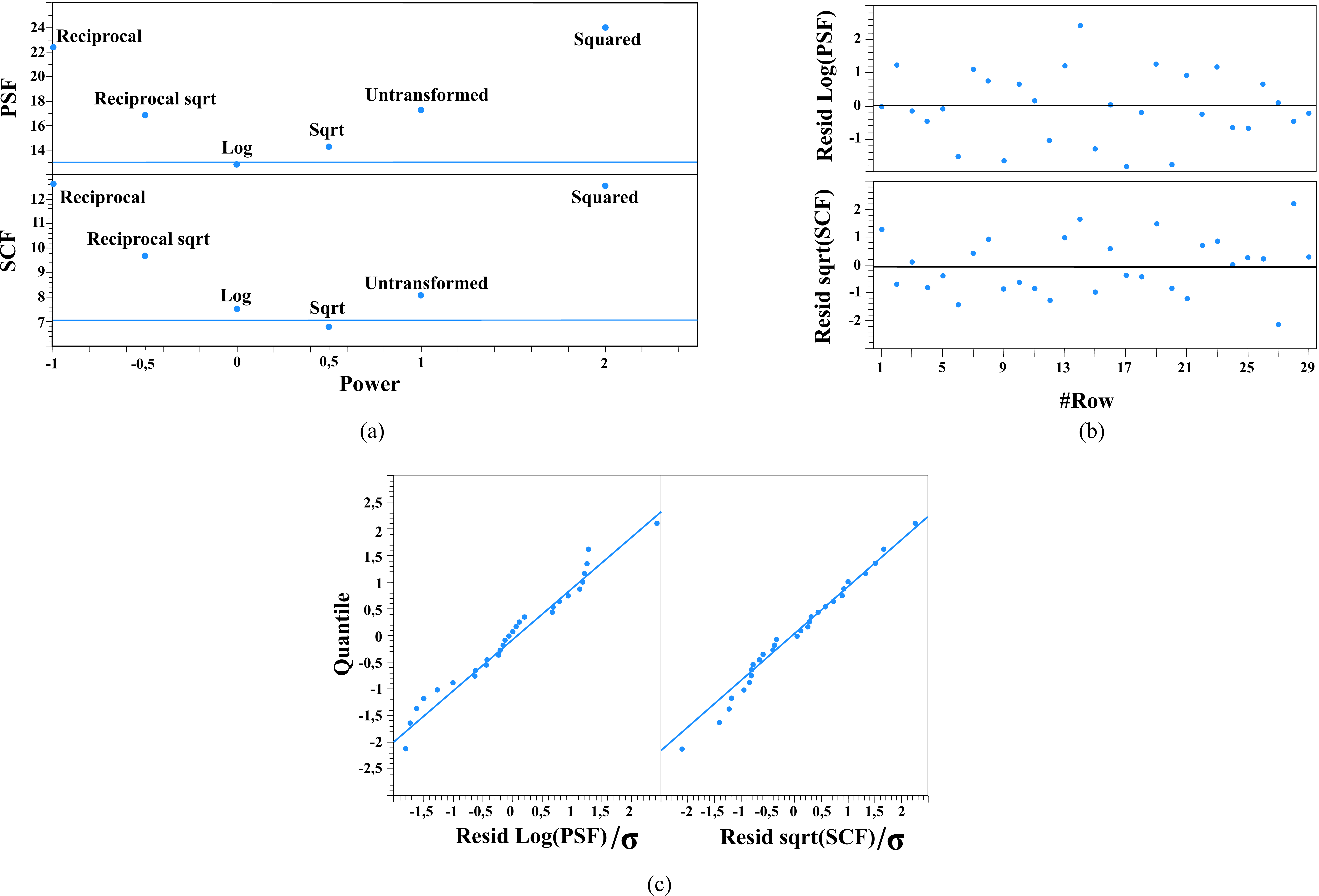}
\caption[DoE Fit quality]{The auxiliary plots that helped to create a reliable fit of the measured values in the DoE. PSF stands for PUF square flat and SCF for surface coverage flat. The Box-Cox plot in (a) serves to transform the variables to obtain a preferably normal distribution. From (b) it can be observed that the residuals are randomly distributed around zero, indicating a decent model. The last graph in (c) is called a normal probability plot. No measurement points fall beyond the interval between -3 and +3 standard deviations.}
\label{fig:5.2_DoE_FitEvaluation}
\end{figure}

Prior to revealing the results of the best process, the quality of the fit and the final model will be discussed. The auxiliary plots created with the help of Cornerstone are presented in figure \ref{fig:5.2_DoE_FitEvaluation}. In subplot (a), a so-called box-cox scheme is depicted. A certain formula is applied to transform non-linear variables into a normal shape to be able to apply more sophisticated analysis methods. The models are transformed by an exponent that takes the values $-1$, $-0.5$, $0.5$, $1$, and $2$. Additionally, the input model is logarithmized. Next, a value is calculated for each case, indicating the extent to which the distribution thus obtained resembles a normal distribution, representing the y-axis. Cornerstone software provides a blue borderline below which the value thus obtained must fall to perform the subsequent analyses. Thus, the resulting values of the PUF square were logarithmized, and the square root was taken from the surface coverage.

In (b), the residuals of the transformed parameters were plotted as a function of process numbers, labeled with rows. The goal is to determine whether the residuals follow a specific dependency or whether the values are randomly distributed around zero, which is the case here. If a dependency were to be seen, this would indicate that a superior model exists that could fit the data more precisely.

The last graph in (c) represents a residual vs probability-, also called a normal probability plot. The residuals are plotted against values of a normal distribution, so-called quantiles. Thus, any deviations from a straight, diagonal line indicate a divergence from normality. It can further be employed to identify outliers. As can be seen, the line is predominantly diagonal, and the data points lie close to it. Furthermore, all points are located in the -3 to 3 sigma range, which indicates a reliable measure for a decent fit.

To establish a trustworthy model of the WS\textsubscript{2} growth, it is necessary to discard the insignificant terms in the fit function, represented by equation \ref{equ:DoE}. A term is called insignificant if it exhibits a p-value of more than $0.1$. A high p-value indicates that the real model is unlikely to depend on this parameter. This correction naturally worsens the fit but enhances the trustworthiness of the model and prevents overfitting. The significant terms that remain are listed in the following equations:

\begin{flalign}
\begin{aligned}
Log(PSF)=\alpha_0 &+ \alpha_1 \, T + \alpha_2 \, H_2 + \alpha_3 \, T_S + \alpha_4 \, t + \alpha_5 \, P && \\
&+ \alpha_{24} \, T \cdot T_S+ \alpha_{25} \, H_2 \cdot P + \alpha_{35} \, T_S \cdot P + \alpha_{45} \, t \cdot P && \\
&+ \alpha_{11} \, T^2 &&
\label{equ:LogPSFDoE}
\end{aligned}
\end{flalign}

\begin{flalign}
\begin{aligned}
Sqrt(SCF)=\beta_0 &+ \beta_1 \, T + \beta_2 \, H_2 + \beta_3 \, T_S + \beta_4 \, t + \beta_5 \, P && \\
&+ \beta_{13} \, T \cdot H_2+ \beta_{25} \, H_2 \cdot P + \beta_{34} \, T_S \cdot t + \beta_{45} \, t \cdot P && \\
&+ \beta_{11} \, T^2 + \beta_{22} \, T_S^2 + \beta_{55} \, P^2 &&
\label{equ:SqrtSCFDoE}
\end{aligned}
\end{flalign}

\begin{minipage}[t]{.5\textwidth}
 \begin{itemize}
  \item $T$: Metal furnace temperature
  \item $H_2$: H\textsubscript{2}/Ar ratio
  \item $T_S$: Sulfur furnace temperature
 \end{itemize}  
\end{minipage}%
\begin{minipage}[t]{.5\textwidth}
  \begin{itemize}
  \item $t$: Process time
  \item $P$: Pressure
 \end{itemize}
\end{minipage}

\vspace*{10px}

The R-squared value before correction amounted to $0.874$ for PUF size and $0.903$ for surface coverage. It was reduced to $0.804$ and $0.830$, respectively.

\begin{figure}[b!]
\centering
\includegraphics[width=13cm]{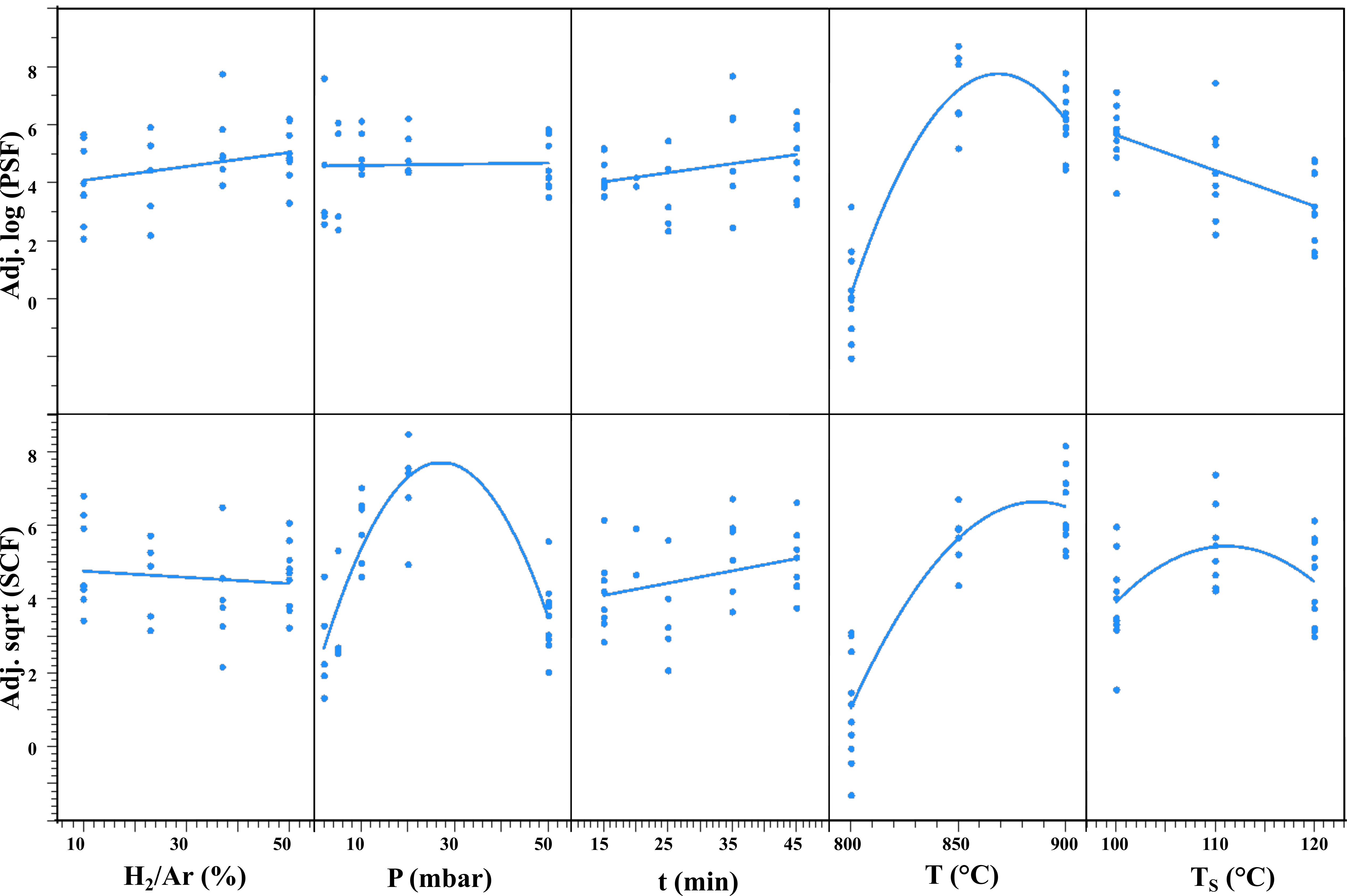}
\caption[Growth model for SCF and PSF in WS\textsubscript{2}]{The final CVD growth model of WS\textsubscript{2} for PSF and SCF, obtained by DoE. The two evaluation parameters were maximized by the software Cornerstone on the basis of the depicted dependencies.}
\label{fig:5.2_DoE_AdjustedResponseGraphs}
\end{figure}

The multidimensional model is finally shown in individual plots in figure \ref{fig:5.2_DoE_AdjustedResponseGraphs}. It can be seen that changing the H\textsubscript{2}/Ar ratio has little effect and that the optimal values in terms of PUF square and surface coverage are contradictory. In addition, it appears that pressure only significantly impacts surface coverage. According to the model, a longer process time has a good but slight influence on the evaluation parameters. However, as the time increases, so do the layer numbers and the impurities in the process.

\begin{table}[b!]
\begin{center}
\begin{tabular}{ccccccc} \toprule
Growth type & Surface coverage & Flake size & PUF square & Layer number \\ \midrule
Flat mean & 83\% & 21µm & 1140µm & 1.25 \\
\rule{0pt}{5ex}
\makecell{Flat \\ standard deviation} & 21\% & 35\% & 14\% & 11\% \\
\rule{0pt}{5ex}
Positive mean & 80\% & 17 & 15 & 2.1 \\
\rule{0pt}{5ex}
\makecell{Positive \\ standard deviation} & 3\% & 22\% & 8\% & 59\% \\ \bottomrule
\end{tabular}
\caption[Optimized process results of SCF and PSF]{Evaluation parameters in the flat region, averaged over two samples and the corresponding standard deviation for the optimized process.}
\label{tab:DoE_ProcessOutput}
\end{center}
\end{table}

The metal furnace temperature seems to have its optimum somewhere between $850$ and $900$°$C$. Nevertheless, it cannot be ruled out that the surface coverage will increase above $900$°$C$. Finally, the sulfur furnace temperature achieves large PUF squares at around $100$°$C$ and the highest surface coverage near $110$°$C$.

\begin{figure}[t!]
\centering
\includegraphics[width=15cm]{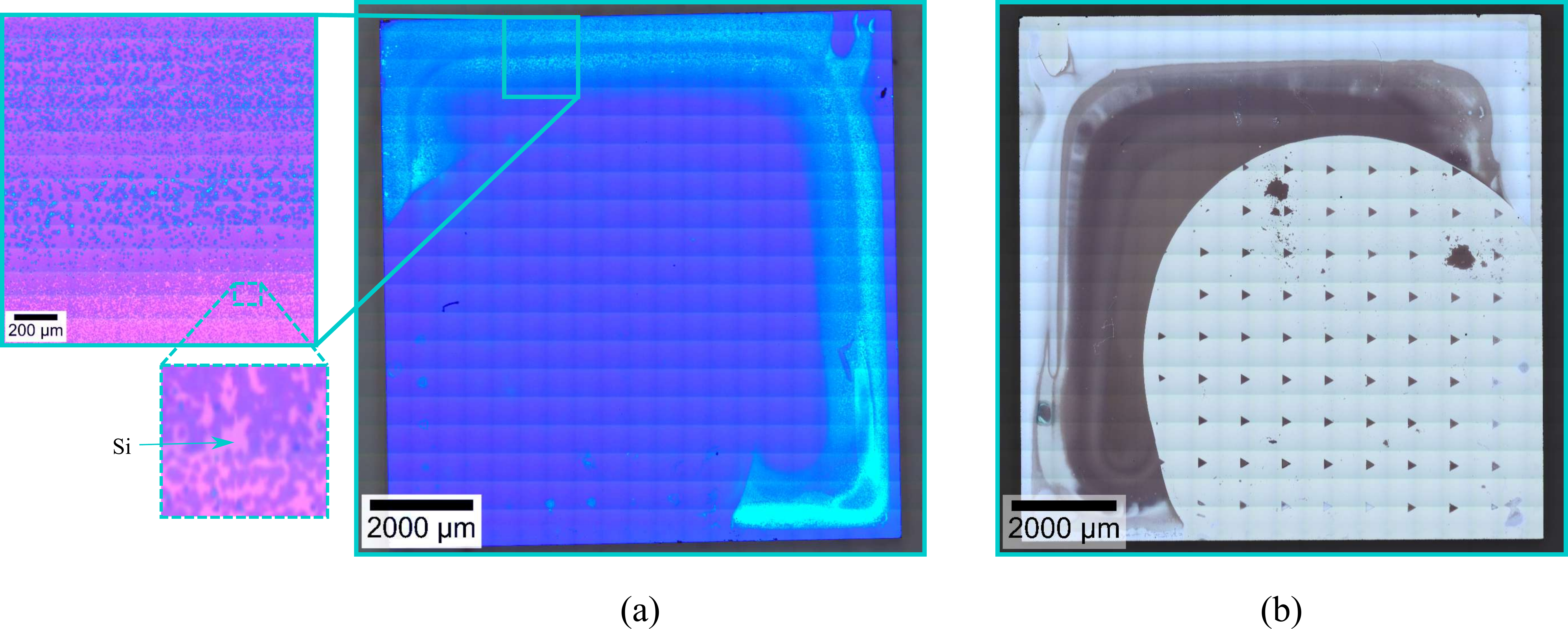}
\caption[DoE optimized process results]{One of two samples of the optimized process with a crop of the PUF square in (a) and the corresponding seed substrate in (b). A high surface coverage with a millimeter-sized PUF square can be identified in the flat region. The DoE was successful.}
\label{fig:5.2_DoE_OptimizedProcessEvaluation}
\end{figure}

Last but not least, the results of the optimal process are provided in table \ref{tab:DoE_ProcessOutput} and figure \ref{fig:5.2_DoE_OptimizedProcessEvaluation}. The process again included two samples and was executed immediately after the DoE. The mean values of the evaluation parameters and their standard deviations are listed for the flat and positive parts. A high surface coverage, large flakes, and millimeter-scale PUF squares were measured. The growth resulted mainly in monolayer in the area of the biggest PUF square. In addition, the overall standard deviation of the optimized process is relatively low compared to the consistency process in figure \ref{fig:5.2_ReproducibilityConsistency} (b). Further, it is astonishing that, despite the substantial degradation of the experimental setup, a process with greatly enhanced growth emerged. An image of one of the two growth samples, including a crop of the PUF square and the associated seed substrate, can be observed in figure \ref{fig:5.2_DoE_OptimizedProcessEvaluation}.

To validate the consistency of the final process, the tube was first replaced, and further processes were carried out. The pre-optimized process and the DoE-optimized process were performed alternately thrice each. The DoE process improved compared to the values in table \ref{tab:DoE_ProcessOutput} and outperformed the pre-optimized process every time.


Finally, the processes with different distances and a continuous precursor were carried out again with the DoE process, similarly to section \ref{sec:DistanceBehaviour}. However, no growth was obtained for any of the distances, and the continuous flat part showed no improvement compared to the pre-optimized process. The DoE was only performed on a non-distance patterned precursor sample. Therefore, it cannot be expected that setups with alternative precursors or distances were optimized simultaneously. One would have to perform a DoE for each approach to optimize each growth genuinely.

In summary, it can be concluded that the DoE has made large-scale growth possible. Unfortunately, the EBL mask did not yield enhanced negative growth. Nevertheless, any number of samples with millimeter-scale WS\textsubscript{2} can now be produced to fabricate PUFs. Yet, the optimal PUF size is still unknown and must be discovered by electrical measurements. Therefore the next chapter contains structuring and contacting of the samples.

\clearpage
\newpage

\subsection{Patterning and Contacting}
\label{sec:5.4_PatterningAndContacting}

\begin{figure}[b!]
\centering
\includegraphics[width=15cm]{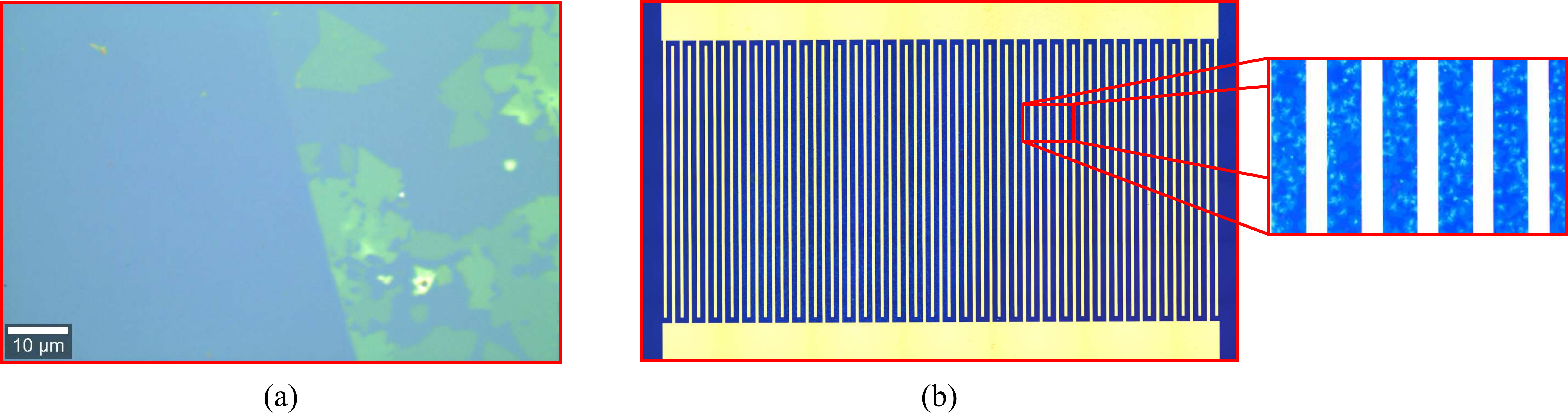}
\caption[A patterned and a contacted substrate]{A patterned substrate, etched with SF\textsubscript{6} in (a), showing a clear boundary between the blue SiO\textsubscript{2} and the light green WS\textsubscript{2}. Another sample, containing light to dark blue WS\textsubscript{2}, contacted with $10\,nm$ Ni + $60\,nm$ Au in an IDE structure in (b). Although the structure is extremely dense, only leakage currents could be measured.}
\label{fig:5.4_PatterningAndContacting}
\end{figure}

Finally, in order to be able to electrically evaluate the PUF, the 2D material has to be brought into a proper shape and contacts have to be deposited. A crop of a substrate etched with SF\textsubscript{6} can be found in figure \ref{fig:5.4_PatterningAndContacting} (a). Etching was performed using a \textit{TEPLA} etcher for $30\,s$ under a gas flow of $200\,sccm$ SF\textsubscript{6}, mixed with $500\,sccm$ O\textsubscript{2}. The power was $1000\,W$ and the pressure amounted to $200\,Pa$. On optical inspection, a clear boundary separating the light green TMD from the underlying blue SiO\textsubscript{2} substrate can be identified. In addition, the Raman signal and the PL signal of WS\textsubscript{2} disappear entirely in the etched area.

Prior to depositing the contacts, a PLT was performed. A similar recipe as in table \ref{tab:LithoMasksRecipes}, with negative ma-N 1420 resist was applied. Finally, $10\,nm$ nickel (Ni) + $60\,nm$ gold (Au) were vapor-deposited in an interdigitated-electrode (IDE) form, as can be observed in Figure \ref{fig:5.4_PatterningAndContacting} (b).

Unfortunately, only leakage currents could be measured, even at disproportionately high drain voltages. Possible reasons for this are high contact resistances, missing gate voltage, a broken material, or poor conductivity of the TMD. Nevertheless, high mobilities in WS\textsubscript{2} structures have already been measured in the past.[contactcite] The experimental part of this thesis ends here. To solve the difficulties encountered in the electrical measurements, further attempts can be performed, which will be discussed in the next and last chapter.

\clearpage
\newpage
\thispagestyle{plain}

\section{Conclusion and Outlook}

In this work, all the basic building blocks have been set to be capable of implementing the first-ever WS\textsubscript{2} PUF prototype. As preliminary work, the characteristic optical, Raman, and PL properties of the TMD were analyzed in chapter \ref{sec:GrowthCharacterization} and assigned to layer numbers one to five using AFM, in order to efficiently determine the average layer number on a vast covered area.

Subsequently, in \ref{sec:GrowthBehaviour}, the close proximity CVD growth of the 2D material was optimized via trial and error, thereby laying the foundation for the subsequent DoE in section \ref{sec:5DesignOfExperiment}. It was further concluded, that a solid metal precursor is more trustworthy than a liquid one and thus it was used for the DoE. Additionally, the behavior of the growth as a function of the distance of the microreactor was investigated and it was found that no distance works best.

In the next step, the reliability of the evaluation parameters for the DoE, defined in \ref{sec:DesignOfExperiment} was confirmed, the evaluation method was introduced, and an input growth model was proposed. Next, the experiments were scheduled and performed one after the other. The evaluation was finally done using the Cornerstone software and resulted in optimized process parameters. Finally, this process was executed several times and successfully resulted in large-scale and strongly interconnected WS\textsubscript{2} films in the mm range, thus setting the foundation for the PUF. Besides optimizing growth, this study also reveals insights into the nature of the CVD process.

Last but not least, a suitable method for structuring is demonstrated in \ref{sec:5.4_PatterningAndContacting}, followed by the first attempts to contact the material, which, however, encountered some problems and merely leakage current could be measured between the metals.

To further explore growth, the next step could be to test several patterns of different scales to yield negative growth and potentially optimize them with a DoE. Furthermore, it might be examined whether the $20\,nm$ distance with the HF dip results in enhanced growth.

To address the contacting problems, one could first employ EBL and start to contact individual flakes or a few coalesced groups and then gradually approach larger structures. After all, it has already been shown in the past that WS\textsubscript{2} is a conductive material and even possesses the highest theoretical mobility among all TMDs.\textsuperscript{\cite{Sebastian.2021}} Once enough current flows, one can start concerning the structure of the PUF. Either simple contact pairs or somewhat more advanced structures can be implemented, as depicted in figure \ref{fig:2.3_CNT_PUF}. If the contacts are not equidistant to each other, the measured amplitudes can be normalized with respect to the distance of the measured contacts to increase the unpredictability. However, one can also design a structure with an arbitrary number of contacts, equidistant from each other. The contacts must be arranged in a circle around a center and the underlying 2D material has to be structured in such a way that the contacts only touch the 2D material at a central point.

Once a functional device has been implemented, it must be examined and evaluated with respect to PUF characteristics. To meet the criteria of a Strong PUF, the CRPs can be additionally upscaled by a gate contact that is expected to additionally influence the channel current. Then, a suitable threshold voltage must be defined from a large number of devices. Thus, it can be benchmarked and compared to the similar designs from table \ref{tab:CNT_vs_2D} and the CNT PUF from \cite{JonasSchroder.2022}. Finally, it may be examined whether it is a worthy candidate to face the challenges of the IoT.

\clearpage
\newpage
\thispagestyle{plain}

\small
\begin{singlespacing}
\bibliography{Literature}
\addcontentsline{toc}{section}{References}
\end{singlespacing}

\clearpage
\newpage
\pagenumbering{alph}
\thispagestyle{plain}

\appendix
\section{Appendix}
\subsection{Scanning Electron Microscopy}

\begin{figure}[h!]
\centering
\includegraphics[width=7cm]{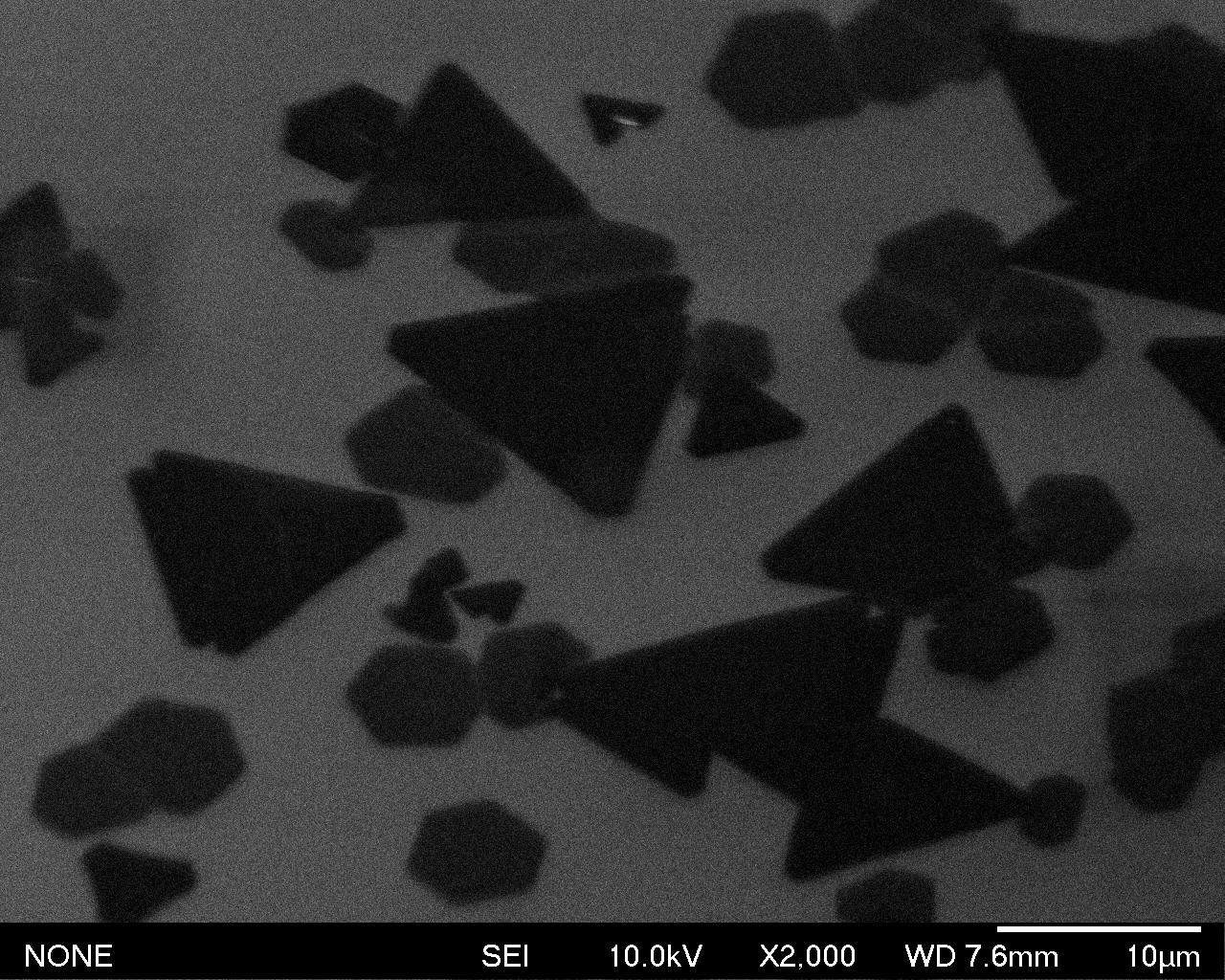}
\caption[A SEM image of WS\textsubscript{2} flakes]{A SEM image of WS\textsubscript{2} flakes.}
\label{fig:A_SEM}
\end{figure}

Figure \ref{fig:A_SEM} provides a section of a sample containing WS\textsubscript{2} acquired with an SEM. Once again, it can be seen that the flakes are present in both triangular and hexagonal shapes. In addition, the hexagonal flakes tend to be brighter and thus possess a lower layer number. An AFM comparison measurement showed that the hexagonal flakes in this image are predominantly monolayer. It can further be observed that a larger flake means a higher layer number. The image was recorded with a \textit{JEOL JSM-6700F} Field Emission SEM.

\end{document}